\documentclass[letterpaper,twoside,english,twocolumn, tighten,appendixfloats,twocolappendix]{aastex631}
\usepackage[T1]{fontenc}
\usepackage{color}
\usepackage{array}
\usepackage{multirow}
\usepackage{amstext}
\usepackage{graphicx}

\makeatletter

\pdfpageheight\paperheight
\pdfpagewidth\paperwidth

\providecommand{\tabularnewline}{\\}

\usepackage{apjfonts}
\usepackage{amsmath}
\usepackage[none]{hyphenat}
\hypersetup{colorlinks=true,citecolor=blue,urlcolor=blue,linkcolor=blue}
\shorttitle{Mg II Doublet Ratio}
\shortauthors{SEON}

\DeclareSymbolFont{CMletters}{OML}{cmm}{m}{it}
\DeclareMathSymbol{\nu}{\mathord}{CMletters}{23}
\DeclareMathSymbol{v}{\mathord}{CMletters}{`v}
\DeclareMathSymbol{w}{\mathord}{CMletters}{`w}
\DeclareMathSymbol{g}{\mathord}{CMletters}{`g}

\makeatother

\usepackage{babel}
\begin{document}

\title{On the Doublet Flux Ratio of \ion{Mg}{2} Resonance Lines in and
Around Galaxies}

\author[0000-0001-9561-8134]{Kwang-il Seon}

\affiliation{Korea Astronomy \& Space Science Institute, 776 Daedeokdae-ro, Yuseong-gu,
Daejeon 34055, Republic of Korea; kiseon@kasi.re.kr}

\affiliation{Astronomy and Space Science Major, University of Science and Technology,
217, Gajeong-ro, Yuseong-gu, Daejeon 34113, Republic of Korea}
\begin{abstract}
Observations of metallic doublet emission lines, particularly \ion{Mg}{2}
$\lambda\lambda2796,\,2803$, provide crucial information for understanding
galaxies and their circumgalactic medium. This study explores the
effects of resonant scattering on the \ion{Mg}{2} doublet lines and
the stellar continuum in spherical and cylindrical geometries. Our
findings show that under certain circumstances, resonance scattering
can cause an increase in the doublet flux ratio and the escaping flux
of the lines beyond what are expected in optically thin spherical
media. As expected, the doublet ratio is consistently lower than the
intrinsic ratio when the scattering medium is spherically symmetric
and dusty. However, if the scattering medium has a disk shape, such
as face-on disk galaxies, and is viewed face-on, the doublet ratio
is predicted to be higher than two. It is also shown that doublet ratios as low as those observed in compact star-forming galaxies cannot be explained solely by pure dust attenuation of intrinsic \ion{Mg}{2} emission lines in spherical models unless dust opacity deviates markedly from that expected based on the dust-to-Mg$^{+}$ gas ratio of our Galaxy.
The importance of the continuum-pumped emission lines and expanding
media is discussed to understand observational aspects, including
doublet flux ratios, which can be lower than 1.5 or higher than two,
as well as symmetric or asymmetric line profiles. It is also discussed
that the diffuse warm neutral medium may be an important
source of Mg II emission. These results provide
insight into the complexity of the shape and orientation of distant,
spatially-unresolved galaxies.
\end{abstract}


\section{INTRODUCTION}

The majority of advances in the circumgalactic medium (CGM) study
have been achieved by observing ultraviolet resonance lines, such
as Ly$\alpha$ $\lambda1216$, \ion{Mg}{2} $\lambda\lambda2796,\:2803$,
and \ion{C}{4} $\lambda\lambda1548,\:1551$, which are some of the
most prominent lines in the spectra produced by the interstellar medium
(ISM) and CGM. \citet{Bahcall1969} put forward the idea that most
quasar absorption lines can be attributed to gas present in the extended
halos of normal galaxies, which has a larger cross-sectional area
indicated by the galaxy's optical or radio appearance. Since then,
resonance lines have been extensively employed for studying the CGM.

In particular, \ion{Mg}{2} absorption or emission lines have been
demonstrated to be effective in tracing gas in galaxies and their
surroundings \citep{Shapley2003,Weiner2009,Rubin2011,Erb2012,Guseva2013,Martin2013,Schroetter2015,Finley2017,Feltre2018,Huang2021,LeeJC2021,Xu2022,Xu2023}.
The studies by \citet{Finley2017} and \citet{Feltre2018} demonstrated
that the manifestation of the \ion{Mg}{2} doublet, either as emission
lines or absorption lines, is dependent on the stellar masses and
ultraviolet (UV) spectral slopes of galaxies. The
studies revealed that galaxies with lower stellar masses and bluer
spectral slopes tend to exhibit \ion{Mg}{2} emission, while galaxies
with higher stellar masses and redder spectral slopes tend to show
\ion{Mg}{2} absorption. \citet{Rubin2011} made the first discovery
of spatially extended \ion{Mg}{2} emission from a bright, starburst
galaxy at $z=0.69$. This discovery was later confirmed by \citet{Burchett2021}
using spatially resolved spectroscopy. In recent years, advances in
integral field unit spectrograph technology have allowed for the measurement
of spatially resolved \ion{Mg}{2} emission in the CGMs of star-forming
galaxies \citep{Rupke2019,Chisholm2020,Burchett2021,Zabl2021,Seive2022,Shaban2022,Dutta2023,Leclercq2024}
and the intragroup medium \citep{Leclercq2022}. The observations
of the extended emission provide strong constraints not only on the
spatial extent of the outflowing gas but also on the mass-outflow
rates of the galaxy when combined with outflow velocity and column
density measurements.

Resonance lines like \ion{Mg}{2} are dispersed in both space and
wavelength as a result of repeated resonance scatterings. Furthermore,
the 2796\AA\ line, with its shorter wavelength, has a higher resonance
scattering cross-section than the 2803\AA\ line. This increases the
probability of the former being absorbed by dust more than that of
the latter. These characteristics make it possible to use the spectral
shape and flux ratio ($F_{2796}/F_{2803}$) of the \ion{Mg}{2} doublet
lines as proxies of the physical conditions of the ISM and CGM. In
particular, it may serve as an indirect indicator of Lyman continuum
(LyC) leakage \citep{Henry2018,Chisholm2020,Katz2022,Xu2023,Leclercq2024}. \citet{Henry2018}
discovered that in green pea galaxies, \ion{Mg}{2} line profiles
tend to be wider and more redshifted when the estimated escape fractions
for Ly$\alpha$ and \ion{Mg}{2}  are low. This suggests that the
escape fractions and line profiles of Ly$\alpha$
and \ion{Mg}{2}  are influenced by resonance scattering. Hence, the
suggestion has been made to utilize the flux ratio between the \ion{Mg}{2}
emission lines as a potential indirect measure for estimating the
escape fraction of LyC in the epoch of reionization. In their study
of the spatially resolved spectroscopic data of \ion{Mg}{2} in a
LyC leaking galaxy located at $z=0.36$, \citet{Chisholm2020} found
that the flux ratio $R=F_{2796}/F_{2803}$ ranges from 0.8 to 2.7
across the galaxy. It was discussed that $R$ would decrease as the
\ion{Mg}{2} optical depth along the line of sight increases, suggesting
that LyC photons escape through regions of high $R$. They also found
that the \ion{Mg}{2} 2796\AA\ line tends to be slightly broader
than the \ion{Mg}{2} 2803\AA, particularly in regions with high
$R$; this observation suggest the involvement of resonance scattering.
They and \citet{Xu2022} also found that the anticipated LyC escape
fraction, derived from the \ion{Mg}{2} emission lines, shows a correlation
with the observed fraction in samples of galaxies exhibiting strong
\ion{Mg}{2} emission lines. More recently, \citet{Xu2023} found
that strong LyC emitters (LCEs) tend to exhibit larger equivalent
widths (EWs) of \ion{Mg}{2} emission, while non-LCEs show evidence
of more scattering and absorption features in \ion{Mg}{2}.
\cite{Leclercq2024} found that extended \ion{Mg}{2} emission tends to be associated with non- or weak LyC leakers.

Achieving precise modeling of resonance radiative transfer (RT) processes
is crucial to gain a proper understanding of the observational results
of \ion{Mg}{2}. \citet{Prochaska2011} used Monte Carlo RT techniques
to study the propagation of metallic resonance doublet lines, specially
the \ion{Mg}{2} $\lambda\lambda2796$, 2803 doublet and \ion{Fe}{2}
UV1 multiplet at $\lambda\approx2600$\AA. \citet{Scarlata2015}
also developed a semi-analytical line transfer model to examine the
absorption and re-emission line profiles from expanding galactic envelopes.
\citet{Michel-Dansac2020} developed a 3D RT code RASCAS, which can
be used to model the propagation of various resonant lines in numerical
simulations of astrophysical objects. \citet{Burchett2021} utilized
the code developed by \citet{Prochaska2011} to compare the RT models
with observational data from a star-forming galaxy. \citet{Katz2022}
investigated the potential of \ion{Mg}{2} as an indicator of LyC
leakage by analyzing cosmological radiation hydrodynamics simulations
with photoionization and resonant-line RT. The study found that a
majority of bright, star-forming galaxies with high LyC escape fractions
are likely to also emit \ion{Mg}{2}. They also found that the \ion{Mg}{2}
doublet flux ratio is more sensitive to the amount of dust than to
neutral hydrogen, which may limit its usefulness as a LyC leakage
indicator to only galaxies in the optically thin regime. \citet{Nelson2021}
studied theoretical predictions for \ion{Mg}{2} emission in the CGM
of star-forming galaxies in the high-resolution TNG50 cosmological
simulations. However, no resonance scattering was considered.

As stated above, despite some theoretical efforts, there has been
sparse fundamental modeling conducted thus far to comprehend the escape
fraction and the line flux ratio of the \ion{Mg}{2} resonance doublet.
The primary aim of the present study is to establish a core understanding
of how to interpret observational data of the Mg II emission line,
especially through a comparison of spherical and non-spherical geometries.
This study can also be extended to similar data from other metallic
doublet lines, such as \ion{C}{4}. The occurrence of absorption or
P-Cygni profiles in relatively luminous galaxies implies that there
is an involvement of resonance scattering from the stellar continuum
radiation near the \ion{Mg}{2} lines. It is therefore investigated
how continuum photons produce the \ion{Mg}{2} emission and absorption
line features. The main objective of this study is not to provide
a detailed comparison between the model predictions and observations,
but rather to present the general properties of \ion{Mg}{2} RT.

This paper is organized as follows. Section \ref{sec:2} describes
the Monte Carlo RT methods, definitions, and assumptions. Section
\ref{sec:3} presents the simulation results for the emission lines
in both spherical and cylindrical mediums. The results for the continuum-pumped
emission and absorption line features are also discussed in both geometries.
The spatial variation of the doublet flux ratio is examined for sample
models in the spherical geometry. Section \ref{sec:4} discusses the
observational implications of the present results, \ion{Mg}{2} emission
mechanisms, other metallic resonance lines, and related subjects.
A summary is provided in Section \ref{sec:5}.

\section{Methods}

\label{sec:2}

\subsection{Definitions}

The fine structures of the $n=3$ quantum state of Mg$^{+}$ resemble
the Ly$\alpha$ doublet state of neutral hydrogen. However, unlike
Ly$\alpha$, their level splittings are significant and, thus, must
not be disregarded. We refer the transition $^{2}S_{1/2}\leftrightarrow{}^{2}P_{1/2}^{{\rm o}}$
(corresponding to the lower frequency) to as ``H'' and the transition
$^{2}S_{1/2}\leftrightarrow{}^{2}P_{3/2}^{{\rm o}}$ (higher frequency)
to as ``K,'' as in the \ion{Ca}{2} $\lambda\lambda3933,3968$ doublet
lines. The frequencies for the H and K transitions are represented
by $\nu_{{\rm H}}$ and $\nu_{{\rm K}}$, respectively. By integrating
over the one-dimensional Maxwellian velocity distribution of the ionic
gas at temperature $T$, the cross section in a reference frame comoving
with the gas fluid results in
\begin{equation}
\sigma_{\nu}=\frac{1}{\sqrt{\pi}\Delta\nu_{{\rm D}}}\left[\chi_{{\rm K}}H(x,a)+\chi_{{\rm H}}H(x+x_{{\rm HK}},a)\right],\label{eq1}
\end{equation}
where $\chi_{{\rm K}}=f_{{\rm K}}(\pi e^{2}/m_{e}c)$ and $\chi_{{\rm H}}=f_{{\rm H}}(\pi e^{2}/m_{e}c)$.
Here, $f_{{\rm K}}=0.608$ and $f_{{\rm H}}=0.303$ are the oscillator
strengths of the \ion{Mg}{2} K and H lines, respectively. The K transition
has a twice-higher oscillator strength than the H transition due to
the difference in the statistical weights of 2J + 1. $H(x,a)$ is
the Voigt-Hjerting function given by
\begin{align}
H(x,a) & =\frac{a}{\pi}\int_{-\infty}^{\infty}\frac{e^{-y^{2}}}{(x-y)^{2}+a^{2}}dy.\label{eq2}
\end{align}
In this paper, $x$ is defined as the relative frequency of the photon
measured from $\nu_{{\rm K}}$ and normalized to the thermal Doppler
frequency width $\Delta\nu_{{\rm D}}=\nu_{{\rm K}}(V_{{\rm th}}/c)$:
\begin{align}
x & =(\nu-\nu_{{\rm {\rm K}}})/\Delta\nu_{{\rm D}},\label{eq:3}
\end{align}
Here, $V_{{\rm th}}=(2k_{{\rm B}}T/m_{\text{Mg}})^{1/2}=8.27(T/10^{5}\ {\rm K})^{1/2}$
km s$^{-1}$ and $a=\Gamma/(4\pi\Delta\nu_{{\rm D}})$ are the thermal
speed of gas and the natural width parameter of $H(x,a)$, respectively.
The damping constant (the Einstein A coefficient) of the \ion{Mg}{2}
transitions is $\Gamma=2.590\times10^{8}$ s$^{-1}$. If an additional
turbulence motion, characterized by $V_{{\rm turb}}$, is taken into
account, the Doppler parameter is given by $\Delta\nu_{{\rm D}}=\nu_{{\rm K}}(b/c)$
where $b=(V_{{\rm th}}^{2}+V_{{\rm turb}}^{2})^{1/2}$. The frequency
difference between the \ion{Mg}{2} K and H lines is $\Delta\nu_{{\rm HK}}=\nu_{{\rm K}}-\nu_{{\rm H}}=2745.2$
GHz, which is equivalent to a Doppler shift of $\sim770$ km s$^{-1}$.
The normalized frequency difference between $\nu_{{\rm K}}$ and $\nu_{{\rm H}}$
is $x_{{\rm HK}}=\Delta\nu_{{\rm HK}}/\Delta\nu_{{\rm D}}\simeq93(T/10^{5}\ {\rm K})^{-1/2}$
for \ion{Mg}{2}. For comparison, it's worth noting that $x_{{\rm HK}}\simeq0.032(T/10^{5}\ {\rm K})^{-1/2}$
for Ly$\alpha$. Therefore, in most cases, the \ion{Mg}{2} doublet
transitions can be treated separately, unless there is a significant
velocity variation in the gas, as considerable as $\sim770$ km s$^{-1}$.

The optical depth $\tau_{\nu}(s)$ of a photon with frequency $\nu$
traveling along a path length $s$ is given by
\begin{equation}
\tau_{\nu}(s)=\int_{0}^{s}\int_{-\infty}^{\infty}n(V_{\parallel})\sigma_{\nu}dV_{\parallel}d\ell,\label{eq:4}
\end{equation}
where $n(V_{\parallel})$ represents the number density of Mg$^{+}$
with the velocity component $V_{\parallel}$ parallel to the photon's
propagation direction. In this paper, the total amount of the Mg$^{+}$
gas is measured using the column density $N_{\text{Mg}^{+}}$ or the
optical depth $\tau_{0}$ at the K line center.

\subsection{Monte Carlo Algorithms}

The RT calculation of \ion{Mg}{2} was carried out by updating LaRT,
which was originally developed for Ly$\alpha$ RT. A detailed description
of the basic RT algorithms employed LaRT can be found in \citet{Seon2020},
\citet{Seon2022} and \citet{Yan2022}. LaRT has been updated to deal
with metallic resonance lines other than Ly$\alpha$ and fluorescence
emission lines caused by resonant absorption, such as \ion{Fe}{2}$^{*}$
2626\AA\ and \ion{Si}{2}$^{*}$ 1533\AA. The RT algorithms are
similar to those of \citet{Prochaska2011} and \citet{Michel-Dansac2020},
except that LaRT can handle scattering and polarization using a quantum-mechanically
correct scattering phase function. In contrast, their codes assume
the scattering phase function to be isotropic. A complete description
of the update will be given elsewhere. In the following, only the
contents relevant to the \ion{Mg}{2} doublet lines are described.

The velocity component $u_{\parallel}=V_{\parallel}/V_{{\rm th}}$
(or $u_{\parallel}=V_{\parallel}/b$) of the scattering atom, which
is parallel to the photon's propagation direction, is sampled from
the following composite distribution function:
\begin{equation}
f_{{\rm FS}}(u_{\parallel}|x)=\mathcal{P}_{{\rm K}}f(u_{\parallel}|x)+(1-\mathcal{P}_{{\rm K}})f(u_{\parallel}|x+x_{{\rm HK}}),\label{eq:5}
\end{equation}
where
\begin{align}
\mathcal{P}_{{\rm K}} & =\frac{2H(x,a)}{2H(x,a)+H(x+x_{{\rm HK}},a)},\label{eq:6}\\
f(u_{\parallel}|x) & =\frac{a}{\pi H(x,a)}\frac{e^{-u_{\parallel}^{2}}}{a^{2}+(x-u_{\parallel})^{2}}.\label{eq:7}
\end{align}
If a uniform random number $\xi$ ($0\le\xi\le1$) is selected and
found to be smaller than $\mathcal{P}_{{\rm K}}$, the photon is scattered
through the K transition; otherwise, it is scattered through the H
transition. The algorithm developed by \citet{Seon2020} is used to
obtain a random parallel velocity component of the scattering atom,
given a specific transition type.

The scattering phase function is given by
\begin{eqnarray}
\mathcal{P}\left(\cos\theta\right) & = & \frac{3}{4}E_{1}\left(\cos^{2}\theta+1\right)+\left(1-E_{1}\right)\label{eq:8}
\end{eqnarray}
as for Ly$\alpha$. Here, $E_{1}$ is the function of frequency given
by

\begin{eqnarray}
E_{1} & = & \frac{2\left(\nu-\nu_{{\rm K}}\right)\left(\nu-\nu_{{\rm H}}\right)+\left(\nu-\nu_{{\rm H}}\right)^{2}}{\left(\nu-\nu_{{\rm K}}\right)^{2}+2\left(\nu-\nu_{{\rm H}}\right)^{2}}\nonumber \\
 & = & \frac{2x(x+x_{{\rm KH}})+(x+x_{{\rm KH}})^{2}}{x^{2}+2(x+x_{{\rm KH}})^{2}}.\label{eq:9}
\end{eqnarray}
The parameter $E_{1}$ is 1/2 for the K and 0 for the H transition.
Random numbers for the scattering angle $\theta$ are obtained following
the method described by \citet{Seon2022}. In this study, we will
not address the calculation of polarization for the \ion{Mg}{2} resonance
lines, although LaRT has the capability to perform such calculations.

To examine optically thin or moderately thick cases similar to those
encountered in the \ion{Mg}{2} lines, the first forced-scattering
algorithm was implemented in LaRT. In regions with low optical depth,
most photon packets will escape from the system without interactions,
resulting in poor statistics of the scattered light. The technique
of forced scattering offers a solution to overcome this low efficiency.
The forced scattering technique has been incorporated into the majority
of Monte Carlo dust RT codes \citep[e.g.,][]{Gordon2001,Baes2011,Steinacker2013,Seon2014}.
This technique employs a photon weight $w$ that is initially set
to 1. Instead of sampling from the standard exponential probability
density function (PDF) $p(\tau)=\exp(-\tau)$, a random optical depth
$\tau$ is generated by following a ``truncated'' exponential distribution.
This distribution is truncated at the optical depth $\tau_{{\rm path}}$,
which is calculated from the current position of the photon to the
system boundary along its trajectory, as follows:
\begin{equation}
p(\tau)=\begin{cases}
e^{-\tau}\left(1-e^{-\tau_{\text{path}}}\right)^{-1} & \tau\leq\tau_{{\rm path}},\\
0 & \tau>\tau_{{\rm path}}
\end{cases}\label{eq:10}
\end{equation}
A random optical depth is sampled as follows:
\begin{equation}
\tau=-\ln\left[1-\xi\left(1-e^{-\tau_{{\rm path}}}\right)\right],\label{eq:11}
\end{equation}
where $\xi$ is a uniform random number between 0 and 1. The truncated
PDF guarantees to produce an interaction before the photon exits the
system. To compensate for this biasing of the PDF, the photon weight
$w$ is reduced to 
\begin{equation}
w'=w\left(1-e^{-\tau_{{\rm path}}}\right),\label{eq:12}
\end{equation}
 where $w$ and $w'$ are the photon weights before and after the
scattering, respectively. \citet{Baes2011} forced the scattering
until the photon weight becomes lower than a predefined critical value.
In the present study, forcing is limited only to the first scattering,
while subsequent scattering is carried out using the standard exponential
PDF without a cutoff.

\subsection{Column Density, Optical Depth, and Doppler Parameter}

The column density of Mg$^{+}$ can be expressed in terms of the hydrogen
column density $N_{\text{H}}$, as done by \citet{Chisholm2020}:

\begin{align}
N_{\text{Mg}^{+}} & =3.08\times10^{13}\left(\frac{N_{\text{Mg}^{+}}}{N_{\text{Mg}}}\right)\left(\frac{\delta_{\text{Mg}}}{0.426}\right)\left(\frac{N_{\text{Mg}}/N_{\text{O}}}{0.0813}\right)\nonumber \\
 & \ \ \ \ \ \times\left(\frac{N_{\text{O}}/N_{\text{H}}}{8.91\times10^{-5}}\right)\left(\frac{N_{\text{H}}}{10^{19}\ \text{cm}^{-2}}\right)\ \text{cm}^{-2}.\label{eq:13}
\end{align}
Unlike the Mg/H abundance ratio, the O/H ratio is more readily observable.
Therefore, the column density of Mg$^{+}$ is parameterized based
on the O abundance. The abundance ratio between Mg and O is expected
to be similar to that of the sun \citep{Asplund2009} and does not
significantly change because both are primarily produced by core-collapse
supernovae \citep{Johnson2019}. The gas phase abundance of Mg is
depleted onto dust and reduced by a factor of $\delta_{\text{Mg}}\sim0.426$
(corresponding to $\sim$0.37 dex) in the warm neutral medium (WNM)
with a temperature of $\sim10^{4}$ K \citep{Jenkins2009}. In this
paper, we use the O/H of a LyC leaker J1503+3644 (hereafter J1503)
from \citet{Izotov2016}, which is $\sim1/5$ of the solar abundance,
to parameterize the Mg/H ratio. We note that the mean O/H abundance
ratio of eleven LyC emitters at $z\sim0.3-0.4$, as discussed in \citet{Ramambason2020},
is $\sim8.13\times10^{-5}$, which is consistent with the value adopted
in this study.

\citet{Chisholm2020} estimated the width of the \ion{Mg}{2} emission
line profile in J1503 to be $\sim90$ km s$^{-1}$ after subtracting
the instrumental effects. The full width of half maxium (FWHM) of
\ion{Mg}{2} 2796 from green pea galaxies, as reported by \citet{Henry2018},
was found to range from approximately 100 km s$^{-1}$ to 300 km s$^{-1},$
corresponding to $b\approx60-180$ km s$^{-1}$. Their research suggests
that the resonance scattering effect has a relatively minor impact,
implying that the observed line widths are primarily attributed to
gas motion, including both ISM turbulence and galactic rotation. \citet{Seon2020}
demonstrated that the prescription of incorporating the trubulent
motion into the thermal motion provides an excellent method for predicting
the Ly$\alpha$ emergent spectrum from a turbulent medium in both
cases of microturbulence and macroturbulence. The results would certainly
be applicable to \ion{Mg}{2} line. This study, therefore, assumes
a Doppler width parameter of $b=90$ km s$^{-1}$ for the Mg$^{+}$
gas unless stated otherwise. The initial line profile adopted in all
our models is assumed to be the Voigt function, with a line width
determined by the same Doppler parameter $b=90$ km s$^{-1}$. Unresolved
turbulence motion will give rise to an effect equivalent to producing
an initial line profile corresponding to the gas motion. Although
not explicitly presented in the paper, additional models were computed
using $b=15$ km s$^{-1}$ (equivalent to $T\sim3.3\times10^{5}$
K) and yielded comparable results to those presented.

The optical depth $\tau_{0}$ of the Mg$^{+}$ gas is then given by
\begin{align}
\tau_{0} & \equiv\sigma_{0}N_{\text{Mg}^{+}}\nonumber \\
 & =\left(\chi_{{\rm K}}/\Delta\nu_{{\rm D}}\right)N_{\text{Mg}^{+}}\phi_{x}(0)\simeq\left(\chi_{{\rm K}}/\sqrt{\pi}\Delta\nu_{{\rm D}}\right)N_{\text{Mg}^{+}},\nonumber \\
 & =0.871\left(\frac{90\ \text{km}\ \text{s}^{-1}}{b}\right)\left(\frac{N_{\text{Mg}^{+}}}{3.08\times10^{13}\ \text{cm}^{-2}}\right),\label{eq:14}
\end{align}
where $\sigma_{0}\simeq\chi_{{\rm K}}/(\sqrt{\pi}\nu_{{\rm D}})$
represents the cross section at the center of the K line, $\phi_{x}=H(x,a)/\sqrt{\pi}$
is the normalized Voigt profile, and $H(0,a)\simeq1$. This definition
of $\tau_{0}$ refers to the monochromatic optical depth measured
at the center of the K line ($x=0$). The optical depth at the H line
center is $\tau_{0}/2$. In this study, the optical depth varies in
the range of $\tau_{0}\approx3\times10^{-3}-10^{3}$ and the column
density $N_{\text{Mg}^{+}}\approx10^{11}-3\times10^{16}$ cm$^{-2}$.

It should be noted that in \citet{Chisholm2020}, the optical depth
is defined as that integrated over the H line profile. The integrated
optical depth ($\tau_{*}$) for the H line is related to the monochromatic
one at the K line center by $\tau_{*}\simeq\sqrt{\pi}\tau_{0}/2$
\citep[e.g.,][]{Seon2020}. Furthermore, it is important to note that,
in their Eq.\ (12), they assumed a Doppler parameter of $b=1$ km
s$^{-1}$. Consequently, their column density is 90 times lower than
ours for a given optical depth.

\subsection{Dust Extinction}

For photons that can be resonantly trapped, such as Ly$\alpha$ and
\ion{Mg}{2}, the influence of dust can be considerably amplified
in comparison to the non-resonance line photons, such as [\ion{O}{3}]
$\lambda5008$. Resonantly trapped photons will travel significantly
greater distances before escaping the medium compared to non-resonance
photons. As a result, they may experience significantly higher opacity
due to dust. In \citet{Prochaska2011}, it was assumed that dust only
absorbs photons without scattering them. However, the present study
also considers the scattering of photons by dust.

The dust effect is examined by assuming the properties of mean Milky
Way dust. The dust scattering albedo and asymmetry factor near the
wavelength of \ion{Mg}{2} are $a=0.57$ and $g=0.55$, respectively
\citep{Weingartner2001,Draine2003}. The scattering angle is sampled
following the Henyey-Greenstein phase function with the asymmetry
factor $g$ \citep{Witt1977}. The dust extinction cross section per
hydrogen atom is $\sigma_{{\rm ext}}/{\rm H}\simeq1\times10^{-21}$
cm$^{-2}$ at the wavelength of \ion{Mg}{2}. The quantity of dust
is assumed to be proportional to the Mg abundance in the same manner
as in the Milky Way. The dust extinction optical depth is then given
by
\begin{align}
\tau_{{\rm dust}} & =1.73\times10^{-3}\left(\frac{\sigma_{{\rm ext}}/\text{H}}{1\times10^{-21}{\rm cm}^{-2}}\right)\left(\frac{1.78\times10^{-5}}{N_{{\rm \text{Mg}}}/N_{{\rm H}}}\right)\nonumber \\
 & \ \ \ \ \ \ \ \times\left(\frac{N_{{\rm \text{Mg}}}}{N_{\text{Mg}^{+}}}\right)\left(\frac{N_{\text{Mg}^{+}}}{3.08\times10^{13}\ \text{cm}^{-2}}\right).\label{eq:15}
\end{align}
in all models, except in some models of Figure \ref{fig03},
where $\tau_{{\rm dust}}$ is three times the value indicated by this
relation to explore the enhanced dust effect. Here, we adopt the
Mg abundance of Mg/H $=1.78\times10^{-5}$, measured in the WNM of
our Galaxy \citep{Jenkins2009}. In the present models, the dust extinction
optical depth varies from $\tau_{\text{dust}}\sim6\times10^{-6}$
to $\sim2$. It is, therefore, anticipated that the impact of dust
extinction would be significant only when the Mg$^{+}$ column density
reaches $\gtrsim10^{15}$ cm$^{-2}$. \ion{Mg}{2} lines observed
from compact galaxies akin to J1503 would likely not have experienced
substantial dust attenuation effects, as will be elaborated upon later.

\begin{figure}[t]
\begin{centering}
\includegraphics[clip,scale=0.6]{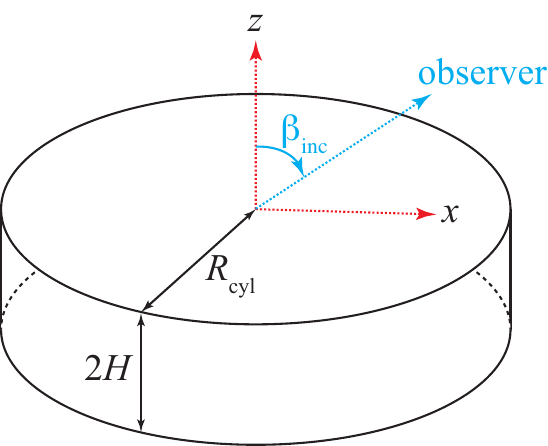}
\par\end{centering}
\medskip{}
\caption{\label{fig01}Cylindrical geometry for the model calculation. Here,
$H$ and $R_{{\rm cyl}}$ denote the height, measured from the center,
and radius of the cylinder, respectively, and $\beta_{{\rm inc}}$
is the inclination angle (or viewing angle). The face-on and edge-on
views correspond to $\beta_{{\rm inc}}=0^{\circ}$ and $90^{\circ}$,
respectively. The optical depth for the cylindrical
model is measured along the vertical direction from the center of
the cylinder and is given by $\tau_{0}=n_{\text{Mg}^+}\sigma_{0}H$,
where $n_{\text{Mg}^+}$ is the number density of Mg$^+$, and $\sigma_{0}$
is the cross section at the center of the K line. The optical depth
along the radial direction is $\tau_{0}(R_{{\rm cyl}}/H)$.}
\end{figure}

\begin{table}[!t]
\caption{\label{tab01}Model types}

\begin{centering}
\begin{tabular}{l|c|c|c}
\hline \hline
\multirow{2}{*}{Geometry} & Velocity & Input & Optical\tabularnewline
 & field & spectrum & depth ($\tau_{0}$)\tabularnewline
\hline 
 & \multirow{2}{*}{static} & line & \multirow{4}{*}{$n_{\text{Mg}^+}\sigma_{0}R_{{\rm sph}}$}\tabularnewline
\cline{3-3} 
spherical medium &  & continuum & \tabularnewline
\cline{2-3} \cline{3-3} 
+ central source & \multirow{2}{*}{Hubble-like} & line & \tabularnewline
\cline{3-3} 
 &  & continuum & \tabularnewline
\hline 
cylindrical medium & \multirow{2}{*}{static} & line & \multirow{2}{*}{$n_{\text{Mg}^+}\sigma_{0}H$}\tabularnewline
\cline{3-3} 
+ uniform source &  & continuum & \tabularnewline
\hline 
\end{tabular}
\par\end{centering}
\tablecomments{The optical depth ($\tau_0$) for the spherical model is measured along the radial diection, but that of the cylindrical model is measured along the vertical direction from the center. The input line spectrum is assumed to have a line width corresponding to a Doppler parameter of $b=90$ km s$^{-1}$. In the models, the input spectrum is either a pure intrinsic emission line or a pure continuum.}
\end{table}

\begin{figure*}[t]
\begin{centering}
\includegraphics[clip,scale=0.5]{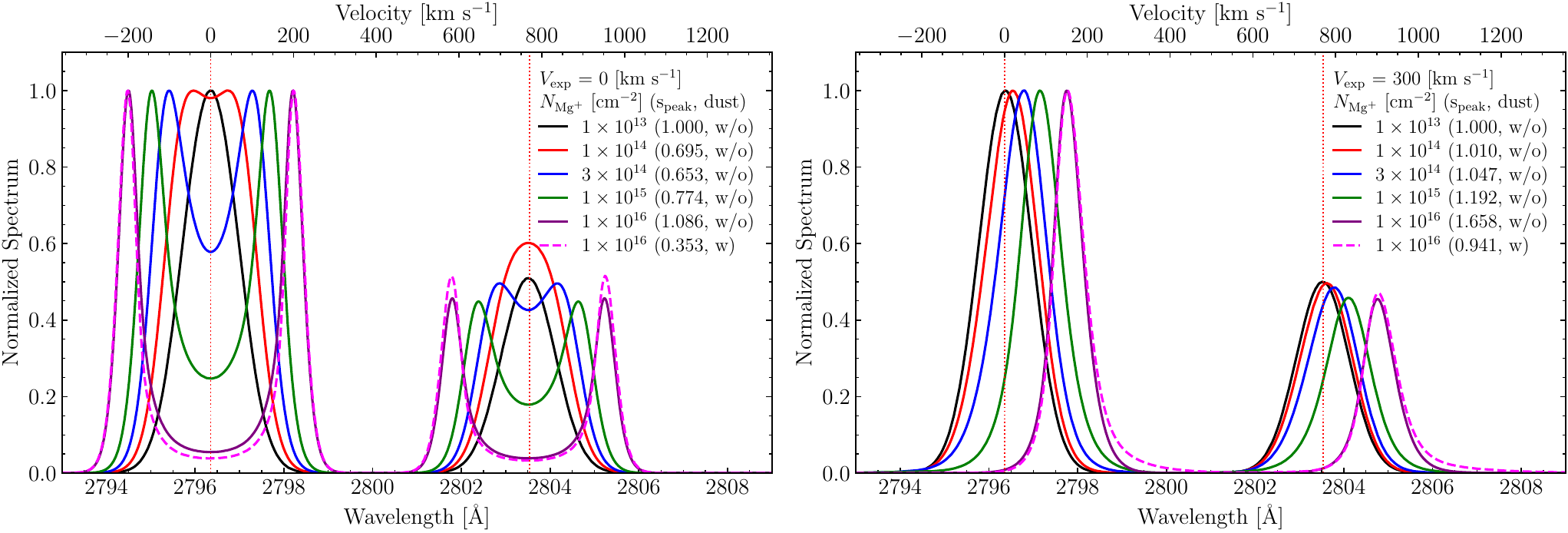}
\par\end{centering}
\begin{centering}
\medskip{}
\par\end{centering}
\caption{\label{fig02} Displayed are the \ion{Mg}{2} emission line spectra
predicted for the spherical models. The left panel exihibits spectra
for a static medium, while the right panel shows those for an expanding
medium with $V_{{\rm exp}}=300$ km s$^{-1}$. In the figures, different
colors denote the varying column density of Mg$^{+}$ ranging from
$10^{^{13}}$ to $10^{16}$ cm$^{-2}$. The peaks of the spectra were
normalized to a value of one. In the parentheses of the legends, the
numbers ${\rm s}_{{\rm peak}}$ represent the peak levels of individual
models measured relative to that of the model with $N_{\text{Mg}^{+}}=10^{13}$
cm$^{-2}$, before normalizing the peaks to 1. The terms ``w'' and
``w/o'' denote the cases with dust and without dust, respectively.
The wavelength is also shown in terms of velocity relative to the
K line center. The red vertical dotted lines denote the line centers
of the \ion{Mg}{2} K and H lines. The spectra obtained for the models
with dust are shown in dashed lines.}
\medskip{}
\end{figure*}

\subsection{Model Types}

This paper explores two fundamental geometries (sphere and cylinder).
The first model to be examined assumes a spherically symmetric medium
with constant density, which can be either static
or expands radially. The radial velocity of a fluid element at a
distance $r$ from the center is assumed to be
\begin{equation}
V(r)=V_{{\rm exp}}\frac{r}{R_{{\rm sph}}},\label{eq:16}
\end{equation}
where $R_{{\rm sph}}$ and $V_{{\rm exp}}$ are the maximum radius
and the velocity at $r=R_{{\rm sph}}$, respectively. In this Hubble-like
expansion model, the maximum velocity varies from $V_{{\rm exp}}=0$
to 300 km s$^{-1}$. The optical depth $\tau_{0}$ and column density
$N_{\text{Mg}^{+}}$ are measured radially from the center of the
sphere to the outer boundary.

In the second model, the scattering medium has a cylindrical shape
with a height of $2H$ and a radius of $R_{{\rm cyl}}$, as illustrated
in Figure \ref{fig01}. A medium characterized by a low height-to-radius
ratio $H/R_{{\rm cyl}}\lesssim0.1$ can be regarded as a disk galaxy,
while a medium with a high $H/R_{{\rm cyl}}\approx1$ may be considered
to represent a relatively round galaxy. Virtual observers are assumed
to measure the system at various inclination angles (or equivalently
viewing angles), denoted as $\beta_{{\rm inc}}$, spanning from $0^{\circ}$
to $90^{\circ}.$ A face-on galaxy corresponds to an inclination angle
of $\beta_{{\rm inc}}=0^{\circ}$, while an edge-on disk galaxy corresponds
to an inclination angle of $\beta_{{\rm inc}}=90^{\circ}$. It is
assumed that the density of the medium is constant. The optical depth
$\tau_{0}$ is measured along the height direction of the cylinder
from the center to the boundary; hence, the total optical depth is
$2\tau_{0}$ when observed face-on.

In the spherical model, photons are emitted from the center unless
otherwise specified. In contrast, photons are spatially uniformly
emitted in the cylindrical model. Additional calculations
were performed for cases in which photons originate from the center
of the cylinder. These cases yielded similar results, although they
are not presented in this paper. The model types investigated
in this paper are summarized in Table \ref{tab01}.

For the present study, at least $10^{8}$ photon packets were used in all models; in some models, either $10^{9}$ or $10^{10}$ packets were utilized to minimize inherent random noises in Monte Carlo simulations. No smoothing was applied to reduce noises. The peeling-off technique, as described by \citet{Seon2022}, was employed to obtain all the results presented in this paper.
The simulations in this paper were conducted on a Cartesian grid of $200^3$ or $300^3$. The spherical and cylindrical models were generated by assigning zero density outside of the respective spherical or cylindrical radii.

\subsection{Definitions of Equivalent Widths and Escape Fraction}

Not only will emission line photons be resonantly scattered by Mg$^{+}$
gas, but continuum photons near \ion{Mg}{2} lines will also undergo
the same scattering process. Resonance scattering of the stellar continuum
produces both emission line-like features and absorption lines. In
order to investigate this continuum effect, the equivalent widths
(EWs) for the emission ($W^{{\rm e}}$) and absorption ($W^{{\rm a}}$)
are calculated for both the K (2796) and H (2803) lines as follows:

\begin{align}
W_{2796,\ 2803}^{{\rm e}} & =\int_{F_{\lambda}>F_{0}}\left(1-\frac{F_{\lambda}}{F_{0}}\right)d\lambda,\nonumber \\
W_{2796,\ 2803}^{{\rm a}} & =\int_{F_{\lambda}<F_{0}}\left(1-\frac{F_{\lambda}}{F_{0}}\right)d\lambda,\label{eq:eq17}
\end{align}
where $F_{0}$ is the initial, flat continuum spectrum. Therefore,
the emission EWs have negative values, while the absorption EWs are
positive. In this paper, even if an emission EW has
a negative value, the terms `high' and `low' will be used
to indicate the magnitude of its absolute value.

This paper also compares the EWs of absorption and emission features with those predicted using the curve-of-growth theory \citep[e.g.,][]{Draine_book2011}, which provides the EW for a pure absorption line. The EW calculated using the curve of growth is referred to as the `reference' EW. The term `relative' EW is used to denote the EW divided by that of the curve of growth. Utilizing the relative EW would be beneficial for quantifying the EWs of absorption and emission features compared to those expected in pure absorption lines.

In the models for the emission line, the escape fraction of \ion{Mg}{2} lines ($f_{{\rm esc}}=F_{{\rm obs}}/F_{0}$) is defined as the ratio of the escaping flux ($F_{{\rm obs}}$) to the intrinsic flux ($F_{0}$) along the line of sight under consideration, including both the K and H lines. The intrinsic flux represents the expected flux when no scattering or absorption occurs.  It's important to note that the definition of the escape fraction of \ion{Mg}{2} is not applicable to the continuum models. The dust attenuation associated with the escape fraction of the continuum and its effect on the EWs are briefly discussed in Sections \ref{subsec:3.3} and \ref{subsec:3.4}.

\begin{figure*}[t]
\begin{centering}
\bigskip{}
\includegraphics[clip,scale=0.5]{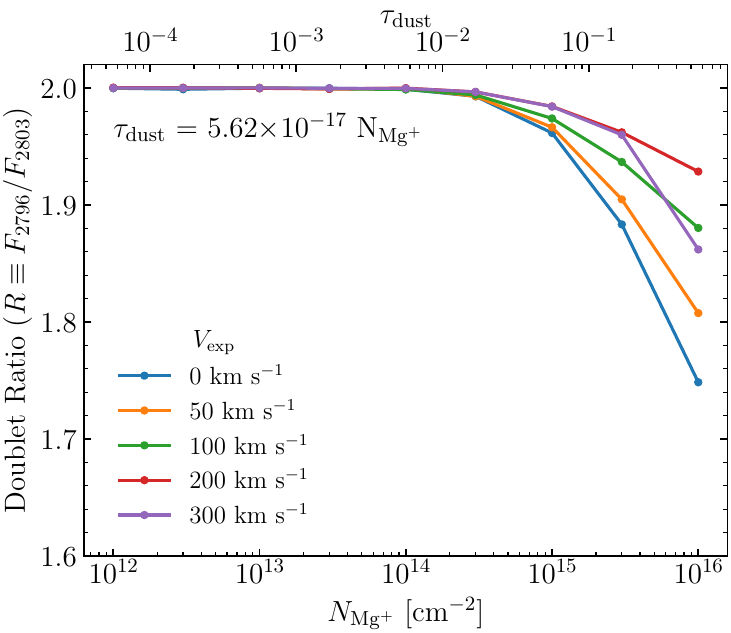}\qquad{}\includegraphics[clip,scale=0.5]{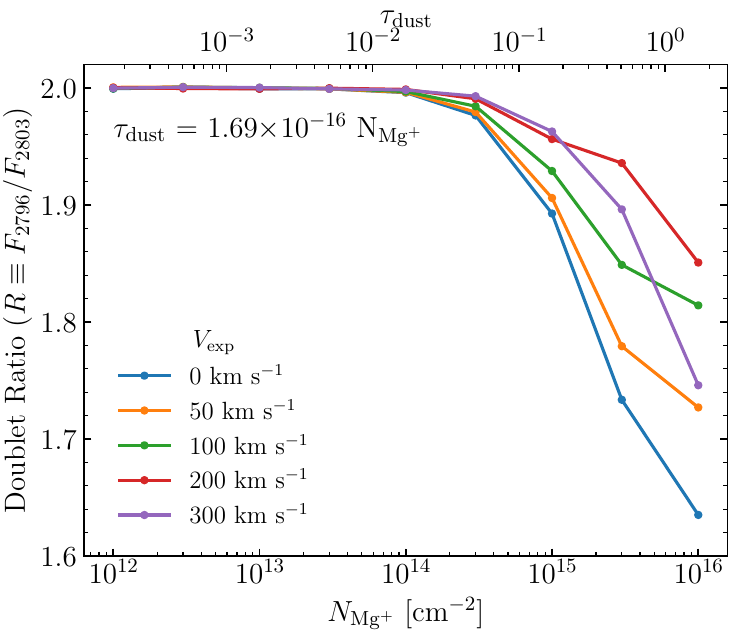}
\par\end{centering}
\medskip{}
\begin{centering}
\includegraphics[clip,scale=0.5]{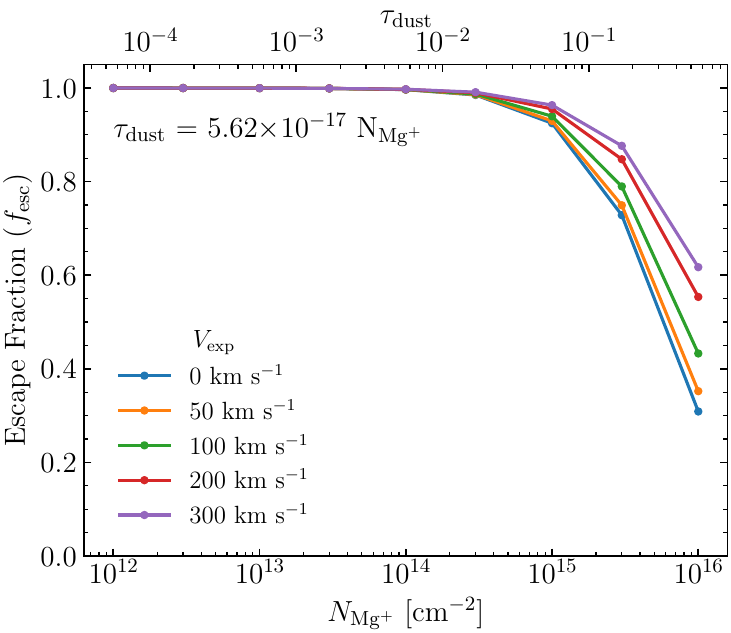}\qquad{}\includegraphics[clip,scale=0.5]{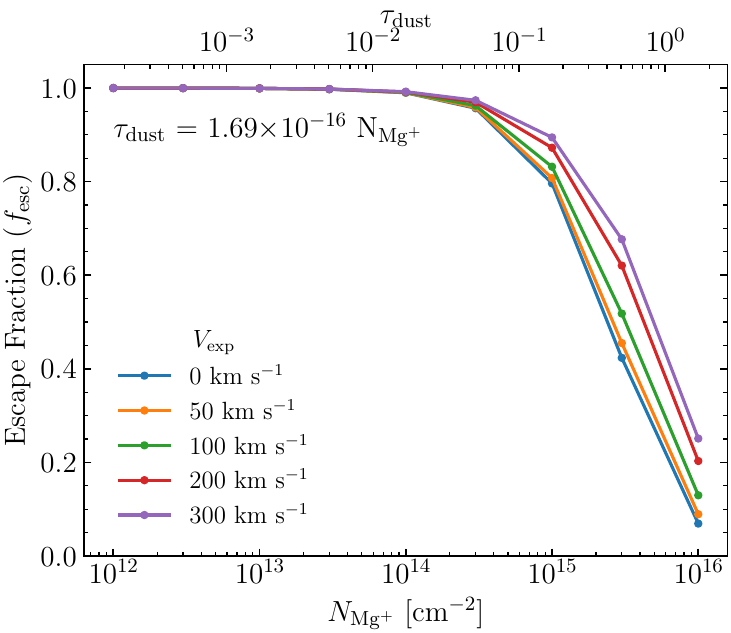}
\par\end{centering}
\begin{centering}
\medskip{}
\par\end{centering}
\caption{\label{fig03}The doublet flux ratio $R$ (top) and escape fraction
$f_{{\rm esc}}$ (bottom) of the \ion{Mg}{2} emission line in the spherical model with dust are shown as functions of the column density of Mg$^{+}$. The medium is assumed to have a constant density and expand radially following the Hubble flow-like law. The maximum expansion velocity $V_{{\rm exp}}$ varies from 0 to 300 km s$^{-1}$. In the left panels, the dust optical depth is assumed to be determined by Equation (\ref{eq:15}). In the right panels, it is assumed to be three times higher than that value. The dust optical depth, proportional to the Mg$^{+}$ column density, is also shown on the secondary $x$-axes.}
\medskip{}
\end{figure*}

\section{RESULTS}

\label{sec:3}

This section begins by describing the doublet ratio and escape fraction
of the \ion{Mg}{2} emission line calculated for the spherical and
cylindrical models. In the cylindrical model, the dependencies of
the doublet ratio and escape fraction on the observer's viewing angle
are also explored. Following that, the emission and absorption features
that arise from resonance scattering of the stellar continuum are
discussed in both geometries. Lastly, this section also explores the
spatial variation of the \ion{Mg}{2} doublet ratio in the spherical
models.

\subsection{Spherical Model - Line Emission}

\label{subsec:3.1}

Some example spectra obtained from the spherical models are shown
in Figure \ref{fig02}. The figure shows spectra for a static medium
(left panel) and an expanding medium (right panel), with various column
densities. The peaks of the spectra are normalized to unity for ease
of comparing spectral shapes. In relatively optically thin ($\tau_{0}\lesssim3$)
and static ($V_{{\rm exp}}=0$) media, the predicted line profiles
of \ion{Mg}{2} emission lines do not exhibit double peaks (equivalently,
there is no absorption at the line center) or show a significant resonance
scattering signature. When $\tau_{0}\gtrsim3$ ($N_{\text{Mg}^{+}}\gtrsim10^{14}$
cm$^{-2}$), the \ion{Mg}{2} line shape starts exhibiting double
peaks in both the K and H lines. As the expanding velocity increases,
the double peaks feature disappears even in models with high column
density. The right panel illustrates the complete disappearance of
double peaks in the spectra for cases with the maximum expansion velocity
of $V_{{\rm exp}}=300$ km s$^{-1}$, irrespective of the gas column
density.

Notably, K-line photons undergo more scatterings than H-line photons,
resulting in a broader peak separation (or line width) of double peak
and a more significant wavelength shift in the K line compared to
the H line. In the static models (left panel), for example, when $N_{\text{Mg}^{+}}=10^{16}$
cm$^{-2}$, the peak separation of the K line is 3.72\AA\ (equivalent
to 400 km s$^{-1}$ in velocity), whereas that of the H line is 3.36\AA\
(361 km s$^{-1}$). In the right panel (expanding medium), the wavelength
(velocity) shifts of the K and H lines in the model with $N_{\text{Mg}^{+}}=10^{16}$
cm$^{-2}$ are 1.41\AA\ (151 km s$^{-1}$) and 1.23\AA\ (132 km
s$^{-1}$), respectively. These differences mainly emerge within the
relatively central region around the source, where most scattering
events take place. In the outer region, photons undergo fewer resonance
scatterings because their wavelengths have already substantially shifted
away from the central wavelength. For instance, the frequency shift
in the central region of an expanding medium becomes more pronounced
for K-line photons due to their experiencing a more significant number
of scatterings than H-line photons. As discussed in Section \ref{subsec:3.5},
this effect can, in some cases, lead to fewer K-line photon scatterings
in the outer region than H-line photons, ultimately resulting in spatial
variation in the doublet flux ratio.

In the presence of dust, the spectral shapes change only slightly,
except for a reduction in the flux level. However, the dust effect
is appreciable only when $N_{\text{Mg}^{+}}\gtrsim10^{15}$ cm$^{-2}$
in a static medium. Therefore, for models that include dust, Figure
\ref{fig02} shows only the spectra for $N_{\text{Mg}^{+}}=10^{16}$
cm$^{-2}$. When $V_{{\rm exp}}=0$ km s$^{-1}$ (300 km s$^{-1}$),
the total fluxes of dusty models are reduced by factors of 0.99, 0.92,
0.73, 0.31, and 0.05 (0.99, 0.96, 0.88, 0.62, and 0.23) for $N_{\text{Mg}^{+}}=3\times10^{14},\ 10^{15},\ 3\times10^{15},\ 10^{16}$,
and $3\times10^{16}$ cm$^{-2}$, respectively, compared to the models
without dust. The overall
spectral shape of the case with dust resembles that of the case without
dust. Nevertheless, there is a significant increase in flux attenuation
near the line center, where resonance trapping is most pronounced.
In addition, the K line experiences stronger dust attenuation than
the H line, which slightly enhances the H line in the normalized spectra
presented in the figure. The spectrum of the model with $V_{{\rm exp}}=300$
km s$^{-1}$ exhibits a slightly more elongated tail due to stronger
suppression at the line center than in the wing.

It is important to note that in spherically symmetric models, the
estimated doublet flux ratio $R=F_{2796}/F_{2803}$, averaged over
all lines of sight, is always equal to the optically thin value of
2 when there is no dust in the medium, except in cases where mixing
of the K and H lines due to fluid motion and line emission due to
continuum pumping occur. The escape fraction also remains at 100\%
due to the absence of photon loss. The variation in the doublet ratio
and escape fraction occurs only in the presence of dust.

Figure \ref{fig03} shows the doublet ratio ($R$, top panel) and
escape fraction ($f_{{\rm esc}}$, bottom panel) as functions of the
column density of Mg$^{+}$ gas for the spherical models in the presence
of dust grains. The left panels show the results for
standard cases where $\tau_{{\rm dust}}$ is determined by Equation
(\ref{eq:15}), while the right panels show the model results when
$\tau_{{\rm dust}}$ is three times higher. In models with $N_{\text{Mg}^{+}}\gtrsim10^{16}$
cm$^{-2}$ and $V_{{\rm exp}}\gtrsim100$ km s$^{-1}$, the K and
H lines are found to merge. In this case, the two lines were considered
to be separated at the wavelength corresponding to the minimum flux,
and the line fluxes were calculated for wavelengths less than or greater
than the wavelength of the minimum flux. The doublet flux ratio and
escape fraction both show a decrease with increasing $N_{\text{Mg}^{+}}$
for a given $V_{{\rm exp}}$. They also decrease in general when $V_{{\rm exp}}$
decreases for a given $N_{\text{Mg}^{+}}$. However, the doublet ratio
$R$ of the model with $V_{{\rm exp}}=300$ km s$^{-1}$ and $N_{\text{Mg}^{+}}=10^{16}$
cm$^{-2}$ shows an abrupt drop, deviating from the trend in other
models. This drop is attributed to the transfer of some of the K line
flux to the H line. It is clear that the maximum deviations from the
dust-free case are found in the static medium with $V_{{\rm exp}}=0$
km s$^{-1}$. As expected, optically thin or moderate cases ($\tau_{0}\lesssim30$,
$N_{\text{Mg}^{+}}\lesssim10^{15}$ cm$^{-2}$, $\tau_{{\rm dust}}\lesssim0.06$
in the left panels and $\tau_{{\rm dust}}\lesssim0.17$
in the right panels) show no significant reduction in the doublet
ratio and escape fraction due to the presence of dust. The effects
of dust become appreciable only in optically thicker cases, even
when $\tau_{{\rm dust}}$ is increased by a factor of 3 in the right
panels. This result indicates that the doublet ratios ($R<2$) found
in compact star-forming galaxies, such as J1503, which exhibits a
median ratio of $R\simeq1.7$ and is supposed to have a low Mg$^{+}$
column density of $<10^{15}$ cm$^{-2}$, cannot be explained solely
by pure dust attenuation in spherical models. The conclusion
remains valid unless dust opacity and Mg$^{+}$ column density are decoupled,
deviating significantly from the relation in Equation (\ref{eq:15}).
Ratios of $R\lesssim1.7$ are achieved only in static models where $N_{{\rm Mg}^{+}}$
exceeds $\sim4\times10^{15}$ cm$^{-2}$, and $\tau_{{\rm dust}}$
is three times as high as that in Equation (\ref{eq:15}).
Therefore, assuming that $\tau_{{\rm dust}}$ does not deviate significantly
from the scaling relation, we need to consider the effects arising
from non-spherical geometry and continuum effects, as described in
the following sections.

\begin{figure*}[t]
\begin{centering}
\includegraphics[clip,scale=0.5]{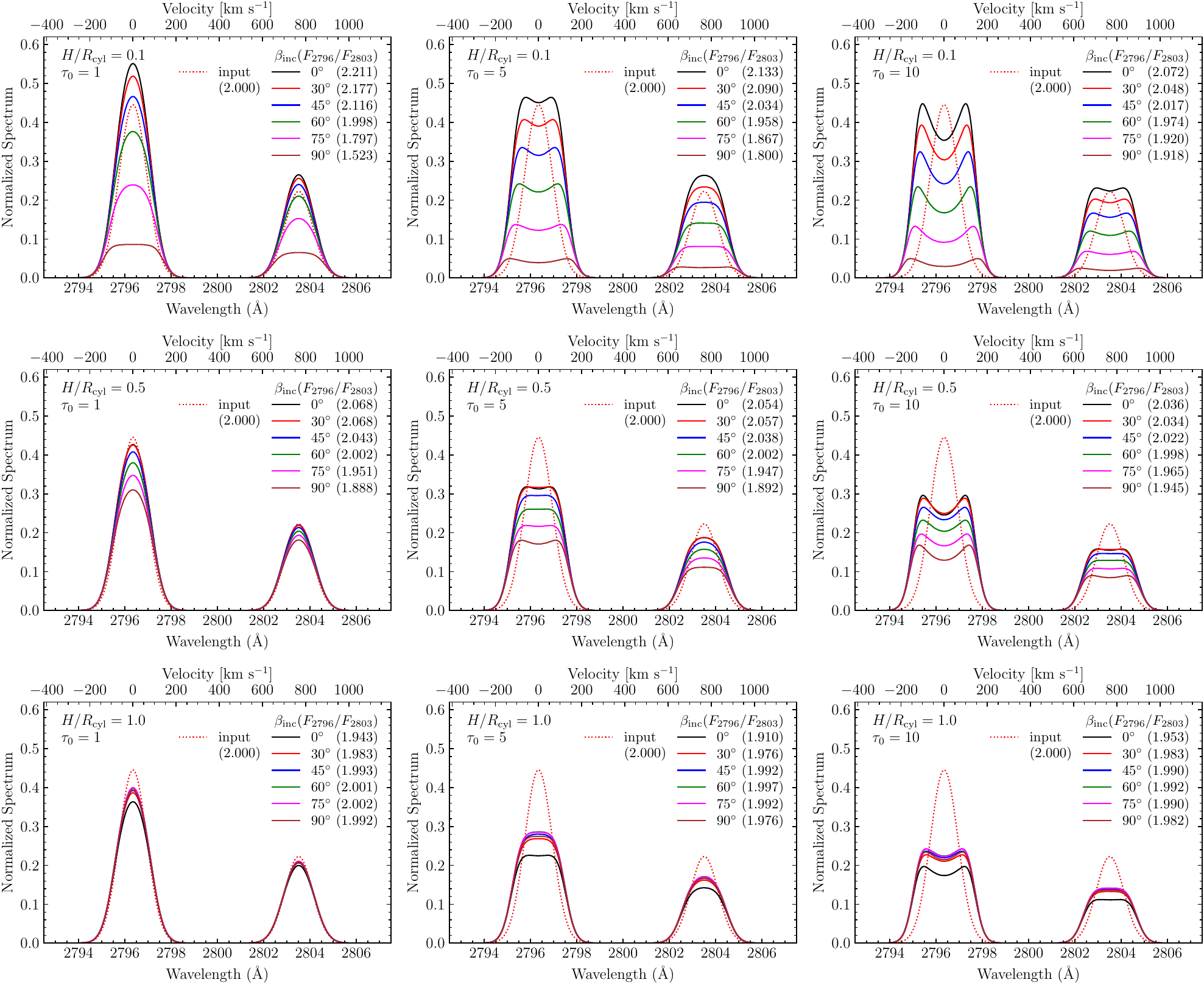}
\par\end{centering}
\begin{centering}
\medskip{}
\par\end{centering}
\caption{\label{fig04} \ion{Mg}{2} emission line spectra predicted for the
cylindrical models. No dust was assumed for these spectra. The height-to-radius
of the cylinder is $H/R_{{\rm cyl}}=0.1$, 0.5, and 1.0 from top to
bottom. The optical depth is $\tau_{0}=1$, 5, and $10$ from left
to right. In each panel, the model spectra for the inclination angle
$\beta_{{\rm inc}}$, ranging from $0^{\circ}$ to $90^{\circ}$,
are shown in different colors. The input spectrum, expected when there
is no Mg$^{+}$ gas, is also shown in the red dotted line in each
panel. The spectra were normalized to ensure that the total integration
of the input spectrum over wavelength becomes unity. The numbers in
the parentheses denote the doublet flux ratios $R=F_{2796}/F_{2803}$.}
\end{figure*}

\begin{figure*}[!tph]
\begin{centering}
\includegraphics[clip,scale=0.46]{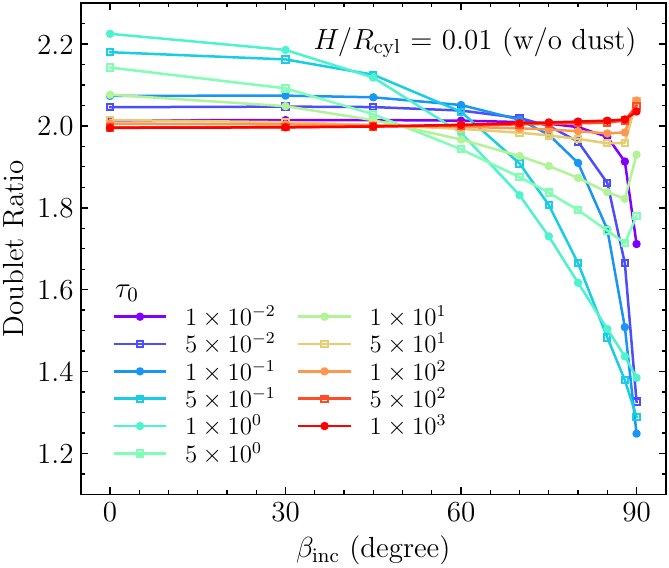}\ \ \includegraphics[clip,scale=0.46]{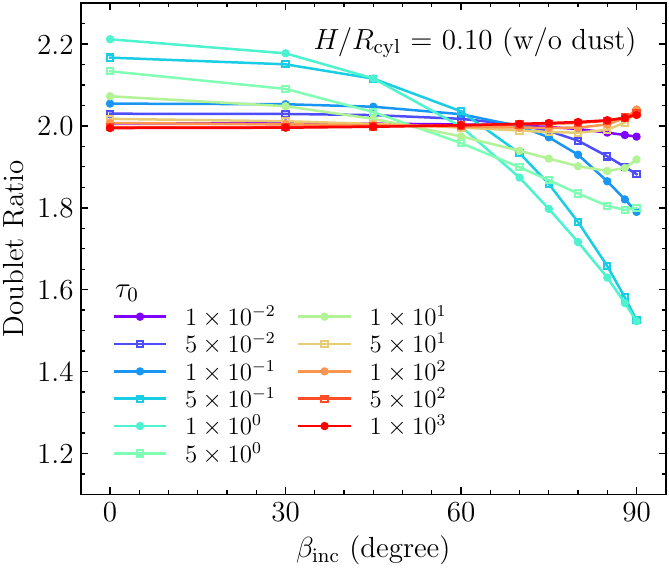}\ \ \includegraphics[clip,scale=0.46]{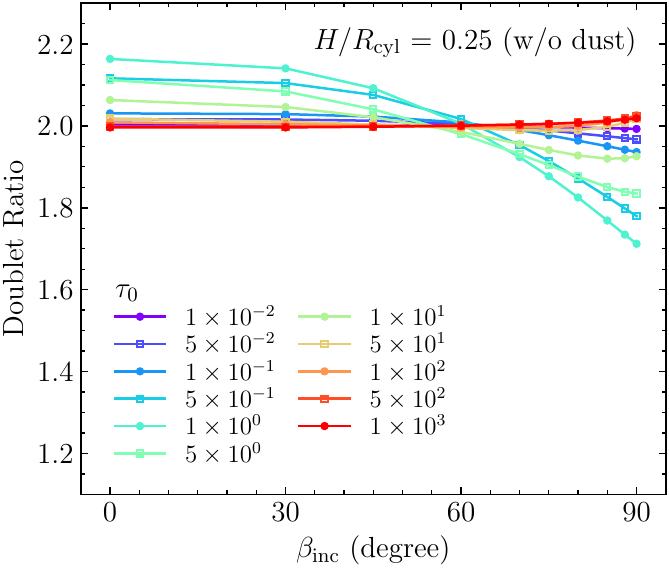}
\par\end{centering}
\begin{centering}
\includegraphics[clip,scale=0.46]{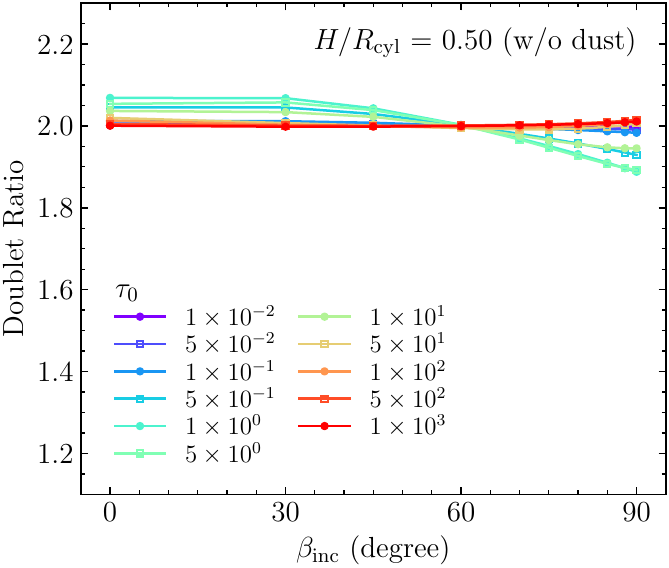}\ \ \includegraphics[clip,scale=0.46]{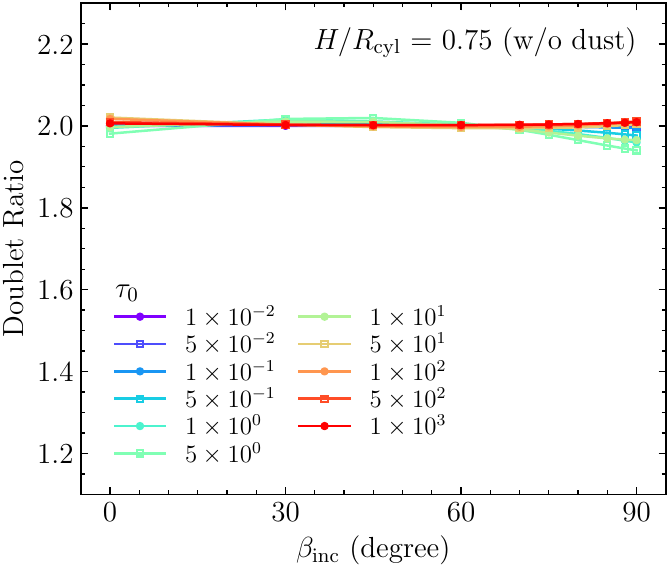}\ \ \includegraphics[clip,scale=0.46]{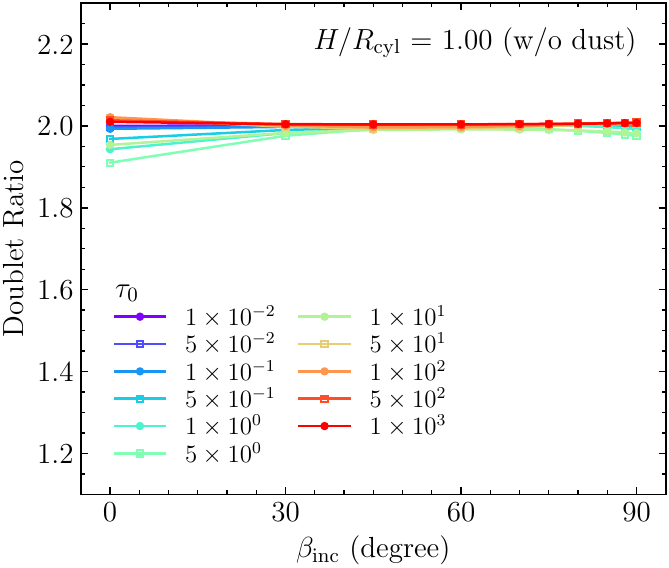}
\par\end{centering}
\begin{centering}
\medskip{}
\par\end{centering}
\caption{\label{fig05} Variation of the doublet flux ratio of \ion{Mg}{2},
in the absence of dust grains, depending on the height-to-radius ratio
$H/R_{{\rm cyl}}$, inclination angle $\beta_{{\rm inc}}$, and optical
depth $\tau_{0}$ of the cylinder. The \ion{Mg}{2} line photons are
assumed to be emitted spatially uniformly within the medium. The optical
depth varies from $\tau_{{\rm 0}}=10^{-2}$ to $10^{3}$ and it is
represented by different colors in the figures. To distinguish between
different values of optical depth $\tau_{{\rm 0}}$, the symbols are
alternated between open squares and filled circles.}
\end{figure*}

\begin{figure*}[!tph]
\begin{centering}
\includegraphics[clip,scale=0.46]{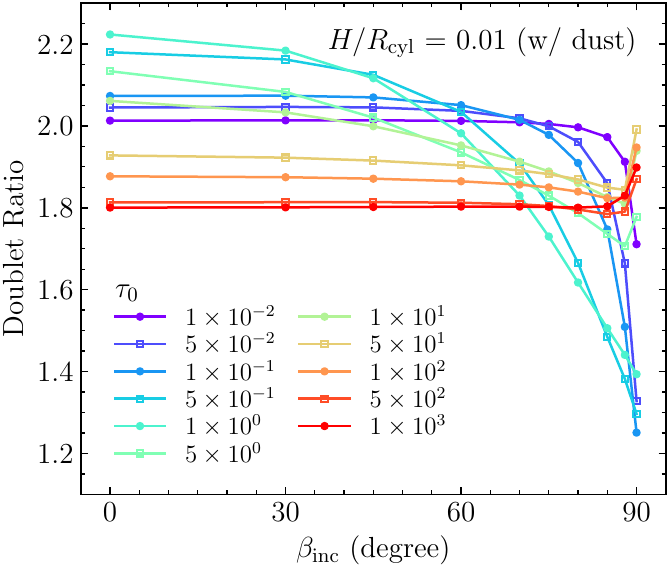}\ \ \includegraphics[clip,scale=0.46]{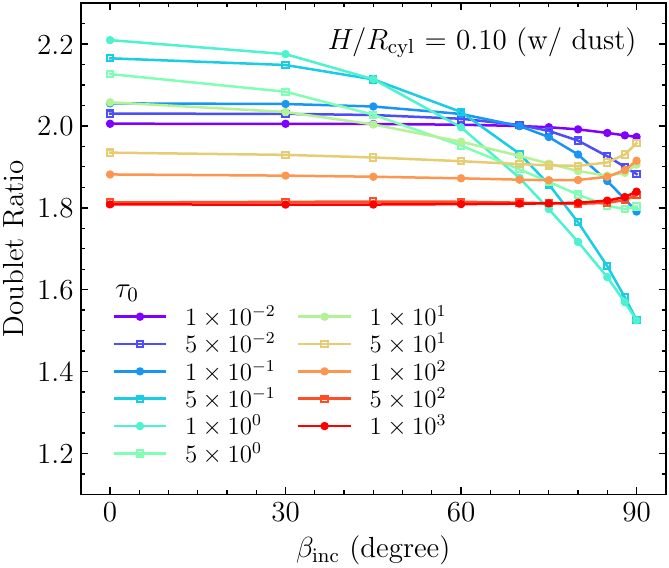}\ \ \includegraphics[clip,scale=0.46]{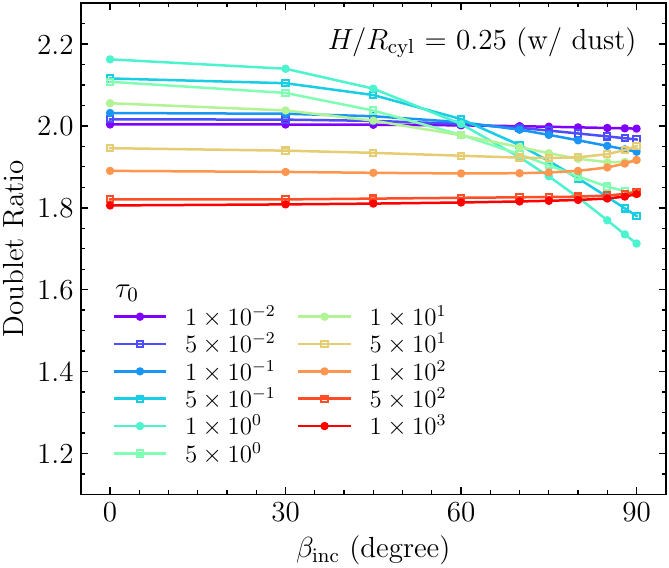}
\par\end{centering}
\begin{centering}
\includegraphics[clip,scale=0.46]{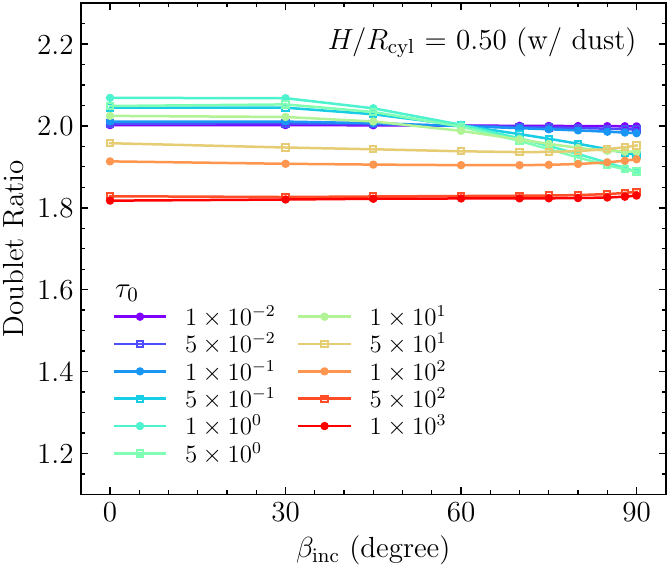}\ \ \includegraphics[clip,scale=0.46]{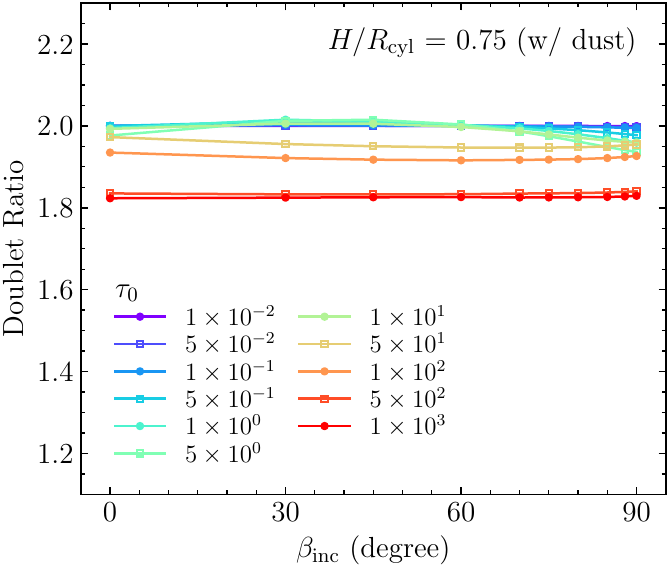}\ \ \includegraphics[clip,scale=0.46]{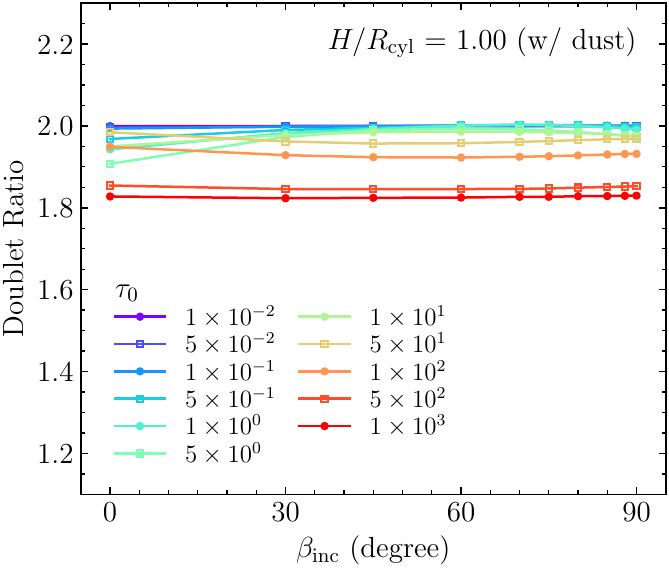}
\par\end{centering}
\begin{centering}
\medskip{}
\par\end{centering}
\caption{\label{fig06} Variation of the doublet flux ratio of \ion{Mg}{2},
in the presence of dust grains, depending on the $H/R_{{\rm cyl}}$
ratio, inclination angle $\beta_{{\rm inc}}$, and optical depth $\tau_{0}$
of the cylinder.}
\end{figure*}

\begin{figure*}[t]
\begin{centering}
\includegraphics[clip,scale=0.46]{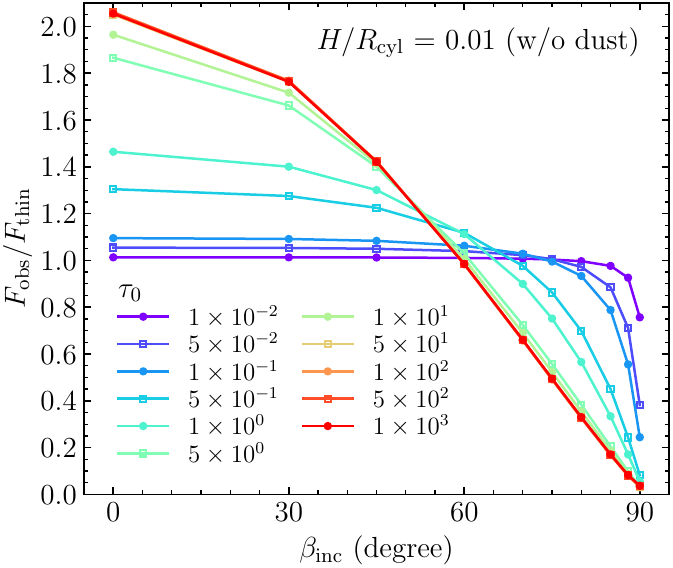}\ \ \includegraphics[clip,scale=0.46]{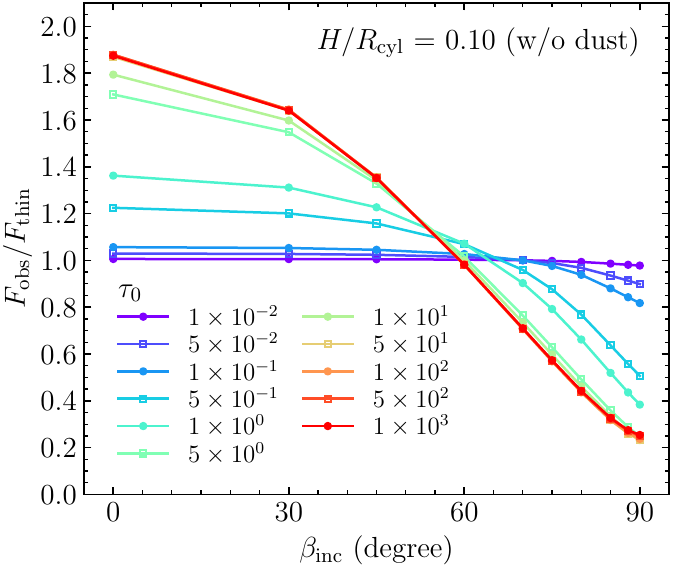}\ \ \includegraphics[clip,scale=0.46]{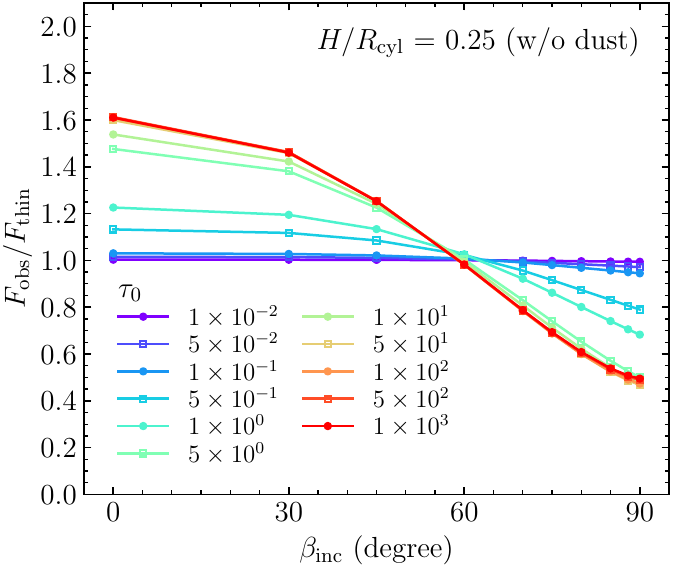}
\par\end{centering}
\begin{centering}
\includegraphics[clip,scale=0.46]{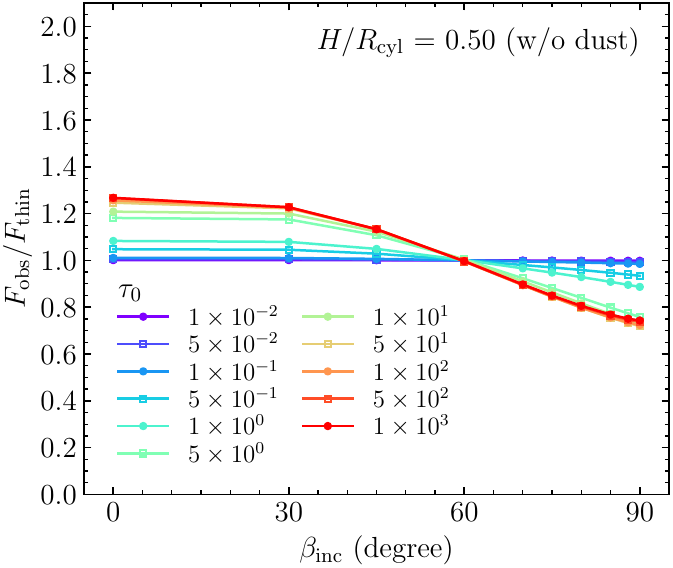}\ \ \includegraphics[clip,scale=0.46]{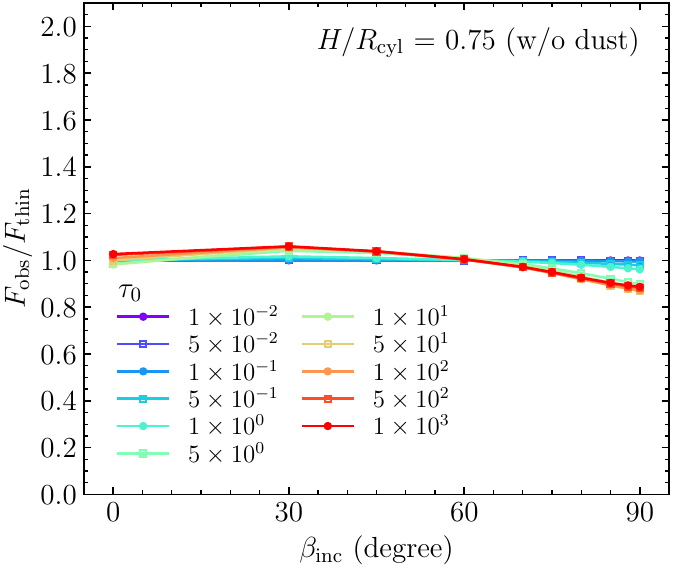}\ \ \includegraphics[clip,scale=0.46]{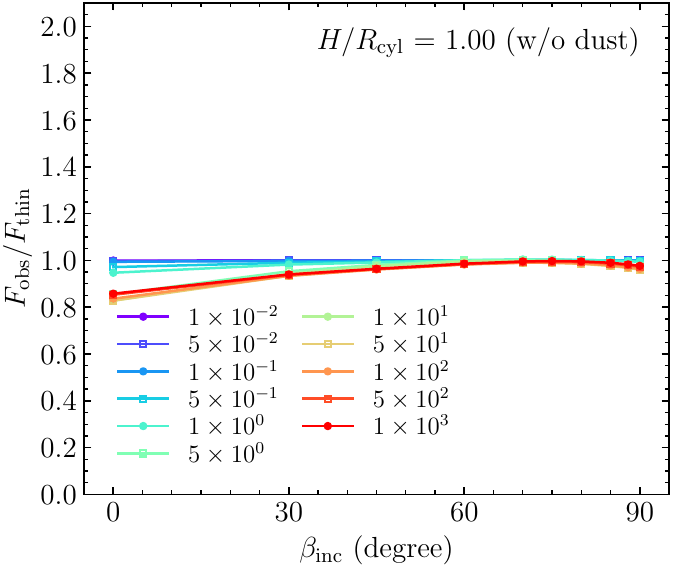}
\par\end{centering}
\begin{centering}
\medskip{}
\par\end{centering}
\caption{\label{fig07} Variation of the escape fraction of \ion{Mg}{2}, in
the absence of dust grains, depending on the $H/R_{{\rm cyl}}$ ratio,
inclination angle $\beta_{{\rm inc}}$, and optical depth $\tau_{0}$
of the cylinder.}
\medskip{}
\end{figure*}

\begin{figure*}[t]
\begin{centering}
\includegraphics[clip,scale=0.46]{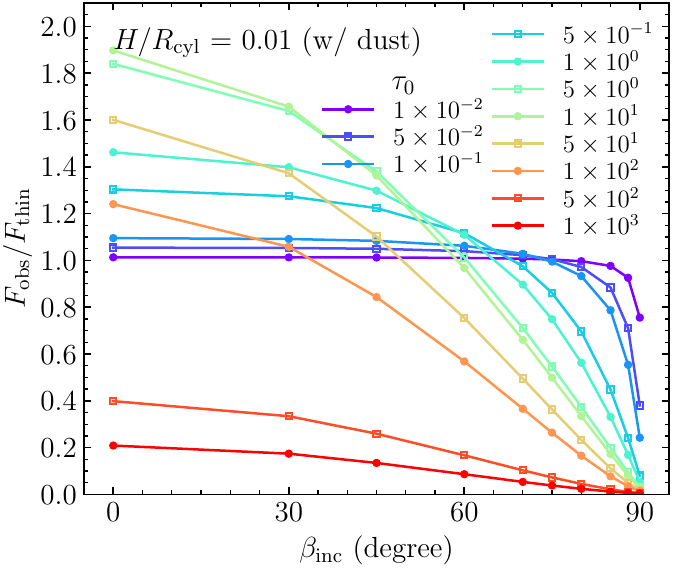}\ \ \includegraphics[clip,scale=0.46]{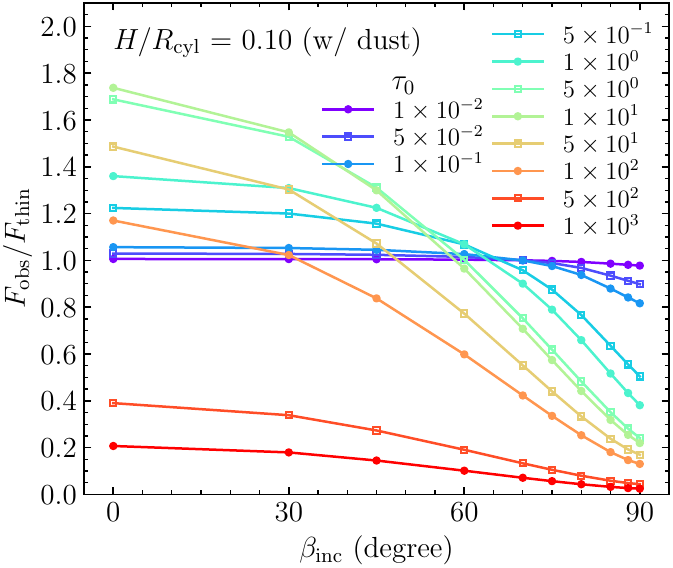}\ \ \includegraphics[clip,scale=0.46]{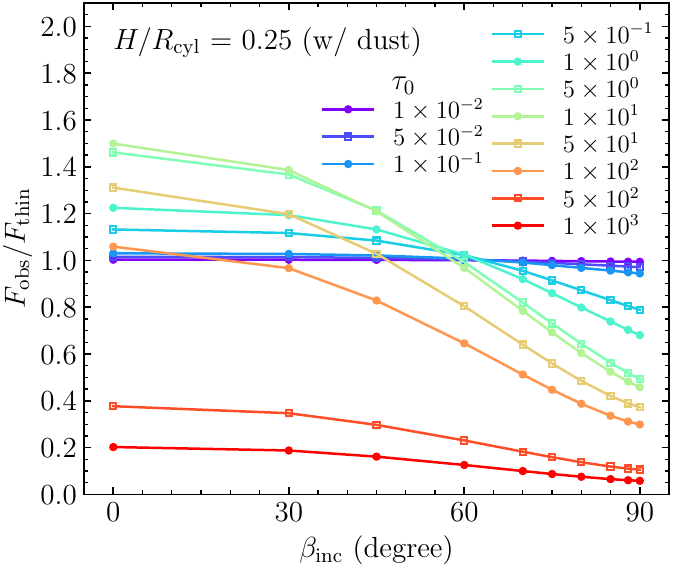}
\par\end{centering}
\begin{centering}
\includegraphics[clip,scale=0.46]{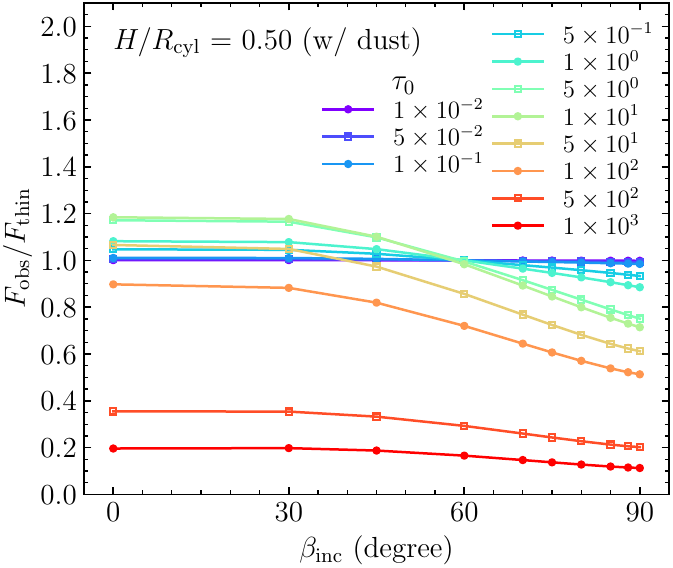}\ \ \includegraphics[clip,scale=0.46]{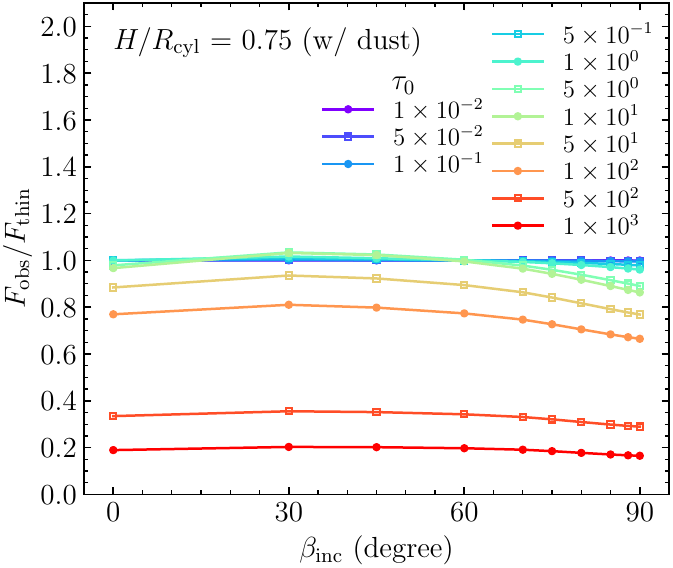}\ \ \includegraphics[clip,scale=0.46]{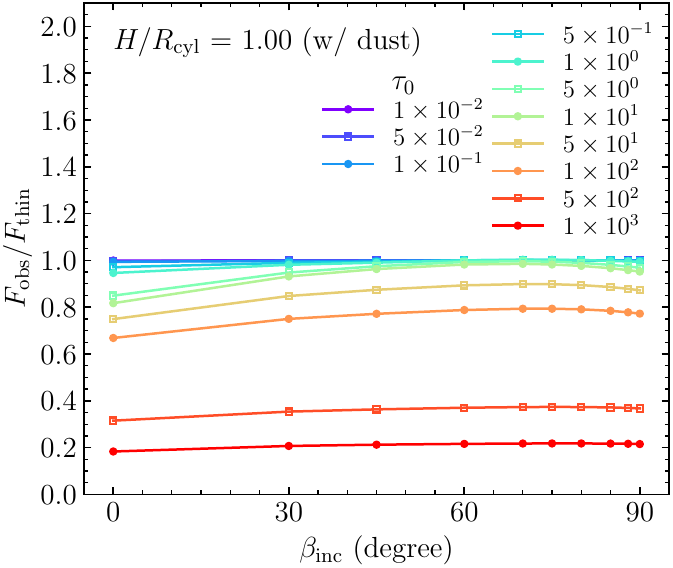}
\par\end{centering}
\begin{centering}
\medskip{}
\par\end{centering}
\caption{\label{fig08} Variation of the escape fraction of \ion{Mg}{2}, in
the presence of dust grains, depending on the $H/R_{{\rm cyl}}$ ratio,
inclination angle $\beta_{{\rm inc}}$, and optical depth $\tau_{0}$
of the cylinder.}
\medskip{}
\end{figure*}

\begin{figure}[t]
\begin{centering}
\includegraphics[clip,scale=0.55]{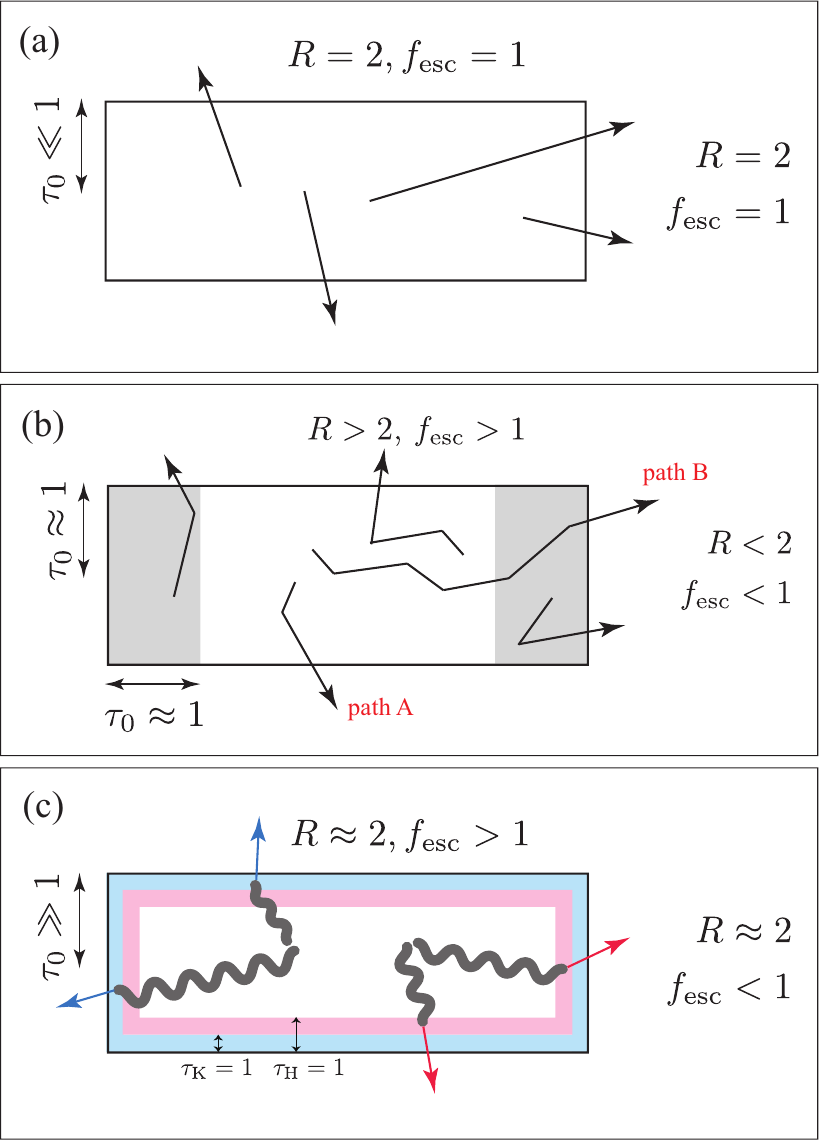}
\par\end{centering}
\begin{centering}
\medskip{}
\par\end{centering}
\caption{\label{fig09}Scattering processes in a relatively flat cylindrical
geometry with a height-to-radius ratio of $H/R_{{\rm cyl}}<1$. The
doublet flux ratio is defined as $R=F_{2793}/F_{2803}$. The escape
fraction, $f_{{\rm esc}}=F_{{\rm obs}}/F_{0}$, represents the ratio
of the escaped flux ($F_{{\rm obs}}$) to the intrinsic flux ($F_{0}$)
of both lines. (a) In an optically thin medium with $\tau_{0}\ll1$,
most photons escape the system without undergoing scattering. (b)
In a medium with $\tau_{0}\approx1$, most photons escape through
a single scattering in the vertical direction, as indicated by path
A. For photons originating from deep within the medium to escape radially,
they must undergo multiple scattering, as illustrated by path B. (c)
In an optically thick medium with $\tau_{0}\gg1$, photons undergo
multiple scattering (denoted by thick gray wiggling lines) before
reaching the outer regions, which are indicated by a bluish color
for K-line photons or a reddish color for H-line photons. These regions
have optical depths of $\tau_{{\rm K}}=1$ and $\tau_{{\rm H}}=\tau_{{\rm K}}/2=1$
from the closest boundaries for K- and H-line photons, respectively.
Photons that have reached the outer regions will escape through a
single scattering.}

\centering{}\medskip{}
\end{figure}

\begin{figure}[t]
\begin{centering}
\includegraphics[clip,scale=0.48]{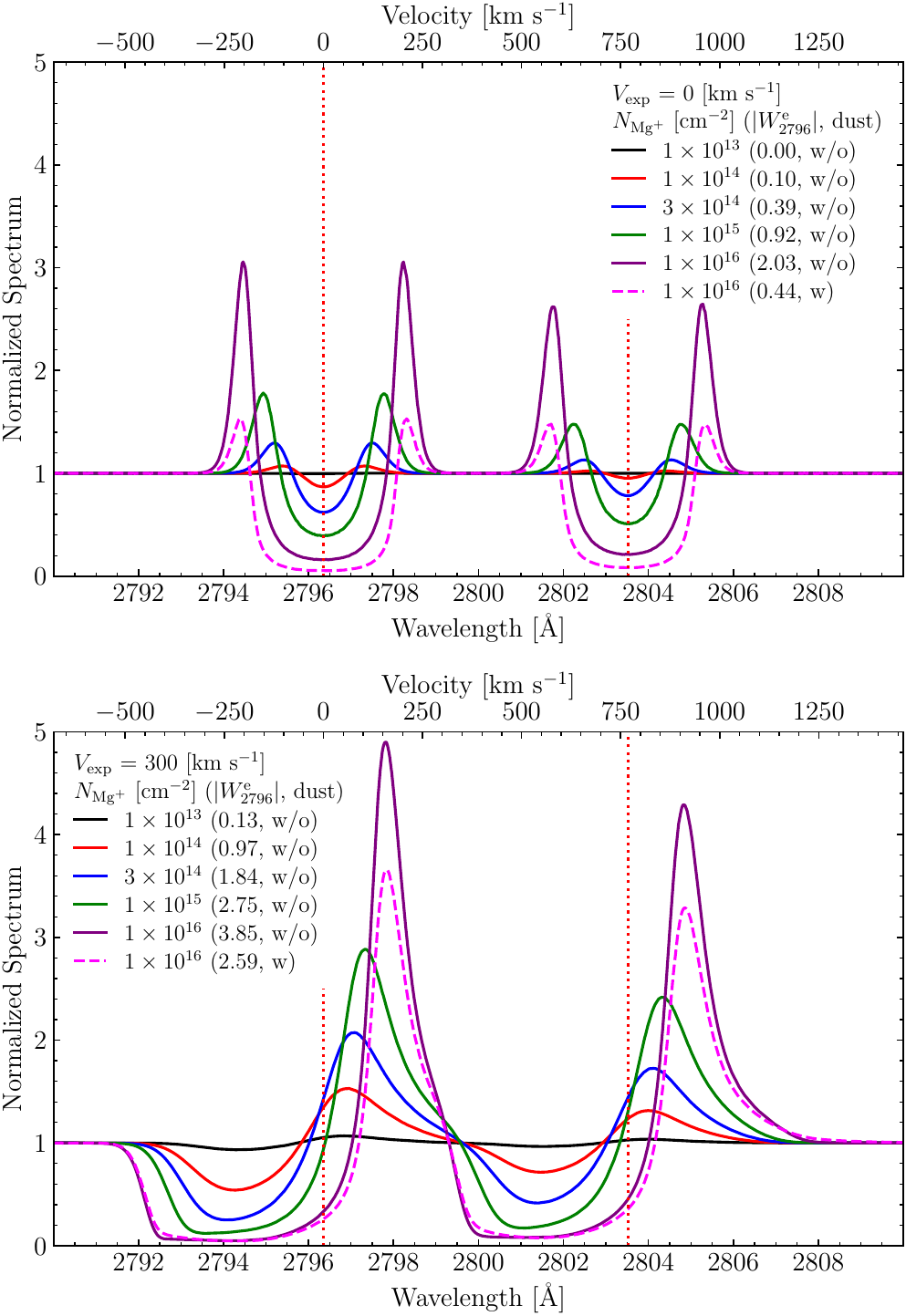}
\par\end{centering}
\begin{centering}
\medskip{}
\par\end{centering}
\caption{\label{fig10} Continuum spectra predicted from the spherical models.
The top panel exhibits spectra for a static medium, while the bottom
panel shows those for an expanding medium with $V_{{\rm exp}}=300$
km s$^{-1}$. In the figures, different colors denote various column
densities of Mg$^{+}$, ranging from $10^{^{13}}$ to $10^{16}$ cm$^{-2}$.
The continuum was normalized to a value of one. In the parentheses
of the legends, $\left|W_{2976}^{{\rm e}}\right|$ represents the
EW of the emission feature in units of \AA; the labels ``w'' and
``w/o'' denote the cases with and without dust, respectively. The
wavelength is also shown in terms of velocity relative to the K line
center. The vertical red dotted lines denote the line centers of the
\ion{Mg}{2} K and H lines. The spectra obtained for the models with
dust are shown as magenta dashed lines.}
\end{figure}

\subsection{Cylindrical Model - Line Emission}

\label{subsec:3.2}

Figure \ref{fig04} shows examples of spectra obtained for nine combinations
of the optical depth ($\tau_{0}=1$, 5, and 10) and the height-to-radius
ratio ($H/R_{{\rm cyl}}=0.1$, 0.5, and 1.0) when observing the cylindrical
models at various inclination angles ($\beta_{{\rm inc}}=0^{\circ}$,
30$^{\circ}$, 45$^{\circ}$, 60$^{\circ}$, 75$^{\circ}$, and 90$^{\circ}$).
The initial spectrum, expected when there is no scattering, is also
shown as the dotted line in the figure. The doublet ratio $(R=F_{2796}/F_{2803})$
for each model is also denoted within parentheses in the figure.
The figure was obtained under the condition when there is no dust.

There are several noteworthy features in Figure \ref{fig04}. (1)
The flux in a disk-like cylinder ($H/R_{{\rm cyl}}<1$) is enhanced
when viewed face-on, compared to the initial input flux, while it
is reduced in the edge-on view. (2) This variation in flux as a function
of the viewing angle tends to be larger in a flatter disk with a smaller
$H/R_{{\rm cyl}}$ ratio. In contrast, when $H/R_{{\rm cyl}}\approx1$,
the line flux becomes relatively independent of the inclination angle
because the system would approximately approach a sphere. We also
note that, in the models with $H/R_{{\rm cyl}}=1$, the line flux
is minimized when measured at $\beta_{{\rm inc}}=0^{\circ}$, as this
direction corresponds to the maximum optical depths across the entire
projected area. (3) The K line begins to show double peaks when $\tau_{0}\gtrsim5$,
whereas the H line exhibits double peaks at a higher optical depth,
$\tau_{0}\gtrsim10$. This difference arises because the K line becomes optically thick earlier as $\tau_{0}$ increases, due to difference in oscillator strengths. (4) As the optical depth increases,
the double peaks in a flat disk model begin to appear at a lower optical
depth compared to a round system. This is because $\tau_{0}$ is defined
to be measured along the vertical direction of the cylinder; as a
result, the optical depth along the radial direction and the total
amount of dust contained in the cylinder are proportional to the inverse of the height-to-radius
($R_{{\rm cyl}}/H$) for a fixed $\tau_{0}$. (5) The doublet ratio
(shown within parentheses) varies depending on the viewing angle.
It can be even larger than the intrinsic value 2, particularly when
a flat disk system ($H/R_{{\rm cyl}}=0.1$) is observed face-on. The
change in the ratio is more significant for smaller $H/R_{{\rm cyl}}$
values, whereas it is negligible for a round model with $H/R_{{\rm cyl}}\approx1$.

The following investigates the properties of the cylindrical model
mentioned above, including the variation of doublet ratio and escaping
flux with viewing angle and the influence of $H/R_{{\rm cyl}}$, in
more detail. Figure \ref{fig05} shows the variation of the doublet
ratio $R$, in the absence of dust, as a function of the inclination
angle $\beta_{{\rm inc}}$ for various combinations of the height-to-radius
ratio $H/R_{{\rm cyl}}$ and optical depth $\tau_{0}$. It is noticeable
that the doublet ratio is more or less close to the optically thin
value of $R=2$ when the medium is round ($H/R_{{\rm cyl}}\gtrsim0.5$).
However, when the medium is disk-like ($H/R_{{\rm cyl}}\lesssim0.5$),
and relatively optically thin or moderate ($0.1\lesssim\tau_{{\rm H}}\lesssim10$,
$4\times10^{12}$ cm$^{-2}$ $\lesssim N_{\text{Mg}^{+}}\lesssim$
$4\times10^{14}$ cm$^{-2}$), the ratio deviates significantly from
$R=2$. In these cases, the ratio $R$ becomes lower than 2 when viewed
edge-on ($\beta_{{\rm inc}}\gtrsim60^{\circ}$), while it becomes
greater than 2 when viewed face-on ($\beta_{{\rm inc}}\lesssim60^{\circ}$).
In the optically thick cases, $R$ is always close to the optically
thin value of $R=2$, irrespective of the inclination angle and height-to-radius
ratio. In very optically thin cases ($\tau_{0}<0.1$), $R$ is close
to 2 when $\beta_{{\rm inc}}\lesssim75^{\circ}$ and becomes lower
when viewed edge-on ($\beta_{{\rm inc}}\gtrsim75^{\circ}$).
The doublet ratio exceeding 2 for large $\tau_0$ when viewed edge-on ($\beta_{\rm inc}=90^\circ$) is an artifact attributed to the discreteness of the Cartesian grid used in this study, which does not perfectly mimic geometrically very thin cylinders.

Figure \ref{fig06} shows the variation of the doublet ratio $R$
when dust is present in the medium. No appreciable dust effect is
found in the optically thin or moderate cases ($\tau_{0}\lesssim10$).
When $\tau_{0}\gtrsim10$ ($N_{\text{Mg}^{+}}\gtrsim4\times10^{14}$
cm$^{-2}$), the doublet ratio is slightly lower compared to the case
with no dust. However, it is worth mentioning that even at its highest
optical depth, the impact of dust is not substantial; this results
in only a slight decrease in $R$ to 1.8. Once again, this finding
indicates that the influence of dust is unlikely to be sufficient
to produce a doublet ratio of $R\sim1.7$ or even lower in compact
star-forming galaxies with $\tau_{0}<10$ unless the
dust optical depth is much higher than that expected in Equation (\ref{eq:15}).
Instead, such low doublet ratios can be explained when the galaxies
are geometrically thin disks ($H/R_{{\rm cyl}}\lesssim0.1$) viewed
edge-on ($\beta_{{\rm inc}}\gtrsim80^{\circ}$) or contain large and
relatively flat Mg$^{+2}$ gas clouds situated edge-on.

Figure \ref{fig07} presents the variation of the escape fraction
of \ion{Mg}{2} as a function of the inclination angle for various
combinations of $H/R_{{\rm cyl}}$ and $\tau_{0}$, in the absence
of dust. At first glance, it is surprising that the escape fraction
can exceed 1 for face-on disk-like media with $H/R_{{\rm cyl}}\lesssim0.75$
and $\beta_{{\rm inc}}\lesssim60^{\circ}$, especially when $\tau_{0}\gtrsim0.1$.
On the other hand, when viewed edge-on ($\beta_{{\rm inc}}\gtrsim60^{\circ}$),
the optically thick media yield $f_{{\rm esc}}<1$. The deviation
from $f_{{\rm esc}}=1$ becomes increasingly significant with increasing
optical depth. The escape fraction of \ion{Mg}{2} in the presence
of dust is shown in Figure \ref{fig08}. When the medium is optically
thin ($\tau_{0}\lesssim10$), the results are consistent with those
shown in the absence of dust. The escape fraction decreases significantly
due to dust only in the optically thick cases ($\tau_{0}\gtrsim10$).
Particularly, when $\tau_{0}\gtrsim500$, the escape fraction consistently
falls below 50\%, irrespective of $H/R_{{\rm cyl}}$ and $\beta_{{\rm inc}}$.

\begin{figure*}[!tph]
\begin{centering}
\includegraphics[clip,scale=0.5]{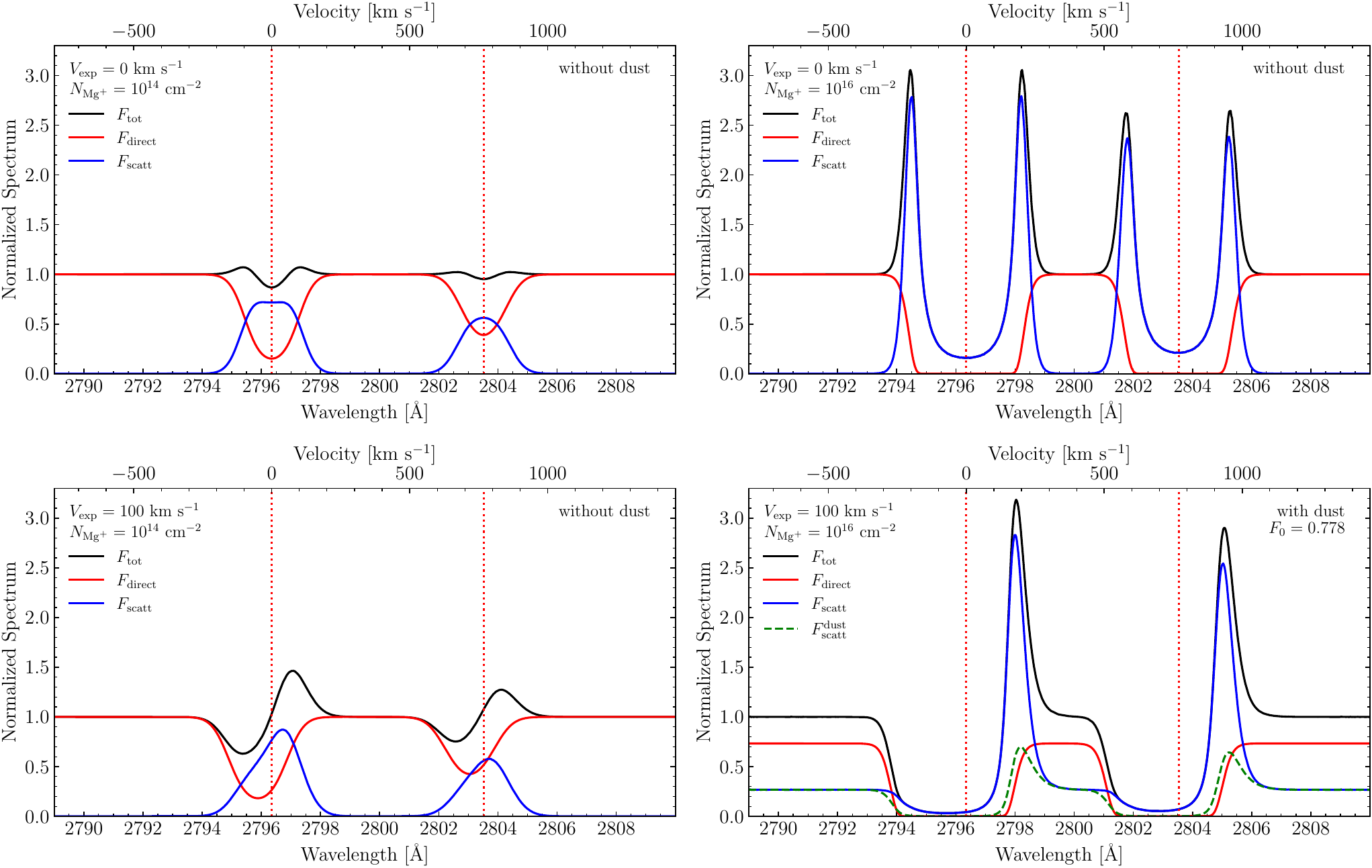}
\par\end{centering}
\begin{centering}
\medskip{}
\par\end{centering}
\caption{\label{fig11} Decomposition of the continuum spectra, predicted from
a spherical medium, into direct and scattered components. The top
panels show the spectra from a static sphere. The bottom panels represent
the spectra from an expanding sphere with $V_{{\rm exp}}=100$ km
s$^{-1}$. The left panels and right panels are for $N_{\text{Mg}^{+}}=10^{14}$
and $10^{16}$ cm$^{-2}$, respectively. The first three models contain
no dust, while the last model contains dust. The red and blue lines
denote the directly escaping ($F_{{\rm direc}}$) and scattered ($F_{{\rm scatt}}$)
spectrum, respectively. The black line represents the total spectrum
($F_{{\rm tot}}=F_{{\rm direc}}+F_{{\rm scatt}}$). In the presence
of dust (the last panel), the scattered spectrum contains both the
dust-scattered component in the continuum and the resonance scattered
component near the \ion{Mg}{2} lines. $F_{{\rm scatt}}^{{\rm dust}}$
represents the component that underwent dust scattering at the last
scattering event and escaped the medium. In this last case, the continuum
level before normalization was $F_{0}=0.779$ relative to the initial
input continuum level.}
\medskip{}
\end{figure*}

It is now discussed why resonance scattering in an asymmetric medium
results in rather unexpected doublet flux ratios of $R>2$ and escape
fractions exceeding 1 ($f_{{\rm esc}}>1$). Three schematic diagrams
in Figure \ref{fig09} illustrate various scattering processes occurring
in a relatively flat cylindrical model, depending on the medium's
optical depth. In an optically thin medium ($\tau_{0}\ll1$), illustrated
in Figure \ref{fig09} (a), both in the vertical and radial directions,
photons will undergo few scatterings, and therefore, the doublet ratio
and escape fraction are not altered significantly from their intrinsic
values. Figure \ref{fig09} (b) shows a case where the optical depth
along the vertical direction is $\tau_{0}\approx1$. In this situation,
photons tend to escape preferentially in the vertical direction due
to the lower optical thickness. Photons originating deep within the
medium find it easier to escape vertically rather than radially. Only
a limited number of photons originating near the boundaries (marked
in gray) can manage to escape radially. Consequently, when observed
face-on, the escape fraction can exceed 1; however, in an edge-on
orientation, it is less than 1. Moreover, K-line photons encounter
an optical depth twice as high as that of H-line photons, leading
to an increased probability of scattering for K-line photons. This
difference in optical depth causes more K-line photons to escape through
scattering in the vertical direction compared to what H-line photons
do. As a result, when viewed face-on, the doublet ratio $R$ appears
higher than 2, whereas it appears lower than 2 when viewed edge-on.
In an optically very thick medium ($\tau_{0}\gg$1), both K- and H-line
photons undergo multiple scattering and become trapped within the
inner region, represented in white in Figure \ref{fig09}(c). Within
this inner region, the radiation field is more or less isotropic,
and the doublet ratio will remain at its intrinsic ratio of 2 (when
there is no dust). Once photons reach the outer regions, shown as
a blueish area for K-line photons and a reddish area for H-line photons
in Figure \ref{fig09}(c), they will predominantly escape through
a single scattering on average. Consequently, the tendency for more
K-line photons to escape than H-line photons in the vertical direction
disappears, resulting in a doublet ratio of $R\approx2$ in both directions.
However, the probability of vertical escape would be much higher,
as photons need to undergo more scatterings to be transferred radially
than when transferred vertically. In addition, the spectrum escaping
in the radial direction will be considerably broader than that escaping
vertically, as shown in Figure \ref{fig04} (for example, refer to
the top right panel with $H/R_{{\rm cyl}}=0.1$ and $\tau_{0}=10$).
This effect is due to a significantly higher number of scatterings
required to escape in that direction.

If dust is present in the medium, it is evident that K-line photons
will be more readily absorbed than H-line photons. Therefore, in an
optically thick medium with dust, the doublet ratio and escape fraction
would be less than their intrinsic values, irrespective of the height-to-radius
ratio and inclination angle. However, this dust effect is negligible
in an optically thin medium.

The orientation effect of favoring face-on escape in asymmetrical geometries was predicted by \citet{Charlot1993} and \citet{Chen1994} in the context of Ly$\alpha$ RT, and it has been demonstrated in many Ly$\alpha$ RT simulations \citep{Laursen2007,Barnes2011,Verhamme2014,Behrens2014,Smith2022}. Ly$\alpha$ photons are typically optically very thick in galactic environments, and thus may correspond to case (c) in Figure \ref{fig09}.

\begin{figure}[!tp]
\vspace{1mm}
\begin{centering}
\includegraphics[clip,scale=0.5]{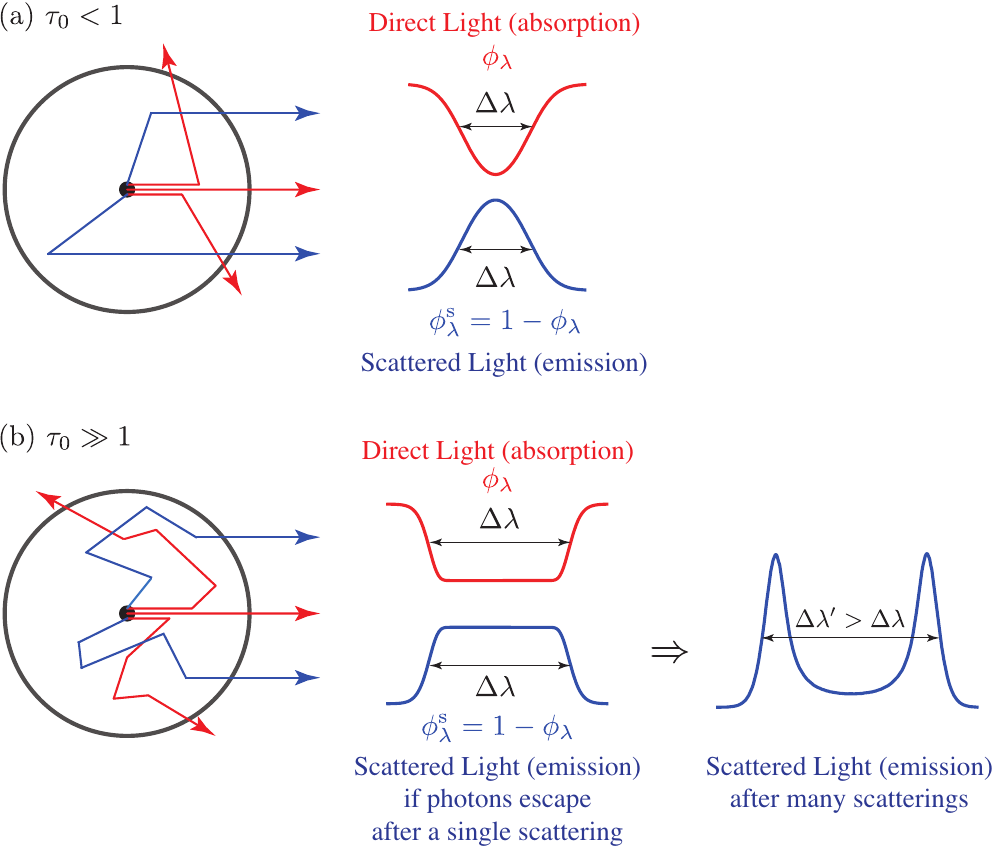}
\par\end{centering}
\begin{centering}
\medskip{}
\par\end{centering}
\caption{\label{fig12} Illustration of the formation of absorption and emission line profiles due to the continuum pumping in a static sphere: (a) optically thin case and (b) optically thick case. In an optically thick case, the emission line profile due to scattering becomes slightly broader and/or doubly-peaked.}
\end{figure}

\begin{figure*}[!t]
\begin{centering}
\includegraphics[clip,scale=0.48]{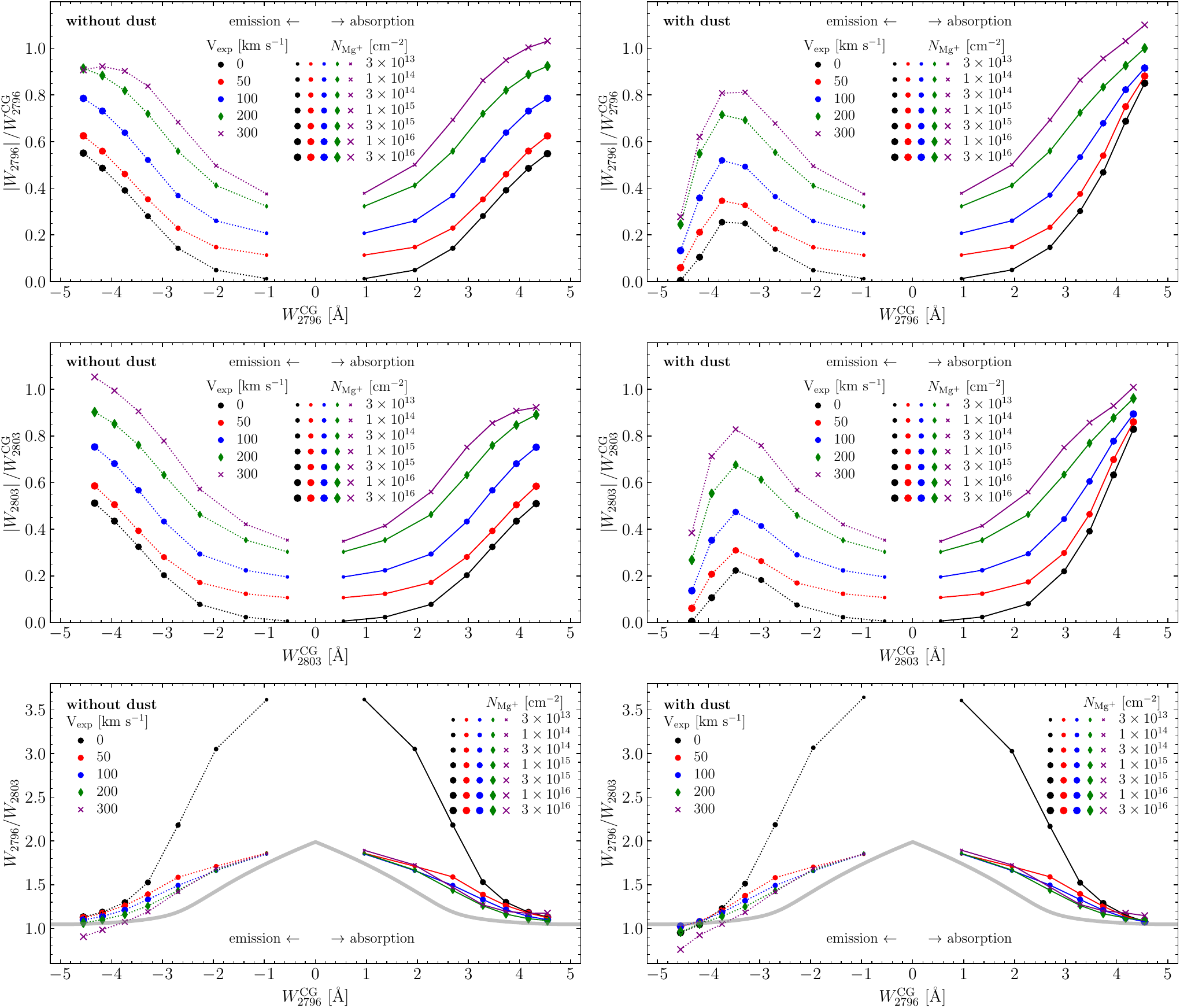}
\par\end{centering}
\begin{centering}
\medskip{}
\par\end{centering}
\caption{\label{fig13}The top and middle panels show the relative EWs, defined as the EWs divided by the reference EWs calculated using the curve of growth (CG). The top panels present the relative EWs of the \ion{Mg}{2} K (2796 \AA) line as a function of its reference EW, and the middle panels show those for the \ion{Mg}{2} H (2803 \AA). The bottom panels show the EW ratio as a function of $W_{2796}^{\rm CG}$. The left and right panels show the results for the cases with and without dust, respectively. Negative and positive values represent the EWs for the emission (dotted) and absorption (solid) lines, respectively. Various symbols and colors correspond to different expansion velocities ranging from $V_{{\rm exp}}=0$ to 300 km s$^{-1}$. The column density of the medium varies from $N_{\text{Mg}^{+}}=3\times10^{13}$ to $3\times10^{16}$ cm$^{-2}$, and this is depicted by the symbol size.}
\end{figure*}

\subsection{Spherical Model - Continuum}

\label{subsec:3.3}

Figure \ref{fig10} shows example spectra calculated with the same parameters as in Figure \ref{fig02}, except that an intrinsically flat continuum spectrum was used in this figure.
The figure also shows the EWs of
the Mg II 2796 line in the parentheses of the legend. In static media
($V_{{\rm exp}}=0$, left panel), the spectra show double peaks and
absorption features caused by resonance scatterings near the line
center. As expected, the absorption depth and the emission height
increase as the Mg$^{+}$ column density increases. The spectra of
expanding media in the right panel show well-known P-cygni profiles
with blueshifted absorption and redshifted emission features. The
figure also compares the spectra with and without dust for the model
with $N_{\text{Mg}^{+}}=1\times10^{16}$ cm$^{-2}$. It is noticeable
that the presence of dust does not significantly alter the absorption
line shape and depth; however, the emission line strength is substantially
reduced by dust.

It should be noted that even in the highest column density models,
the spectra are not entirely carved out at the \ion{Mg}{2} line centers.
This phenomenon is attributed to the ``filling-in'' effect caused
by resonance scattering. The filling-in of the resonance absorption
feature by the resonance scattering itself was also discussed in \citet{Prochaska2011}
and \citet{Scarlata2015}. If there were no filling-in
effect, the continuum near \ion{Mg}{2} for models with $N_{\text{Mg}^{+}}\gtrsim3\times10^{14}$
cm$^{-2}$ ($\tau_{0}\gtrsim9$) would have been completely removed.

Figure \ref{fig11} demonstrates the impact of the filling-in effect
on the EWs of absorption and emission features by decomposing the
spectra obtained from various models into direct and scattered components.
In a relatively optically thin static medium with $\tau_{0}\lesssim1$,
as shown in the top left panel, most of the direct (absorption)
line feature (denoted in red) is compensated by the scattered (emission) line
(in blue), and no significant line features are evident in the final
spectrum. However, as the optical depth increases (top right panel) or
the medium expands (bottom left panel), the absorption line is not entirely
filled, and the absorption and emission lines begin to be separated.

The above trend can be understood as illustrated in Figure \ref{fig12}. The figure depicts the formation mechanism of absorption and emission features by continuum scattering in a static sphere. When continuum photons emitted toward an observer are scattered, they are lost out of that direction, resulting in the formation of an absorption line. In a spherical medium with no dust, the absorbed flux should be equal to the flux from scattered light because of the conservation of photon number and the symmetry. If photons escape after a single scattering, the scattered line profile will be $\phi_\lambda^{\rm s}=1-\phi_\lambda$, assuming the absorbed line profile is $\phi_\lambda$ (the middle column in Figure \ref{fig12}). In an optically thin medium, photons will be at most singly scattered, and thus the line profile of absorbed photons will almost exactly match that of scattered photons (the top left panel of Figure \ref{fig11} and Figure \ref{fig12}a). However, as the optical depth increases, multiple resonance scatterings cause a diffusion in wavelength space. This results in a broadened (and double-peaked) profile of scattered (emission) light compared to that of absorption line (the top right panel of Figure \ref{fig11} and the rightmost column of Figure \ref{fig12}b). This broadening effect leads to a mismatch between absorption (direct light) and emission (scattered light) profiles, becoming more pronounced as the optical depth increases. Similar effects also take place in an expanding media.

The K line is optically thicker than the H line, resulting in a broader and deeper double-peaked profile. Its higher optical depth causes a more significant mismatch between the profiles of direct (absorption) and scattered (emission) light compared to the H line. As a result, the K line exhibits more noticeable `net' absorption and emission features than the H line, as shown in Figure \ref{fig11}. In other words, the H line displays weaker net absorption and emission than the K line.
Expansion of the medium, as shown in the bottom panels of Figure \ref{fig11}, also yields a discrepancy between the absorption and emission profiles by redshifting the absorption and blueshifting the emission features.
Contraction of the medium would also give rise to a similar
mismatching effect, but by blueshifting the absorption and redshifting
the emission features. This mismatch between the absorption and emission
line profiles manifests as the appearance of the ``net''
absorption and emission features in continuum spectra. As the absorption and emission profiles match -- meaning that the absorption is filled in by emission due to scattering -- the EWs for the ``net'' absorption and emission would be very small. On the other hand, the EWs would be relatively large if the emission and absorption profiles do not match.

The bottom right panel of Figure \ref{fig11} illustrates that if dust optical depth is high enough, dust absorption and scattering also play significant
roles in the continuum level and, thus, the EWs of absorption and
emission features. In the figure, to distinguish between the resonantly
scattered and dust-scattered components, the dust-scattered spectrum,
shown by the dashed green line ($F_{{\rm scatt}}^{{\rm dust}}$),
is obtained by collecting photons that underwent dust scattering at
the last scattering event before they escaped. The continuum outside
the \ion{Mg}{2} lines is affected by dust absorption and scattering
but not by resonant scattering. In other words, the continuum denoted
in blue in the figure is solely due to dust scattering. The dust-scattered
continuum makes up approximately 27\% of the total continuum level
in this example. On the other hand, near \ion{Mg}{2} lines, the resonance
scattering predominantly gives rise to both the absorption and emission
features. The filling-in in the absorption troughs is primarily due
to resonant scattering, as evidenced by the lack of or minimal presence
of the dust-scattered component in those regions. Regarding the emission
line feature, its significant portion appears to be attributed to
dust scattering, as indicated by the green dashed line at first glance.
However, this component arises from photons that experienced resonant
scatterings (and additional dust scatterings) before ultimately escaping
through dust scattering. Photons are unlikely to undergo only pure
dust scattering without being resonantly scattered (or such cases
would be extremely rare).

The extinction of continuum photons reduces the continuum level while
scattering by dust from other directions into the line of sight enhances
the continuum. Consequently, the final continuum level becomes higher
than expected as the dust scattering effect is ignored. Neglecting
this enhancement due to dust scattering could lead to overestimating
emission EWs and underestimating absorption EWs. Indeed, a substantial
portion of the UV radiation in our Galaxy and external galaxies is
known to originate from dust scattering of starlight \citep[e.g.,][]{Seon2011,Seon2016}.
Nevertheless, the dust scattering effect is expected to be relatively
weak in compact star-forming galaxies due to their low column density
of Mg$^{+}$ and, consequently, low dust optical depth.

The filling-in effect could be quantified by comparing the EWs with those expected in the curve-of-growth theory for the pure absorption lines. The curve of growth is applicable in situations where there is no filling-in by scattering and the emission and absorption profiles are distinct.
Contrarily, in the present case, absorption lines are filled in through scattering. Nevertheless, as the optical depth increases, the emission component begins to separate from the absorption component, as illustrated in Figure \ref{fig12}. Then, the EWs will eventually tend to approach to those of pure absorption lines but not perfectly so.

The top and middle panels in Figures \ref{fig13} illustrate the variation of the `relative' EWs, which refer to the EWs divided by those of the curve of growth, as the column density varies. The bottom panels show the EW ratios between the doublet. The left panels show the results in the case of no dust, whereas the right panels display the results when dust is included in the model.  In the figures, $W_{2796}^{\rm CG}$ and $W_{2803}^{\rm CG}$ represent the `reference' EWs for the K and H lines, respectively, calculated using the curve of growth in a static medium. These serve as references for the EWs and as the proxies for $\tau_0$ or $N_{\text{Mg}^+}$. The column density of the medium varies from $N_{\text{Mg}^{+}}=3\times10^{13}$ cm$^{-2}$ to $3\times10^{16}$ cm$^{-2}$ and is represented by symbol size. Different symbols and colors are used to denote the expansion velocities of the medium.

In the absence of dust (left panels), the relative EWs increases in both the absorption ($W_{2796}^{{\rm a}}$ and $W_{2803}^{{\rm a}}$) and emission ($\left|W_{2796}^{{\rm e}}\right|$ and $\left|W_{2803}^{{\rm e}}\right|$) components as the column density of Mg$^{+}$ increases. This trend results from the increased number of resonance scatterings. They also increase with expanding velocity, regardless of the presence of dust, due to the enhanced separation between the absorption and emission components. The relative EWs of the H line are generally smaller than those of the K line, except for the fastest expanding model with $V_{\rm exp}=300$ km s$^{-1}$, because the H line is less scattered than the K line.

 One would expect the EW of absorption to ideally match that of emission in a spherical model unless the fluid velocity and line broadening cause a mixing of the K-line emission with the H-line absorption (or a mixing of the K-line absorption with the H-line emission in an infalling medium). This property is confirmed in static and slowly moving models, as shown in the left panels of Figure \ref{fig13}.
It is also expected that in the medium with the fastest expanding velocity and the highest column density, the absorption and emission EWs of both lines will exceed the reference EWs. This is because the rapidly expanding outer part of the medium can scatter much bluer photons than a static medium can. In other words, the continuum in a rapidly expanding medium would be much more carved out compared to a slowly expanding or static medium. Thus, if the K and H lines are well-separated, both their EWs could be enhanced compared to the reference EWs.  However, in the fastest expanding model, the K-line emission is redshifted, partially filling the absorption trough of the H line (see the lower panel of Figure \ref{fig10}). This mixing between the lines results in the reduction of the K-line emission and H-line absorption. Therefore, the relative EW for K-line emission (H-line absorption) is less than 1, whereas that for absorption (emission) is higher than 1. The transfer of red-wing photons in the K line to the H line, leading to the filling of the H-line absorption, was also found in \citet{Prochaska2011}. Note that this mixing effect between lines in a rapidly expanding medium is distinct from the filling-in of absorption by the emission feature within each line.

\begin{figure*}[!t]
\begin{centering}
\includegraphics[clip,scale=0.48]{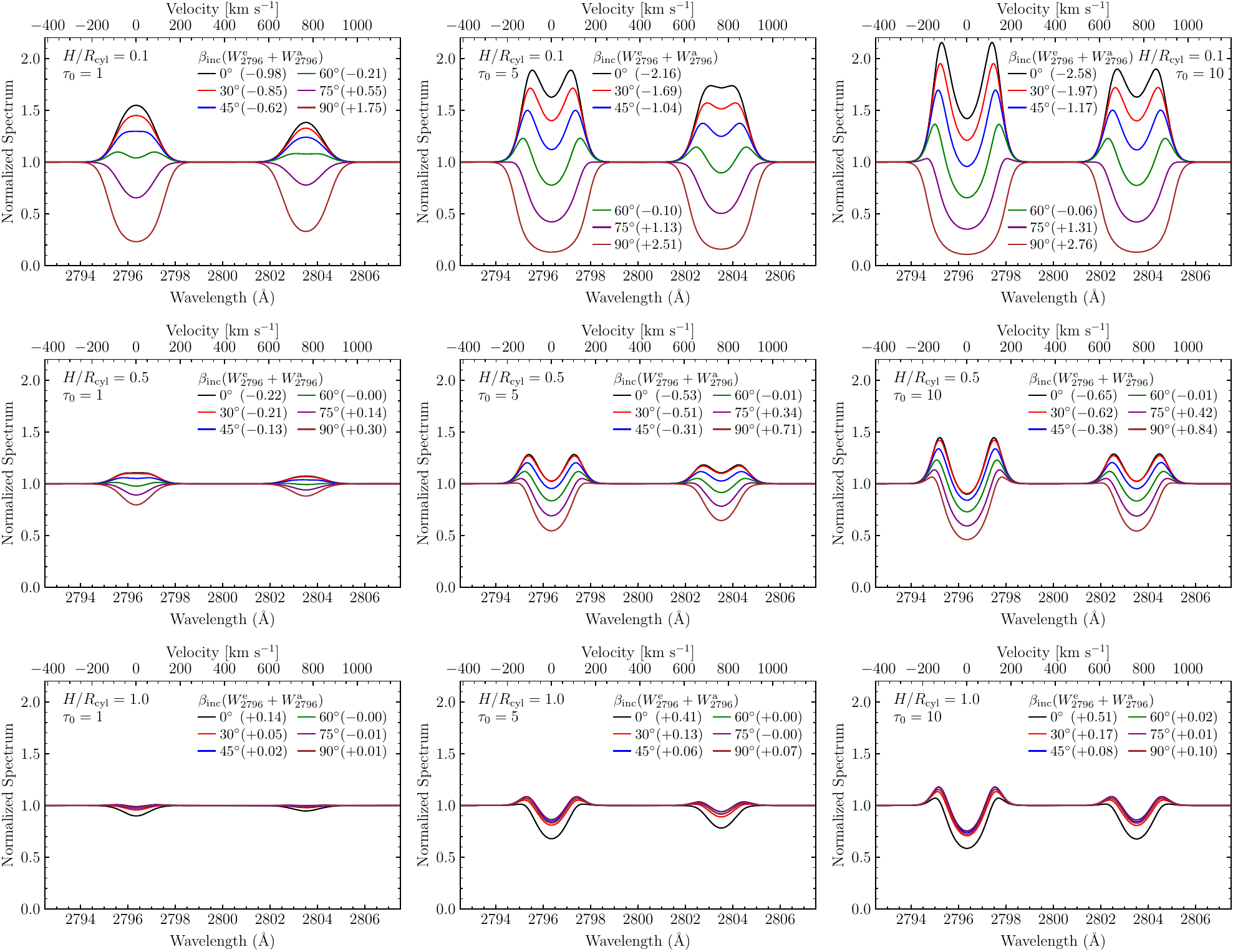}
\par\end{centering}
\begin{centering}
\medskip{}
\par\end{centering}
\caption{\label{fig14}Continuum spectra near \ion{Mg}{2} emission line spectra
predicted for the cylindrical models. The height-to-radius of the
cylinder varies from $H/R_{{\rm cyl}}=0.1$ to $H/R_{{\rm cyl}}=1.0$
from top to bottom. The optical depth varies from $\tau_{0}=1$ to
$\tau_{0}=10$ from left to right. In each panel, the model spectra
for the inclination angle $\beta_{{\rm inc}}$, ranging from $0^{\circ}$
to $90^{\circ}$, are shown in different colors. The numbers in the
parentheses represent the net EWs for the K line ($W_{2796}^{{\rm e}}+W_{2796}^{{\rm a}}$)
in units of \AA.}
\end{figure*}

The bottom panels in Figures \ref{fig13} display the doublet EW ratio between the K and H lines for both emission and absorption as a function of $W_{2796}^{\rm CG}$. The figures also present the prediction by the curve of growth in a static medium, illustrated by the thick gray line. The curve-of-growth theory predicts that in the linear regime ($\tau_0 \lesssim 1$), the EW ratio of a doublet is approximately equal to the ratio of their oscillator strengths ($\sim2$ for \ion{Mg}{2}), and it subsequently decreases with increasing optical depth, becoming approximately unity in the saturated regime ($\tau_0\gtrsim 10$). In the figures, the EW ratios are generally higher than those of the curve of growth. This is due to a more effective separation between emission and absorption features in the K line compared to the H line.
The doublet EW ratio shows a decreasing trend and eventually tends to approach a constant value ($\approx1$) with increasing $N_{\text{Mg}^{+}}$.

The figure also shows that the EW ratio for absorption ($W_{2796}^{{\rm a}}/W_{2803}^{{\rm a}}$) is approximately equal to that for emission ($W_{2796}^{{\rm e}}/W_{2803}^{{\rm e}}$) in static and slowly moving models.
In the model with the fastest expansion velocity of $V_{{\rm exp}}=300$ km s$^{-1}$, the transfer of the K-line emission flux leads to decreases in both $\left|W_{2796}^{{\rm e}}\right|$ and $W_{2803}^{{\rm a}}$, resulting in a decrease of $W_{2796}^{{\rm e}}/W_{2803}^{{\rm e}}$ and an increase of $W_{2796}^{{\rm a}}/W_{2803}^{{\rm a}}$. This trend causes an asymmetry of the EW ratio plot around $W_{{\rm 2796}}=0$, as indicated by purple cross symbols.

The EW ratios for both absorption and emission lines are within the range
of 1 to 2, except in cases of a static medium with a relatively low
column density of $N_{\text{Mg}^{+}}\lesssim3\times10^{14}$ cm$^{-2}$,
where the ratios exceed 2. This result is attributed to the effective cancellation of absorption and emission features in the H line, causing its EWs (denominator of the EW ratio) approach zero more rapidly than those of the K line as the optical depth decreases.  The excess above 2 is not caused by numerical noise during the calculation of EWs in the denominator. To validate the calculation, the number of photon packets was increased from $10^8$ to $10^{10}$, and no significant variation in the ratios was found as the number of photons increased.
In optically thick cases, the EW ratios
agree with the curve of growth theory because the absorption
and emission profiles are well separated.
It is also noteworthy that models with a higher expansion velocity, but not too fast, tend to align more closely with the results of the curve of growth than slower models. This is
because expanding media produce relatively well-distinct absorption
and emission line profiles.

\begin{figure*}[!t]
\begin{centering}
\includegraphics[clip,scale=0.48]{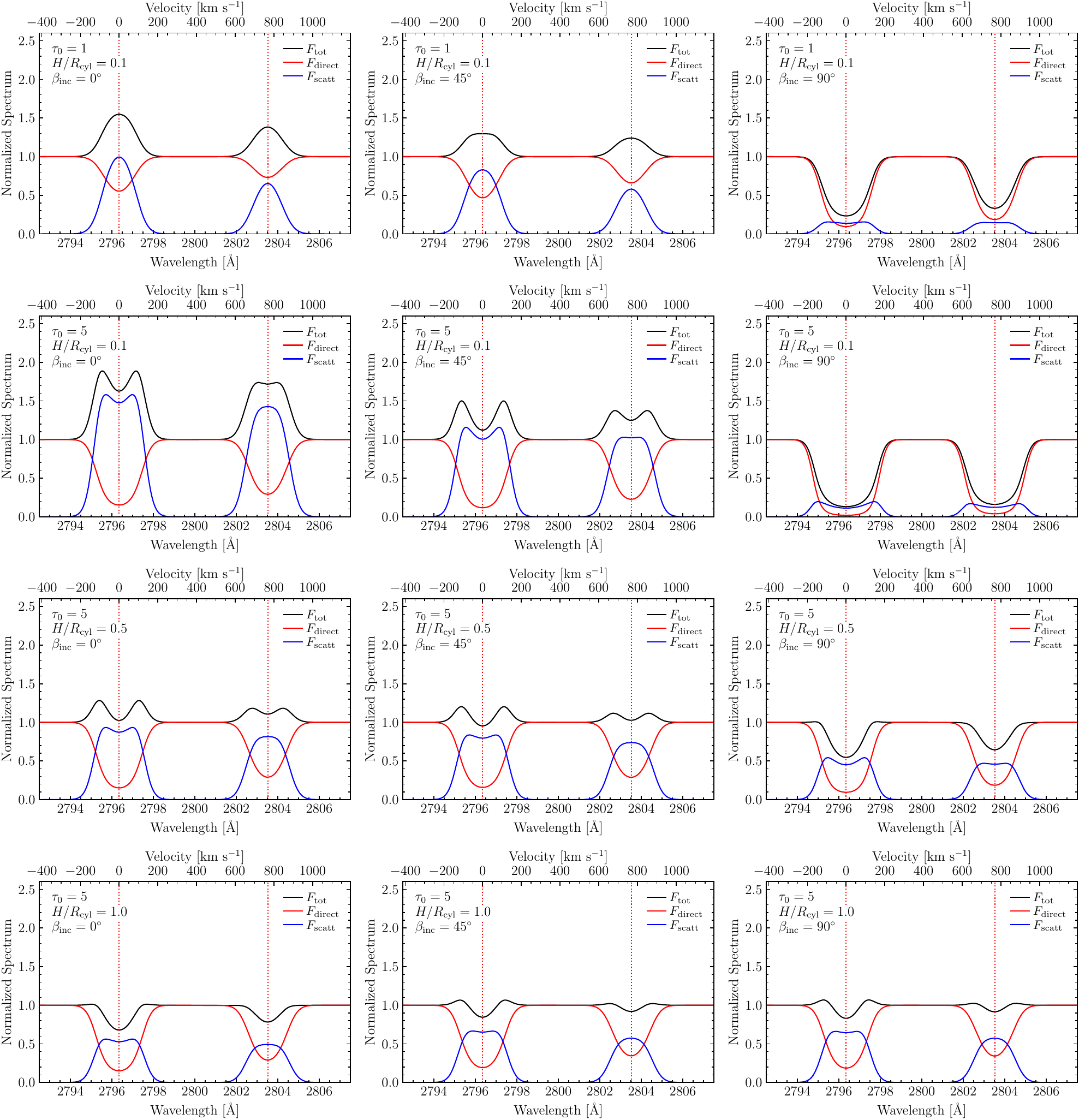}
\par\end{centering}
\caption{\label{fig15}Decomposition of the continuum spectra, predicted from the cylindrical models, into direct (absorption) and scattered (emission) components. The first and second rows show the dependence on the optical depth ($\tau_0=1, 5$) for a fixed value of $H/R_{\rm cyl}=0.1$. The second, third, and fourth rows display the variation of spectra depending on the $H/R_{\rm cyl}$ (0.1, 0.5, and 1.0) for a fixed value of $\tau_0=5$. The left, middle, and right panels show the decomposed spectra for the viewing angle $\beta_{\rm inc}=0^\circ$, $45^\circ$, and $90^\circ$, respectively. The total, direct, and scattered spectra are shown in black, red, and blue, respectively.}
\end{figure*}

The right panels in Figure \ref{fig13} shows the variation of the relative EWs and the doublet EW ratio in the presence of dust. Dust destroys continuum photons
near \ion{Mg}{2} line centers more effectively than those far from
the lines due to the trapping by multiple resonance scattering. This
effect increases the absorption line depth and reduces the emission
line strength. However, attenuation of the continuum by dust tends
to restore the strength of absorption EW, making its reduction less
significant. As a result, the emission EWs are substantially reduced
for high column density models with $N_{\text{Mg}^{+}}\gtrsim3\times10^{15}$ cm$^{-2}$, while the absorption EWs are less altered, as seen in Figure \ref{fig13} (see also the bottom panel of Figure \ref{fig10}).
Comparing the left and right panels in Figures \ref{fig13}, obtained before and after dust is included, clearly shows that the emission EWs are more significantly affected by dust than the absorption EWs. Thus, the relative EWs for absorption remain more or less unaltered, whereas those for emission lines are significantly changed. The EW ratios for emission are not significantly altered compared to cases with no dust because the emission EWs for both the K and H lines are simultaneously reduced.

\begin{figure*}[!t]
\begin{centering}
\includegraphics[clip,scale=0.465]{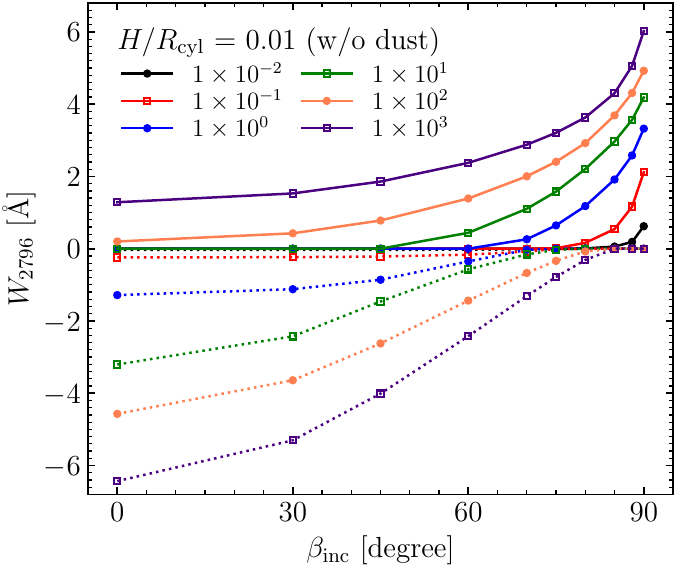}\ \ \includegraphics[clip,scale=0.465]{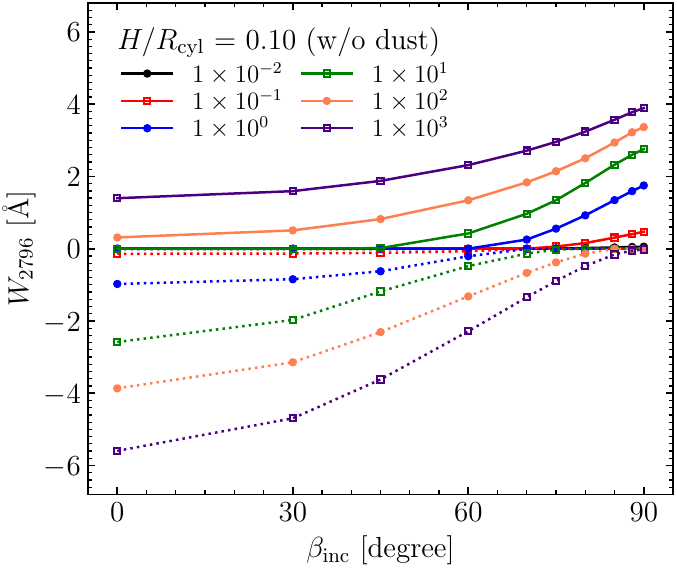}\ \ \includegraphics[clip,scale=0.465]{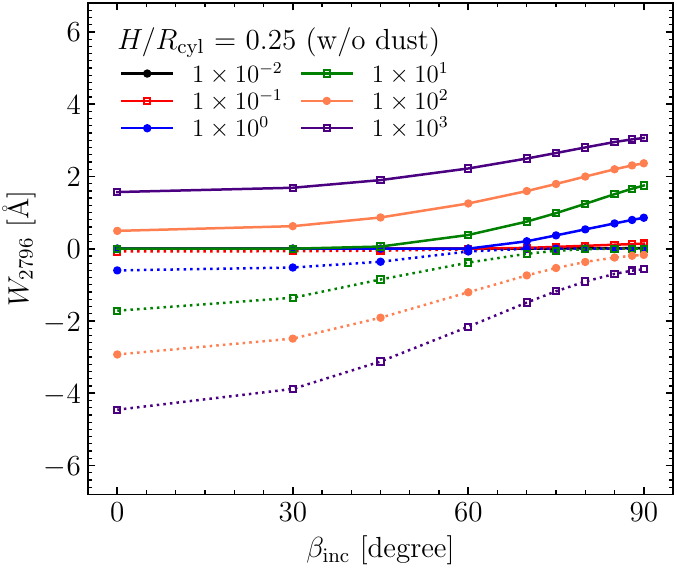}\smallskip{}
\par\end{centering}
\begin{centering}
\includegraphics[clip,scale=0.465]{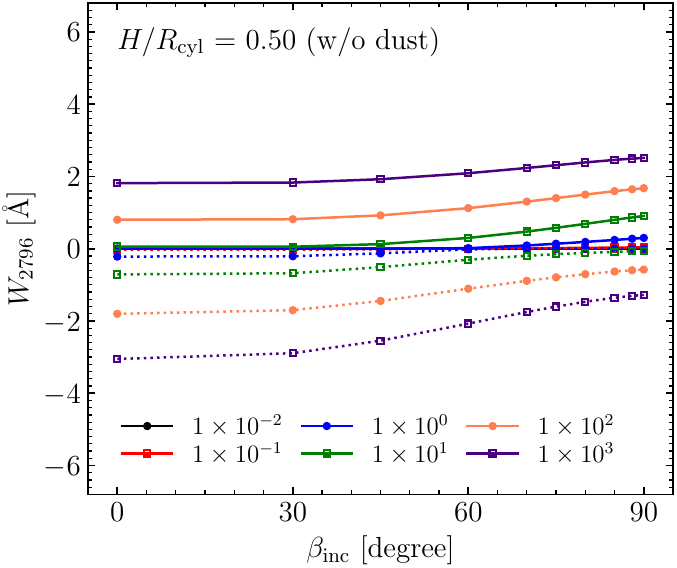}\ \ \includegraphics[clip,scale=0.465]{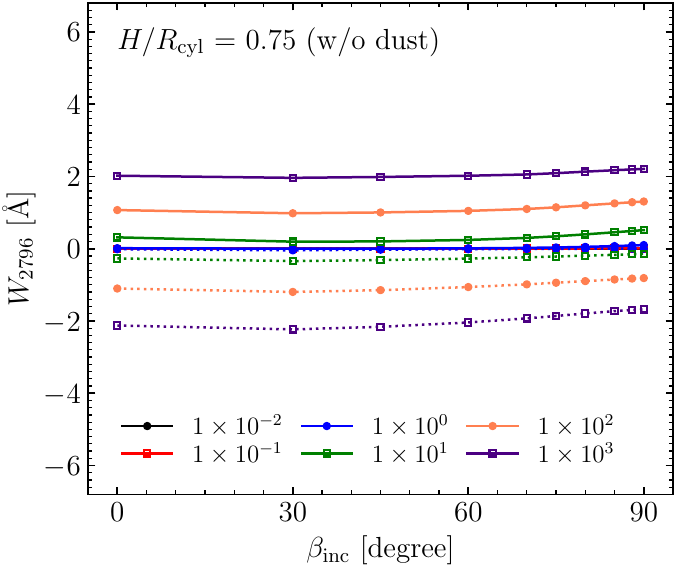}\ \ \includegraphics[clip,scale=0.465]{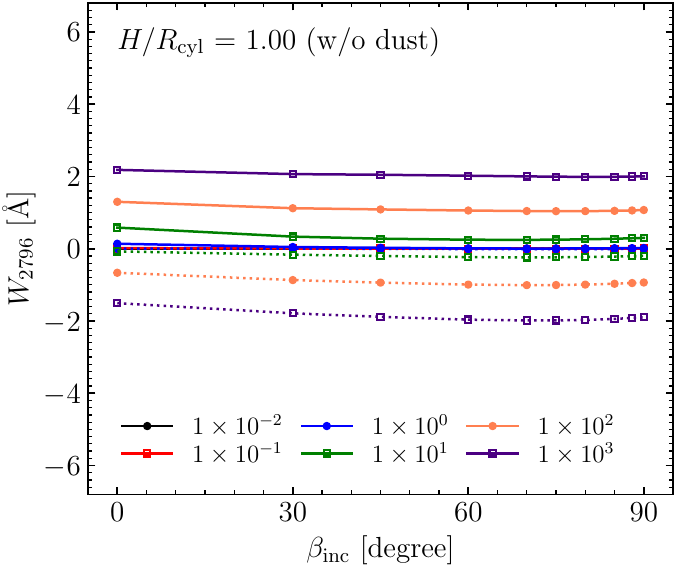}
\par\end{centering}
\caption{\label{fig16} Variation of the EWs of \ion{Mg}{2} $\lambda2796$
absorption and emission lines produced by the continuum, in the absence
of dust, depending on the $H/R_{{\rm cyl}}$ ratio, inclination angle
$\beta_{{\rm inc}}$, and optical depth $\tau_{0}$ of the cylinder. The numbers represent the optical depth $\tau_0$.
The solid lines with positive EWs denote the absorption features and
the dotted lines with negative EWs represent the emission features.}
\vspace{2mm}
\end{figure*}

\begin{figure*}[!t]
\begin{centering}
\includegraphics[clip,scale=0.46]{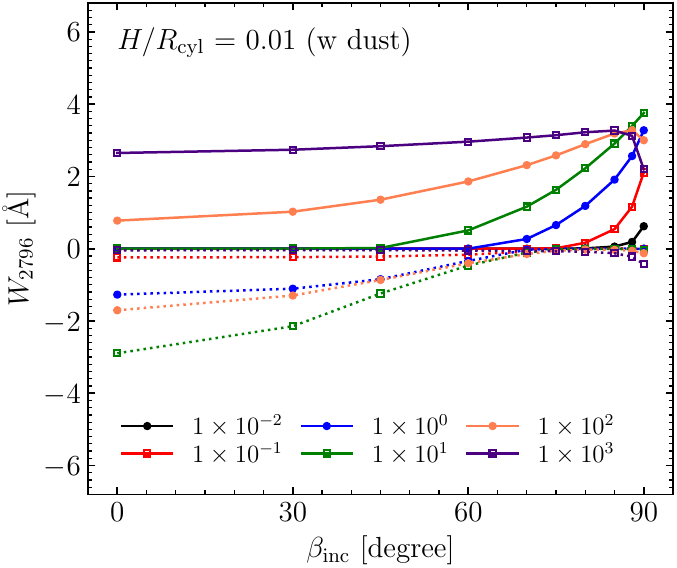}\ \ \includegraphics[clip,scale=0.46]{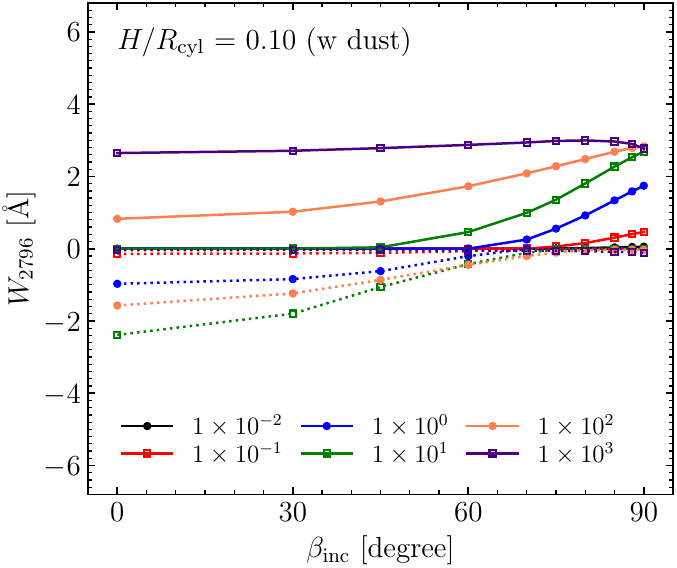}\ \ \includegraphics[clip,scale=0.46]{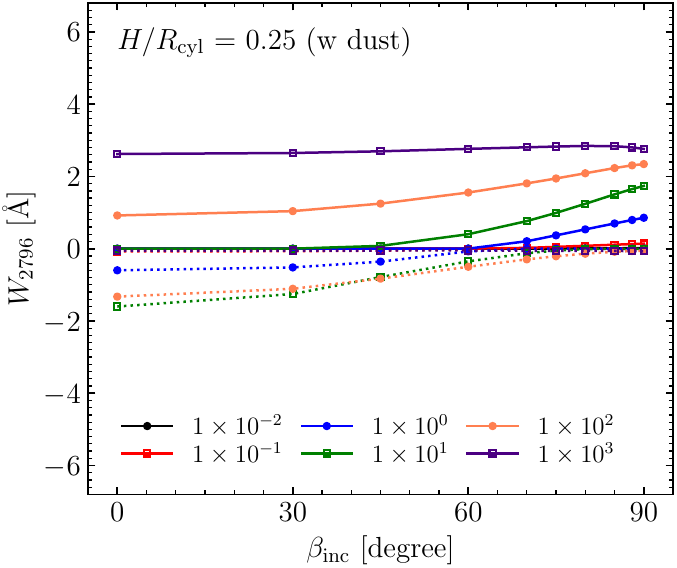}\smallskip{}
\par\end{centering}
\begin{centering}
\includegraphics[clip,scale=0.46]{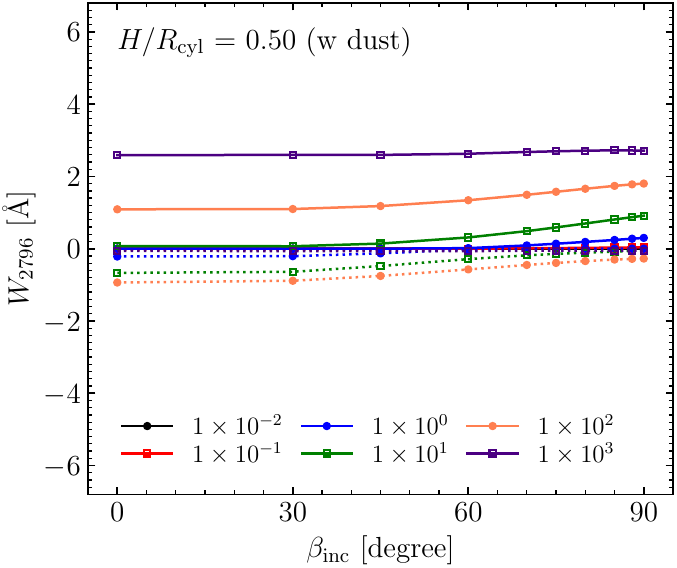}\ \ \includegraphics[clip,scale=0.46]{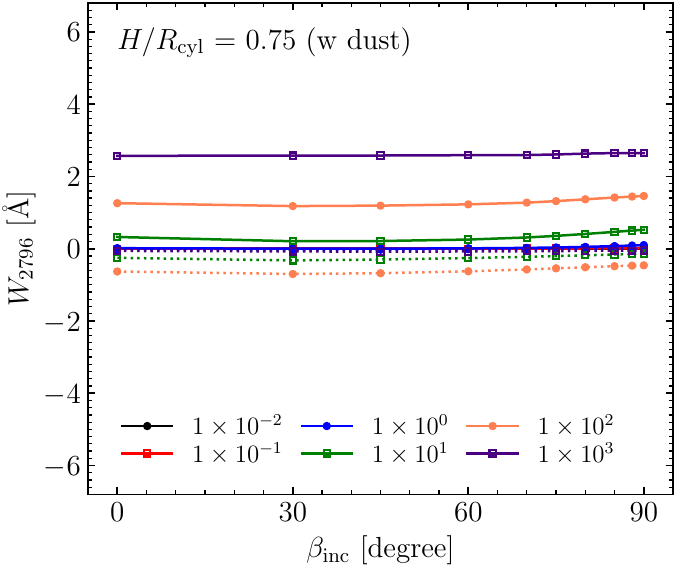}\ \ \includegraphics[clip,scale=0.46]{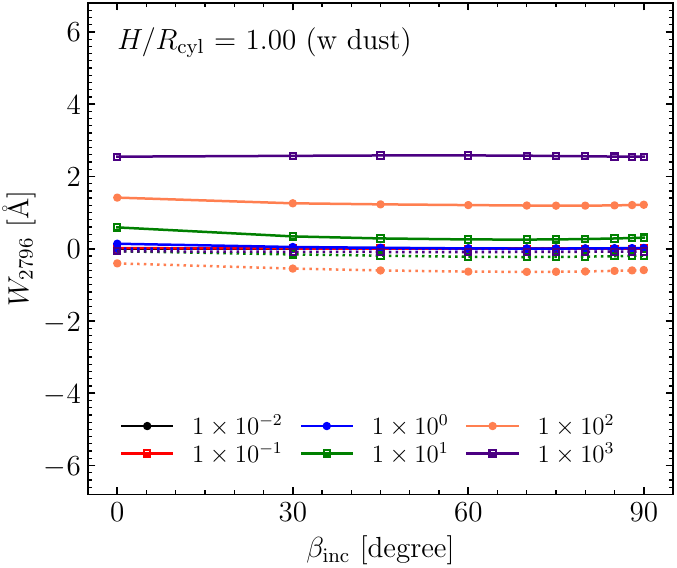}
\par\end{centering}
\caption{\label{fig17} Variation of the EWs of \ion{Mg}{2} $\lambda2796$
absorption and emission lines produced by the continuum, in the presence
of dust, depending on the $H/R_{{\rm cyl}}$ ratio, inclination angle
$\beta_{{\rm inc}}$, and optical depth $\tau_{0}$ of the cylinder. The numbers represent the optical depth $\tau_0$.
The solid lines with positive EWs denote the absorption features and
the dotted lines with negative EWs represent the emission features.}
\vspace{1mm}
\end{figure*}

\begin{figure*}[t]
\begin{centering}
\medskip{}
\includegraphics[clip,scale=0.46]{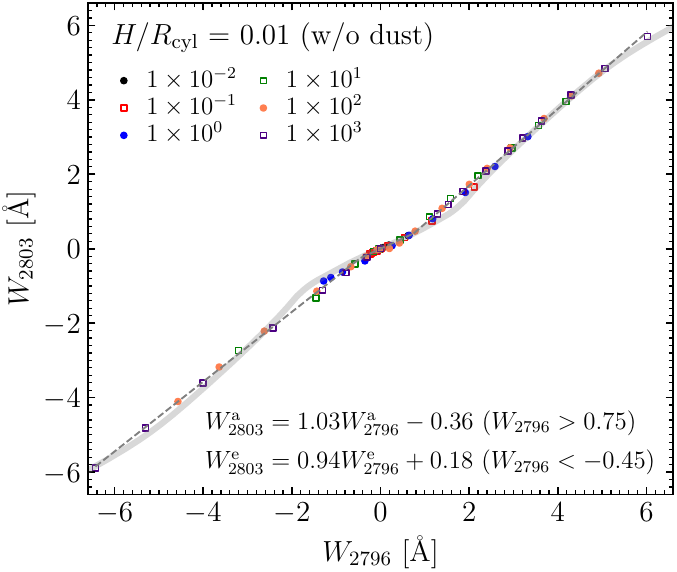}\ \ \includegraphics[clip,scale=0.46]{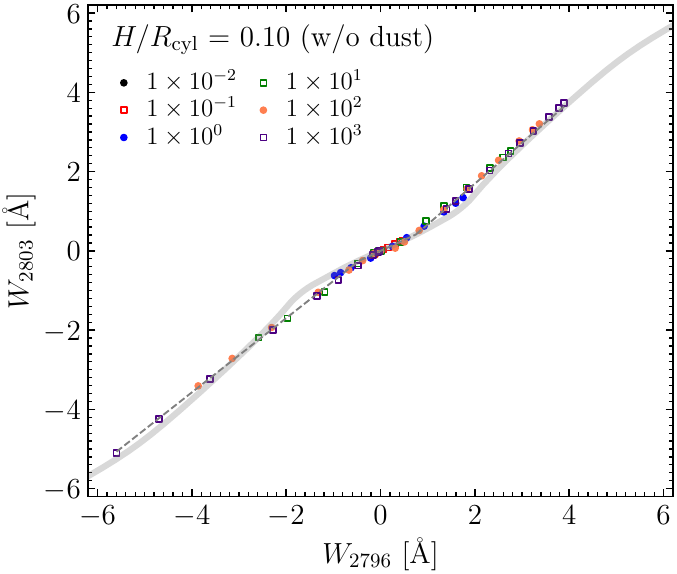}\ \ \includegraphics[clip,scale=0.46]{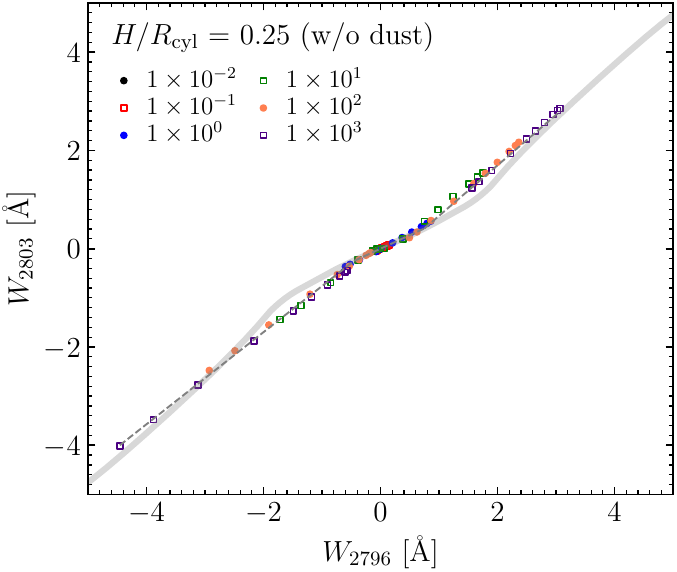}
\par\end{centering}
\begin{centering}
\includegraphics[clip,scale=0.46]{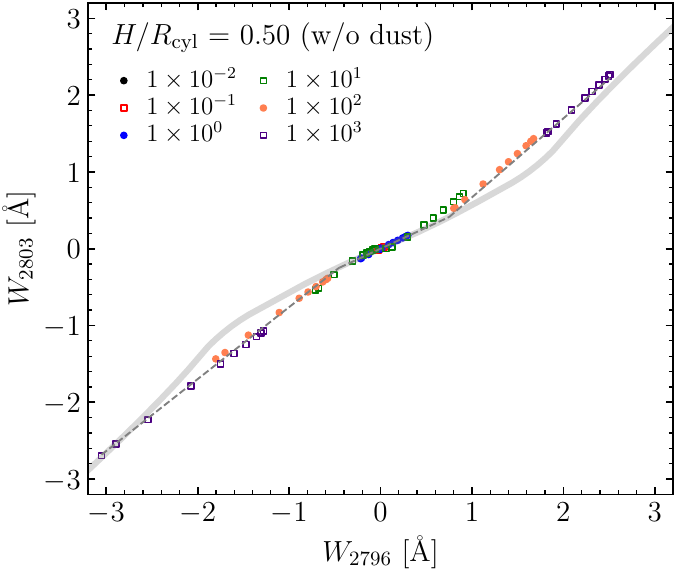}\ \ \includegraphics[clip,scale=0.46]{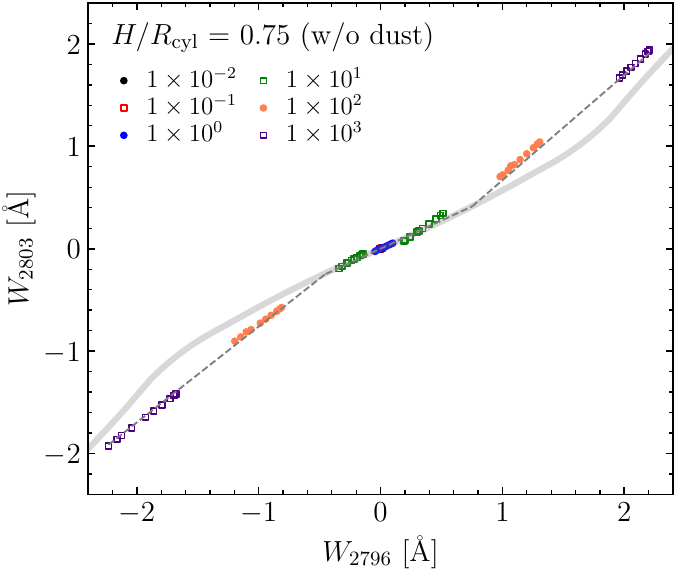}\ \ \includegraphics[clip,scale=0.46]{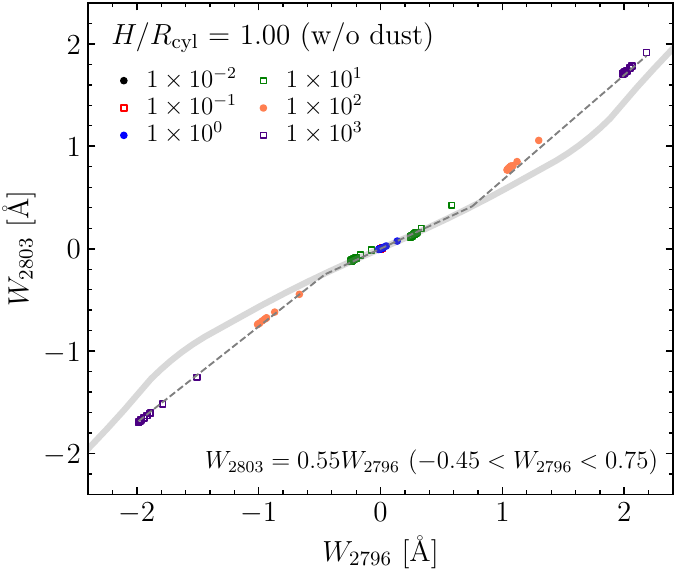}
\par\end{centering}
\medskip{}
\caption{\label{fig18} EW of \ion{Mg}{2} $\lambda2803$ vs. EW of \ion{Mg}{2} $\lambda2796$ produced by the continuum in the cylinder model, in
the absence of dust grains. The symbols represent different optical depth $\tau_0$, indicated by numbers. 
The thick gray curve shows the relation
predicted by the curve of growth for absorption lines. Three thin
gray lines in each panel represent the best-fit linear functions to
the simulation results in three regimes, divided by $W_{2796}=-0.45$
\AA\ and $W_{2796}=0.75$ \AA. The equation representing the middle
regime ($-0.45$\AA\ $<W_{1796}<0.75$\AA) is shown in the last
(bottom right) panel. The two other equations representing the outside
of the middle regime are shown in the first (top left) panel.}
\medskip{}
\end{figure*}

\subsection{Cylindrical Model - Continuum}

\label{subsec:3.4}

Figure \ref{fig14} shows example spectra obtained from resonance
scattering of a flat continuum in cylindrical models. The figure illustrates
the cases with $\tau_{0}=1$, 5, and 10, and $H/R_{{\rm cyl}}=0.1$,
0.5, and 1.0, which are the same as those in Figure \ref{fig04}.
Dust is assumed to be absent. The net EWs for the K line, defined
as the sum of emission and absorption EWs ($W_{2796}^{{\rm e}}+W_{2796}^{{\rm a}}$),
are also shown in parentheses. Surprisingly, unlike the spherical
models, these non-spherical models can give rise to pure absorption
or pure emission spectra, depending on the height-to-radius ratio
and the inclination angle. In an edge-on view, the spectra tend to
show pure absorption, particularly in optically thin and geometrically
thin models. On the other hand, in a face-on view, the spectra exhibit
pure emission, except for the models with $H/R_{{\rm cyl}}\approx1$.
These properties can be understood as in Figure \ref{fig09}. Photons
will easily escape in the vertical direction while they experience
resonance scatterings. In contrast, photons scattered radially will
have to undergo much more scattering before escaping, and thus, few
photons will escape in the radial direction. These trends give rise
to pure absorption spectra in an edge-on view, and pure emission spectra
in a face-on view.

The absorption and emission EWs both reach their minima at $\beta_{{\rm inc}}\approx60^{\circ}$
(denoted in green) for a given $H/R_{{\rm cyl}}$ and $\tau_{0}$,
indicating that they are effectively mixed and canceled. Similar to
the intrinsic emission line model discussed in Section \ref{subsec:3.2},
spectra obtained from a round cylinder ($H/R_{{\rm cyl}}=1$) show
relatively insensitivity to variations in the viewing angle. In addition,
the spectra exhibit minima both in absorption depths and emission
peaks when compared to flat models. In the bottom panels ($H/R_{{\rm cyl}}=1$),
the spectra observed at $\beta_{{\rm inc}}=90^{\circ}$ appear slightly
different from those obtained at other angles. This difference occurs
because, at this particular angle, the column density of Mg$^{+}$
gas is highest, resulting in the strongest absorptions. Dust scattering and absorption effects occur similarly to those in the spherical model; therefore, spectra of the models with dust are not shown in this paper.

To examine the spectral shape in more detail, spectra decomposed into direct and scattered components for a few models are shown in Figure \ref{fig15}, similar to the results presented in Figure \ref{fig11} for the spherical models. In the figures, the black, red, and blue lines denote the total, direct, and scattered spectra, respectively. As the viewing angle $\beta_{\rm inc}$ increases (from face-on to edge-on), the contribution of scattered light decreases. Comparing the first and second rows shows that the scattered (emission) spectrum becomes broader than the direct (absorption) spectrum and reveals double peaks as the optical depth increases. The direct and scattered profiles are similar in shape, except for being upside-down from each other, when the optical depth is low; consequently, the total spectrum exhibits relatively weak `net' absorption and emission features. The second to fourth rows reveal a decrease of scattered light flux as the medium approaches a round shape and is viewed face-on ($\beta_{\rm inc}=0^\circ$ and $45^\circ$). However, when viewed edge-on ($\beta_{\rm inc}=90^\circ$), the contribution of scattered light increases, but the scattered line profile narrows as the cylinder becomes round. This occurs because the optical thickness along the edge-on line of sight decreases as the cylinder becomes more rounded, according to the vertical definition of $\tau_0$. The right panels ($\beta_{\rm inc}=90^\circ$) in the first and second rows also show that the scattered light profile is broader in a higher optical depth model. These trends are essentially consistent with those of the spherical model but are slightly complicated by the non-spherical geometry effect.

The EW of the \ion{Mg}{2} K line, in the absence of dust, is shown
as a function of $\beta_{{\rm inc}}$ for various combinations of
$H/R_{{\rm cyl}}$ and $\tau_{0}$ in Figure \ref{fig16}. The figure
illustrates that the EW for emission is highest when viewed face-on with the lowest $H/R_{{\rm cyl}}$ ratio, while the EW for absorption
is highest in the edge-on view (with the lowest $H/R_{{\rm cyl}}$).
This property was also found in Figures \ref{fig14} and \ref{fig15}. The highest achievable
EW for emission in the parameter space studied in this paper is $\left|W_{2796}^{{\rm e}}\right|\approx6.5$
\AA, while the highest EW for absorption reaches $W_{2796}^{{\rm a}}\approx6$
\AA. When the medium becomes much round, both absorption and emission
EWs are confined within the range of $0\lesssim\left|W_{2796}^{{\rm e,a}}\right|\lesssim2$
\AA. As $H/R_{{\rm cyl}}$ approaches 1, the EWs tend to be independent
of the viewing angle due to the system's increased sphericity, as
discussed with in Figure \ref{fig14}. For a fixed $H/R_{{\rm cyl}}$
and $\beta_{{\rm inc}}$, the EWs for both absorption and emission
increase as $\tau_{0}$ increases. This is not only because absorption
increases with higher optical depth but also because the absorption
and emission profiles become more distinct at greater optical depths.

In the presence of dust, the absorption and emission EWs of the K
line are presented as a function of $\beta_{{\rm inc}}$ in Figure
\ref{fig17}. The figure shows decreases in emission EWs by dust,
as compared to Figure \ref{fig16}. Dust attenuation causes a reduction
in the continuum level, which would increase the emission EW if the
emission line's strength remained constant. However, dust more effectively
destroys photons near \ion{Mg}{2} resonance wavelengths due to their
elongated path lengths caused by resonances compared to continuum
photons. This effect outweighs the reduction in the continuum, resulting
in an overall decrease in the final emission EW. This effect becomes
most noticeable when $\tau_{0}\gtrsim10$. In particular, the emission
EWs for the cases with the highest optical depths become lower than
those with lower optical depths. Similar results were also found in
spherical models shown in Figure \ref{fig13}.

\begin{figure*}[t]
\begin{centering}
\medskip{}
\includegraphics[clip,scale=0.5]{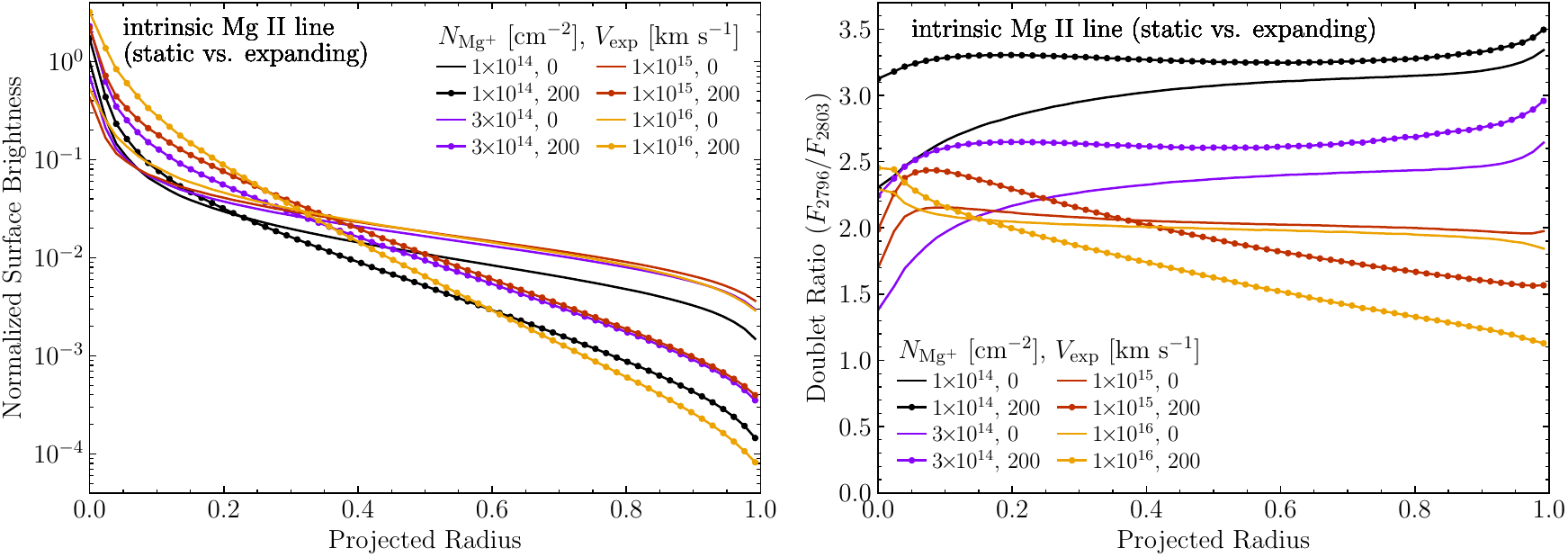}\medskip{}
\includegraphics[clip,scale=0.5]{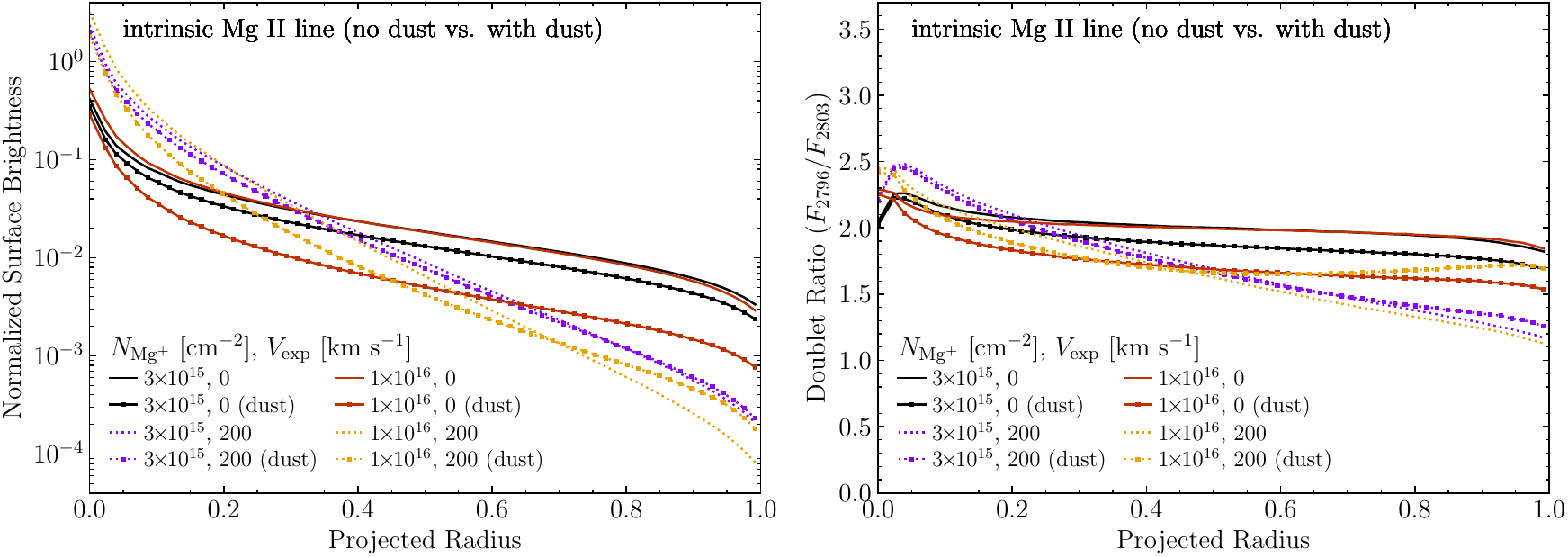}
\par\end{centering}
\medskip{}
\caption{\label{fig19} Radial profiles of the surface brightness (left panels) and doublet flux ratio (right panels) of the intrinsic \ion{Mg}{2} emission line in spherical models. The \ion{Mg}{2} photons are emitted from the center. Direct light from the center is not shown, and the figures display the results for scattering halo only. The surface brightness profiles were constructed by summing up both the K and H lines. The top panels compare the results for static media and expanding media in the absence of dust. The bottom panels compare the results between cases with no dust and cases with dust in optically thick media. Different colors denote various column densities. In the top panels, solid lines without circles represent static models, while solid lines with circles represent expanding models. In the bottom panels, solid lines represent static models, dotted lines represent expanding models, and models that include dust are represented by lines with square symbols. The radius is normalized to the maximum radius of the sphere. All surface brightnesses are normalized to have the same unit.}
\medskip{}
\end{figure*}

As opposed to the emission EWs, absorption EWs generally tend to increase,
except in some instances. As previously described in the spherical
model, the presence of dust causes the absorption features to become
deeper, leading to an overall enhancement in the absorption EW. However,
this enhancement is less substantial than the decrease in the emission
EW. In exceptional cases, when geometrically thin (but optically thick)
flat cylinders are viewed edge-on ($H/R_{{\rm cyl}}\lesssim0.25$,
$\tau_{0}\gtrsim50$, and $\beta_{{\rm inc}}\gtrsim75^{\circ}$),
the resonantly scattered photons escape quickly perpendicular to the
line of sight before being destroyed by dust. Dust-scattered continuum
photons will also escape predominantly in the vertical direction.
Therefore, continuum photons experience significant extinction due
to dust along the line of sight without being compensated by dust-scattered
light. As a result, in these particular cases, the absorption EW decreases
due to dust, contrary to the general trend. Finally, since the absorption
EW increases when viewed face -on ($\beta_{{\rm inc}}\approx0^{\circ}$),
its inclination angle dependence in Figure \ref{fig17} is significantly
reduced for the optically thick flat cylinders ($H/R_{{\rm cyl}}=0.01$
and 0.1).

Figure \ref{fig18} compares the EWs of the \ion{Mg}{2} K line with
those of the H line. In the figure, optical depths are denoted using
different colors and symbols. The prediction by the curve of growth
theory for pure absorption lines is also shown in a thick gray curve.
In this curve, the emission EWs are assumed to be equal to the negative
of the absorption EWs. The EWs differ from those predicted by the curve of growth due to the filling-in of absorption by emission, as previously discussed. It is noteworthy that three piecewise linear
functions can well represent their relationship. When the EWs of the
K and H lines are small ($-0.45$\AA\ $<W_{2796}<0.75$\AA), their
relation is well described by $W_{2803}=0.55W_{2796}$, which is consistent
with the prediction of the curve of growth in an optically thin limit,
as shown in the last panel of the figure. Outside of this regime,
the equations representing the relationships between the EWs of the
K and H lines are presented in the first panel. The curve of growth
reproduces the best-fit linear functions, and the simulation results
within an error of at most 10\% for ranges of $W_{2796}^{{\rm e}}\lesssim-2.0$\AA\
and 2.2\AA\ $\lesssim W_{2796}^{{\rm a}}\lesssim$ 7.3\AA.

\subsection{Radial Variation in Spherical Model}

\label{subsec:3.5}

\begin{figure*}[t]
\begin{centering}
\medskip{}
\includegraphics[clip,scale=0.5]{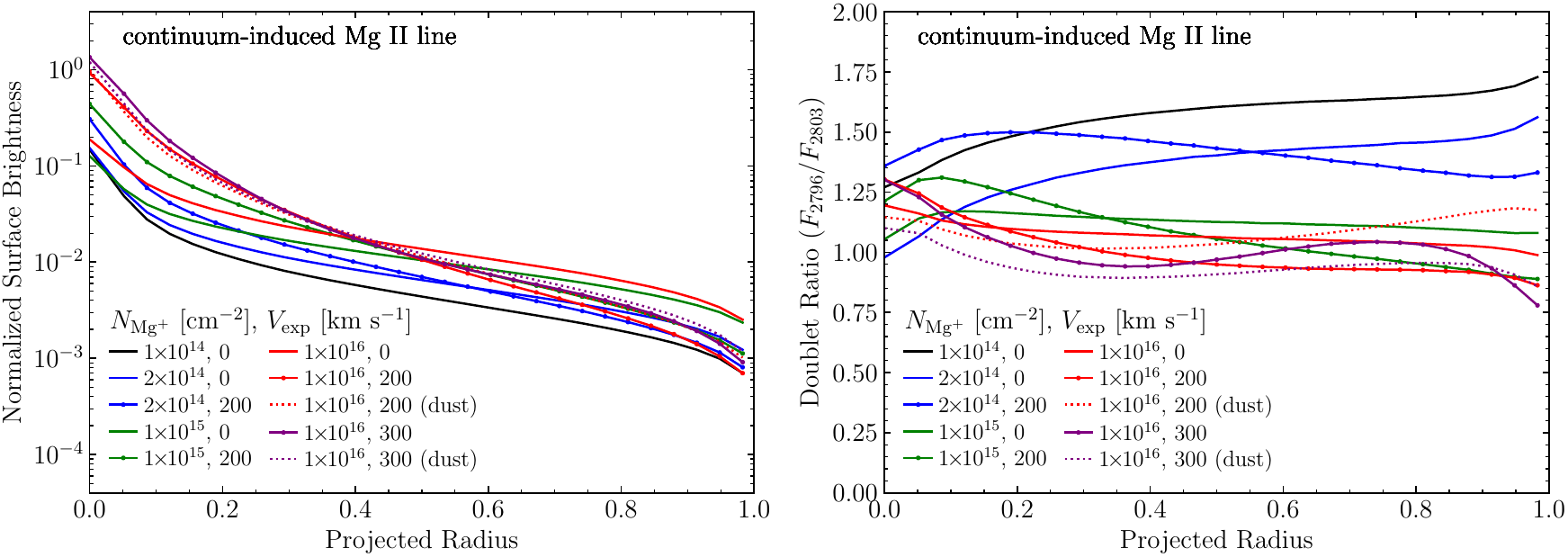}
\par\end{centering}
\medskip{}
\caption{\label{fig20} Radial profiles of the surface brightness (left panel)
and doublet flux ratio (right panel) of the continuum-pumped
\ion{Mg}{2} line in spherical models with a continuum source at the center. The radial profiles are constructed from the \ion{Mg}{2} emission line originating in the continuum scattering process, with no contribution of direct emission from the center. Different colors denote various column densities. Solid lines without circles represent static models, while solid lines with circles represent expanding models. Dotted lines denote models that include dust, as shown in legends. The radius is normalized to the maximum radius of the sphere. All surface brightnesses are normalized to have the same unit.}
\medskip{}
\end{figure*}

In this section, we discuss the radial profiles of surface brightness and doublet flux ratio for the \ion{Mg}{2} emission line halos caused by the scattering of the intrinsic \ion{Mg}{2} emission line (Figure \ref{fig19}) and continuum (Figure \ref{fig20}). In both models, photons originate from the center of a sphere; therefore, the extended surface brightness profiles shown in the figures are solely due to scattered light. The surface brightness profile of the scattered \ion{Mg}{2} line is relatively easy to interpret. However, understanding the spatial variation of the doublet flux ratio is quite complex due to the differences in the line width and frequency shift of the K and H lines arising from the difference in the number of scatterings they experience. These differences between the K and H lines are primarily established in the central region near the source, where most scattering events occur.
Consequently, the differences
originating from the central region subsequently influence the number
of scatterings occurring in the outer region, eventually affecting
the doublet ratio in the outer region. The presence of dust further
complicates the interpretation of results because dust scattering
operates independently of wavelength, while resonance scattering depends
strongly on the wavelength shift from the line center.

\subsubsection{Line Emission Model}

The left and right panels in Figure \ref{fig19} show the radial profiles of surface brightness and the doublet ratio, respectively, for the \ion{Mg}{2} emission line halo produced by a central, intrinsic emission line source. In the figure, the solid lines with no circles represent static models with various column densities ranging from $10^{14}$ to $10^{16}$ cm$^{-2}$. The solid lines with circles denote expanding media, and the dotted lines represent the models with dust. In the following, we first discuss the surface brightness and then address the doublet ratio.

The top panels show the results obtained when there is no dust. The surface brightness profile of the scattered line always exhibits a peak in the central region and declines in the outer region. The peak intensity and slope of the profile depend on the column density and expanding velocity. The directly escaping component causes a strong peak at $r=0$ in the surface brightness profiles of low column-density models (left panels), but this component is not shown in the figures. In static media, as the column density increases, the central peak tends to decrease unless $N_{\text{Mg}^+}$ is not too high. Simultaneously, the extended, scattered component in the outer region is enhanced, leading to a flatter surface brightness profile with increasing column density. The decrease of intensity in the inner region and the increase in the outer region result from the increased number of scatterings and subsequent spatial dispersion of photons as the column density increases.

However, the trend becomes reversed as the column density increases further. In the model with $N_{\text{Mg}^{+}} = 10^{16}$ cm$^{-2}$, the peak at $r = 0$ is slightly lower, and the intensity at $r \approx 0.1$ is enhanced, while it decreases slightly in the outer region ($r \gtrsim 0.5$) compared to the model with $N_{\text{Mg}^{+}} = 10^{15}$ cm$^{-2}$, although this difference is not easily visible in the figure at first glance.
This trend occurs because photons are strongly trapped in the inner region ($r\lesssim0.2$) and undergo a large number of scatterings there.
In such a high optical depth medium, photons will escape the trapped
region through a `single longest flight' or `excursion' when their
frequencies are shifted to a critical frequency at which the optical
depth of the medium is approximately unity, similar to the Ly$\alpha$
RT process \citep{Adams1972}. Consequently, the radial profile for
$N_{\text{Mg}^{+}}=10^{16}$ cm$^{-2}$ shows a higher intensity in the inner region
and a slightly steeper slope in the outer region than that of $N_{\text{Mg}^{+}}=10^{15}$ cm$^{-2}$ because photons with frequencies that have been significantly shifted in the inner region are scattered less in the outer region.

In the case of an expanding medium, the central region is enhanced, and the outer region is lowered compared to its corresponding static medium. This is because most photons undergo scattering only near the central region, where the medium is almost static, and they tend to quickly escape without undergoing further scattering in the outer region. This property arises from a reduction of the effective optical depth in the outer region of expanding media and is thus pronounced when $V_{{\rm exp}}$ is higher. The qualitative dependence of the surface brightness profile on the column density in expanding media is similar to that in static media.

The lower panels of Figure \ref{fig19} compare the results in the presence of dust with those obtained without dust. Only high column density models are presented in the figure. In static models, the surface brightness is reduced due to dust extinction, as anticipated. However, in expanding models represented by dotted lines, the profiles exhibit more complex behavior. For instance, in the models expanding with $V_{{\rm exp}}=200$ km s$^{-1}$, the brightness of the central region is reduced because a substantial fraction of core photons near the line center is absorbed by dust in that region. However, it is found  that the outer region becomes slightly brighter than a medium with no dust. This unexpected enhancement in the brightness of the outer region is caused by dust scattering, which operates independently of the photon frequency and, therefore, the fluid velocity.
Resonance scattering is the only process that creates an outer halo when there is no dust. However, in the presence of dust, dust scattering can dominate over resonance scattering because the latter rarely occurs in rapidly expanding outer regions due to the frequency mismatch. Consequently, the outer region in a rapidly expanding high-density medium becomes brighter when dust is present.

Now, we discuss the radial profile of the doublet ratio, as shown in the right panels of Figure \ref{fig19}. It has been previously mentioned that the doublet ratio is always
2 in spherical media with no dust when averaged over the whole system.
However, it is found that the ratio can vary spatially. This spatial variation
is due to the difference in locations where K- and H-line photons
are primarily scattered and the dependence of these locations on the
optical depth.

In the right upper panel, the doublet ratio tends to progressively increase with radius when the column density is low ($N_{\text{Mg}^+}<10^{15}$ cm$^{-2}$). In models with intermediate column density ($N_{\text{Mg}^+}\approx 10^{15}$ cm$^{-2}$), however, the ratio increases near the central region and then declines at larger distances. In the model with the highest density ($N_{\text{Mg}^+}= 10^{16}$ cm$^{-2}$), the ratio attains its peak at the center and tends to decrease as the radius increases.

The dependences of the doublet ratio in the central and outer regions on the column density can be understood as follows. As the column density increases, the doublet ratio at the central region first decreases ($N_{\text{Mg}^{+}}\lesssim3\times10^{14}$
cm$^{-2}$) and then increases ($N_{\text{Mg}^{+}}\gtrsim3\times10^{14}$
cm$^{-2}$). When $N_{\text{Mg}^{+}}=10^{14}$ cm$^{-2}$, most K-line
photons are scattered once or twice, while H-line photons mostly escape
without being scattered. Thus, the total flux of scattered H-line
photons is much lower, and the surface brightness of the H-line halo
drops very quickly with radius compared to the K-line halo, resulting
in the doublet ratio in the halo always being larger than two and
increasing with radius. In the model with $N_{\text{Mg}^{+}}=3\times10^{14}$
cm$^{-2}$, K-line photons are multiply scattered, whereas H-line
photons are singly scattered on average. Then, the K-line halo is
more spatially extended than the H-line halo, leading to $R<2$ in
the central region and $R>2$ in the outer region. When the column
density is higher, both K- and H-line photons are multiply scattered,
and their surface brightness profiles tend to become flat, with their
ratio approaching $R\approx2$. When the column density is even higher
enough, K-line photons are trapped in a smaller region than H-line
photons (but with the same optical thickness). Thus, the central regions
have a doublet ratio slightly larger than, but not too much larger than, 2.
In the outer region ($r\gtrsim0.25$), the doublet ratio
tends to be higher than $2$ when $N_{\text{Mg}^{+}}<10^{15}$ cm$^{-2}$. 
This is because K-line photons are more spatially extended in relatively
optically thin models due to their higher optical depth than H-line
photons. However, the doublet ratio approaches $R\approx2$ for optically
thick models as both K- and H-line photons experience enough scatterings.
Consequently, the doublet ratio in the outer region decreases with
increasing $N_{\text{Mg}^{+}}$ until it approaches 2.

As the medium expands, the effective optical depth decreases, leading to a slight decrease in scattered halo. The brightness of direct emission at the center increases, but this is not shown in the figure. In optically thin cases, the K-line are scattered more frequently than the H-line. Therefore, in a low column density case of $N_{\text{Mg}^{+}}=3\times10^{14}$ cm$^{-2}$, the central region of the expanding medium with $V_{{\rm exp}}=200$ km s$^{-1}$ shows higher doublet ratios than the static model.
The outer region of this
model also exhibits higher doublet ratios due to the velocity-induced
reduction in optical depth. On the other hand, for high optical depth
models, the outer region tends to show $R<2$. This is because K-line
photons undergo more scattering and experience greater frequency shifts
than H-line photons in the inner regions, resulting in less scattering
of K-line photons in the outer regions. This effect gives rise to
doublet ratios of $R<2$.

In the presence of dust, as shown in the bottom right panel of Figure \ref{fig19}, dust scattering occurs regardless of whether it is K- or H-line and independently of the frequency shift. This effect not only reduces the dependence of the doublet ratio on the optical depth but also diminishes the radial variation of the ratio. Ultimately, dust scattering tends to cause the doublet ratio to approach 2. Note, however, that this effect occurs only when $N_{\text{Mg}^+}$ is high.

\subsubsection{Continuum Model}

Figure \ref{fig20} shows the radial profiles of the surface brightness
(left panel) and the doublet ratio (right panel) of the continuum-pumped \ion{Mg}{2} emission line.
In the models, a continuum source is located at the sphere center, resulting in halo emission solely composed of the pure \ion{Mg}{2} line. The figures do not show peaks arising from direct emission at the center.
Compared to the intrinsic \ion{Mg}{2} line case, the most significant difference is that the doublet ratio always falls within the range of 0.75 to 1.75, and no line ratio was found to exceed 2. Similarly to the intrinsic \ion{Mg}{2} line case, the surface brightness profiles of the continuum-pumped emission in expanding media are steeper than those of static cases. However, the degree of steepening is less pronounced compared to the intrinsic \ion{Mg}{2} line case. This is because the continuum source supplies photons capable of resonant scattering, even in the fastest expanding regions, regardless of fluid velocity.
In contrary, intrinsic \ion{Mg}{2} line photons undergo resonance much less frequently.
Therefore, the intrinsic \ion{Mg}{2} line source produces much
steeper profiles in expanding media than the continuum source.
In the presence
of dust, the continuum photons in fast-moving regions can be scattered
not only by dust but also by resonance, whereas the intrinsic \ion{Mg}{2} line
photons are primarily scattered by dust alone. This difference results
in a relatively small change in the surface brightness profile, independent
of the medium's velocity, even when dust is included.

The intrinsic \ion{Mg}{2} line source produces K- and H-line photons with a flux ratio of 2:1, while the continuum source can supply photons with equal fluxes at the K- and H-line frequencies due to our assumption of a flat spectrum independent of wavelength. This condition causes the doublet ratio to start at a ratio of 1:1. The difference in optical depths at the K- and H-line frequencies causes the ratio in the scattered halo to vary, but it always remains less than two.
In Figure \ref{fig20}, the doublet flux ratio for most models, except for models with the lowest column densities ($N_{\text{Mg}^{+}}\lesssim3\times10^{14}$ cm$^{-2}$), is found to be approximately $1$. In relatively low column-density models, photons with H-line frequency are rarely scattered and thus produce a relatively weaker scattering halo than K-line photons, yielding higher doublet ratios. In contrast, when the column density is high enough, both K- and H-line photons are scattered sufficiently, producing doublet ratios $\sim1$ (i.e., the initial ratio from the continuum).

\section{DISCUSSION}

\label{sec:4}

This section begins by discussing the observational implications of
the present results, specifically regarding the doublet ratio and
escape fraction. It then discusses the \ion{Mg}{2} emission mechanisms
and the sites from which the \ion{Mg}{2} emission lines originate.
Furthermore, it highlights the importance of distinguishing the RT
effects in understanding \ion{Mg}{2} lines from those obtained using
a simple foreground screen model. Additionally, this section covers
other resonance lines with atomic-level structures resembling the
\ion{Mg}{2} doublet.

\subsection{Doublet Flux Ratio and Escape Fraction}

In this paper, the Doppler parameter was assumed to be $b=90$ km
s$^{-1}$ based on observations of compact star-forming galaxies \citep{Chisholm2020,Henry2018}.
Regarding the line width, it is noteworthy that the double peaks in
the \ion{Mg}{2} emission line profiles have not been clearly detected
in most of the galaxies that exhibit the \ion{Mg}{2} emission. However,
the non-detection of double peaks does not necessarily imply their
absence. It was found, though not presented in this paper, that weak
double peaks in models with $N_{\text{Mg}^{+}}\lesssim2\times10^{14}$
cm$^{-2}$ disappeared after convolution with a Gaussian function
having a spectral resolution of $R=8000$ (equivalent to 37 km s$^{-1}$),
as configured for the observation of J1503 by \citet{Chisholm2020}.
Therefore, the absence or weakness of double peaks in most observations
of galaxies exhibiting the \ion{Mg}{2} emission implies that $N_{\text{Mg}^{+}}\lesssim2\times10^{14}$
cm$^{-2}$ ($\tau_{0}\lesssim5$) if the medium is close to being
static. The line broadening due to resonance scatterings is also insignificant
at such a relatively low optical depth. Thus, after considering the
instrumental line spread function, the estimated line width from observational
data would not significantly differ from the intrinsic width. The
results, therefore, indicate that the input line width of \ion{Mg}{2}
adopted in this paper is a reasonable choice.

In the study of ten green pea galaxies by \citet{Henry2018}, it was
observed that the \ion{Mg}{2} lines are systematically redshifted
by an average of 70 km s$^{-1}$, with line widths (FWHM) ranging
from 100 to 300 km s$^{-1}$. These galaxies were also found to have
a doublet flux ratio ranging from $\sim0.3$ to $\sim2.7$, with a
median of $\approx1.7$. There was no evidence of \ion{Mg}{2} extending
beyond the continuum. Similarly, \citet{Chisholm2020} also found
no spatially extended \ion{Mg}{2} emission beyond the continuum.
However, in contrast to \citet{Henry2018}, they found no strong line
profile asymmetries. The spatially resolved map of the doublet flux
ratio of J1503 by \citet{Chisholm2020} shows a rather patchy pattern.
Meanwhile, its Gaussian-smoothed image shows two blobs with a doublet
ratio of $R\approx1.8-2$, between which relatively lower ratios are
found.

The doublet flux ratio as low as $R\lesssim1.5$ and its spatial variation,
as observed in J1503, cannot simultaneously be explained by spherical
models. Doublet ratios of $R\approx1.8-2$ may be reproduced when
dust is included; however, ratios as low as $R\lesssim1.5$ cannot
be explained. In the spatially resolved radial profiles, such low
doublet flux ratios can be obtained in its outer regions if the medium
expands and has a high column density (Figure \ref{fig19}). However,
such high column densities and velocity redshifts are inconsistent
with the observed spectra. Another option is that if the continuum
and emission line sources coexist, and thus the continuum-pumped \ion{Mg}{2}
emission is combined with the `intrinsic' emission line, the doublet
flux ratio for emission may become much lower than two. However, in
static media with a low column density, the combined doublet ratio
would not be much different from those obtained from the emission
line model alone because the continuum-pumped emission feature is
very weak, as demonstrated in the upper left panel of Figure \ref{fig11}.

The present study presented models expanding with $V_{\text{exp}}=300$
km s$^{-1}$ or slower. An optically thick, expanding medium with a velocity of $V_{\text{exp}} = 300$ km s$^{-1}$ is capable of producing a line separation corresponding to that between the \ion{Mg}{2} K and H lines.
As shown in Figure \ref{fig10}, as
the expanding velocity increases, the redshifted K line (or blueshifted
H line in the case of a contracting medium) begins to overlap with
the H line (or the K line), altering the doublet flux ratio of the
continuum-pumped emission lines. This effect reduces the doublet flux
ratio even lower than one, as demonstrated in Figure \ref{fig13}
when a substantial amount of dust is present. However, such highly
expanding models are inconsistent with the observational data of J1503
\citep{Chisholm2020}, which shows no signature of velocity shifts.
Nevertheless, galaxies exhibiting asymmetric line profiles or absorption
features, as seen in the samples of \citet{Henry2018} and \citet{Xu2023},
could, at least qualitatively, be explained by combinations of the
models for the intrinsic emission lines and continuum.

Instead of simple geometries considered in this paper, more complicated
geometries may be necessary to explain the observational results.
For example, cylindrical models can yield such low doublet ratios
when a relatively flat medium is viewed edge-on (Figures \ref{fig05}
and \ref{fig06}). The present study assumed a simple cylindrical
shape. However, in reality, many different geometrical shapes and
densities may coexist. If an elongated or flat patch of the medium
is observed in a face-on-like direction, it will give doublet flux
ratios of $R\gtrsim2$ along that line of sight. In this context,
it should not be ruled out that the doublet flux ratios $R\gtrsim2$
found in \citet{Chisholm2020} might genuinely reflect the phenomenon
rather than arising from statistical fluctuations. They considered
values above 2 to be statistically consistent with the intrinsic value
of 2 at the $1\sigma$ significance level. Conversely, if it happens
to be oriented in a highly inclined (edge-on) direction, the line
of sight would result in doublet ratios much lower than 2. Even lower
values could arise when the resonantly-scattered continuum plays a
role in the doublet ratios.

\subsection{Foreground Screen vs. Radiative Transfer Effects}

As pointed out by \citet{Katz2022}, the escape fraction and doublet ratio maintain their intrinsic values within spherical configurations. It is also important to note that the escape fraction predicted in non-spherical, cylindrical media can exceed unity depending on the viewing angle.
This implies that the optical depth and escape fraction estimated
using the foreground screen model adopted by \citet{Chisholm2020}
could provide a misleading impression when estimating the actual escape
fraction of \ion{Mg}{2} in galaxies. In their model, a background
source is assumed to impinge upon a foreground screen of Mg$^{+}$
gas, with no consideration of scattered components directed toward
an observer. Given that resonantly scattered \ion{Mg}{2} emission
from different lines of sight contributes to the observed \ion{Mg}{2}
fluxes, the observationally estimated optical depth and escape fraction
should be considered as effective values that incorporate the scattered
flux. As demonstrated in the next section, \ion{Mg}{2} photons originating
from a spatially extended \ion{Mg}{2} source will experience relatively
weaker resonance effects than those from \ion{H}{2} regions. Thus,
in this case, using the foreground screen assumption will also lead
to a somewhat smaller amount of Mg$^{+}$ gas.

The situation is similar to distinguishing between `attenuation' and
`extinction' to understand the dust effects on spectral energy densities
or spectra of galaxies. As explained in \citet{Calzetti1994} and
\citet{Seon2016}, extinction refers to the disappearance of light
from a line of sight when observing a point-like source. In contrast,
attenuation refers to a situation where the spatially extended emission
source and scattering material are well mixed, and scattered light
partially compensates for extinction in a spatially extended system.
The same distinction should be applied when analyzing the \ion{Mg}{2}
emission lines. The optical depth estimated using the foreground screen
model is not an actual value but an effective one. The effective optical
depths estimated from the `attenuation' situations are always smaller
than the real ones. Systematic studies using complex, coherent, and
clumpy media, as demonstrated by \citet{Seon2016} in their investigation
of dust attenuation curves, may help disentangle the observed doublet
ratios and escape fractions of \ion{Mg}{2} from geometrical and RT
effects. Research on \ion{Mg}{2} RT, similar to the work of \citet{Seon2016}
for dust RT, is deferred to future studies.

\subsection{\ion{Mg}{2} Emission Mechanisms and Sites}

In theory, there are two intrinsic mechanisms that can create \ion{Mg}{2}
K and H emission lines: (1) recombination of Mg$^{+2}$ and (2) collisional
excitation of Mg$^{+}$ in the ground state, followed by radiative
decay. Photoionization of Mg$^{+}$ atoms requires an energy of 15.035
eV or higher ($\lambda\lesssim824.64$\AA). Therefore, doubly ionized
Mg$^{+2}$ atoms will only be present near the central star(s) in
\ion{H}{2} regions. Consequently, the \ion{Mg}{2} recombination
line is expected to be produced primarily in the central part of \ion{H}{2}
regions, and the total luminosity of the \ion{Mg}{2} recombination
line is likely negligible due to the relatively small volume occupied
by Mg$^{+2}$ gas. The majority of \ion{Mg}{2} will be produced through
collisional excitation of Mg$^{+}$ followed by radiative decay.

Indeed, the photoionization code Cloudy, which is last described in
\citet{Chatzikos2023}, predicts only the collisionally excited \ion{Mg}{2}
emission line. \citet{Erb2012} and \citet{Jaskot2016} calculated
photoionization models for the \ion{Mg}{2} K and H emission lines
originating from \ion{H}{2} regions. Their findings revealed that
a significant amount of \ion{Mg}{2} line is emitted from \ion{H}{2}
regions. \citet{Nelson2021} and \citet{Katz2022} utilized the Cloudy
code to calculate \ion{Mg}{2} line emissions originating from \ion{H}{2}
regions in galaxies in cosmological simulations. However, the \ion{Mg}{2}
emission calculated using the Cloudy code originates predominantly
from the transition region between the fully ionized \ion{H}{2} region
and the neutral outer region. \ion{Mg}{2} emission is produced in
the boundary region where the photoionizing radiation field with energies
$E\gtrsim$ 15 keV ($\lambda\lesssim825$\AA) is fully attenuated,
and the gas temperature is high enough to excite Mg$^{+}$ atoms collisionally.
Similarly, {[}\ion{S}{2}{]} $\lambda$6716 and {[}\ion{N}{2}{]}
$\lambda6583$ lines are also mainly emitted from the boundary \citep[e.g.,][]{Seon2012}.
Detailed studies of this property of \ion{Mg}{2} are beyond the scope
of this paper and will be presented elsewhere.

Additionally, it should be noted that \ion{Mg}{2} can also be created
in the diffuse WNM with a temperature of $\sim10^{4}$ K, which has
often been overlooked in the literature. The diffuse far-ultraviolet
(FUV) continuum radiation field at $\lambda\sim1620$\AA, which is
composed of direct starlight (and the radiation from AGN if present)
and its dust-scattered component, can produce Mg$^{+}$ gas. The diffuse
FUV radiation field at $\lambda\sim1620$\AA\ in the neutral ISM
will singly ionize Mg atoms because the ionization energies of Mg$^{0}$
and Mg$^{+}$ are 7.646 eV and 15.035 eV (corresponding to $\sim8.6\times10^{4}$
K and $\sim1.7\times10^{5}$ K), respectively. As a result, unless
the stellar FUV radiation is significantly attenuated by dust, Mg$^{+}$
is expected to be the predominant form of Mg in both the cold neutral
medium (CNM) and WNM. Collisions with electrons with temperatures
of $\approx10^{4}$ K will excite the Mg$^{+}$ ions in the WNM, and
subsequent radiative decay to the ground state will produce \ion{Mg}{2}
emission. Therefore, it may be essential to consider the diffuse \ion{Mg}{2}
emission, which is not directly associated with \ion{H}{2} regions.
\ion{Mg}{2} emission from \ion{H}{2} regions would be confined to
relatively compact volumes, whereas emission from the WNM will be
distributed widely throughout galaxies.

As a result, the observed \ion{Mg}{2} emission lines in galaxies
(including the CGMs) would arise from a combination of three components:
(1) \ion{Mg}{2} originating from the outer boundaries of \ion{H}{2}
regions, where the fully ionized region meets with the ambient CNM
or WNM, (2) \ion{Mg}{2} originating from the diffuse WNM, which widely
spreads throughout and around galaxies, and (3) \ion{Mg}{2} emission
pumped by the resonance scattering of the continuum radiation. The
relative importance between \ion{H}{2} regions and the WNM will depend
on their total luminosity. The luminosity of the \ion{Mg}{2} line
is proportional to a product of the emissivity, the volume of the
emission site, and the densities of Mg$^{+}$ and electrons. \ion{H}{2}
regions occupy a relatively small volume but have high density, while
the WNM occupies a relatively large volume with low density. Therefore,
understanding the factors that determine the relative importance of
these mechanisms and how they are interconnected becomes essential.

The extended \ion{Mg}{2} line emission has been observed in the CGMs of star-forming galaxies and in intragroup medium \citep{Zabl2021,Burchett2021,Leclercq2022,Dutta2023}. These detections indicate no direct association of \ion{Mg}{2} emission line with \ion{H}{2} regions. On the other hands,
\citet{Henry2018} and \citet{Chisholm2020} found no evidence of \ion{Mg}{2} extending beyond the stellar continuum in their observations of star-forming galaxies. These observations do not necessarily indicate that most \ion{Mg}{2} emission is directly associated with \ion{H}{2} regions in these galaxies. The diffuse Mg$^{+}$ gas is also likely to be dominantly produced by bright young stars, which emit most of the FUV continuum radiation at $\lambda\lesssim1620$\AA. Therefore, \ion{Mg}{2} from the WNM can have a similar spatial extent to the stellar continuum.
\begin{figure}[t]
\begin{centering}
\includegraphics[clip,scale=0.5]{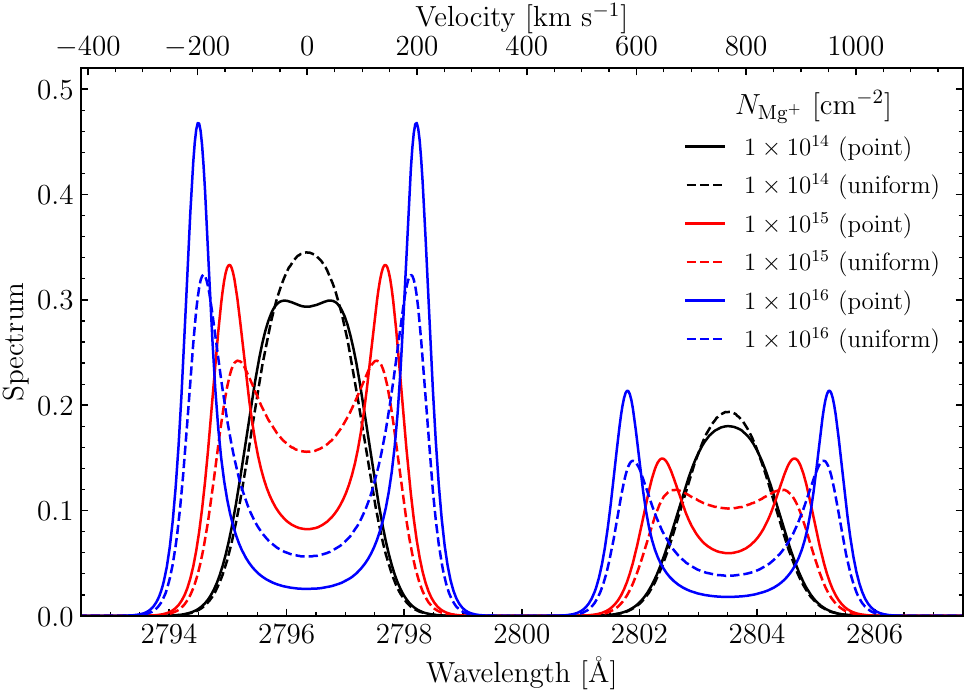}
\par\end{centering}
\begin{centering}
\medskip{}
\par\end{centering}
\caption{\label{fig21} Comparison of the central point source model and the
uniform source model. The solid lines represent the spectra obtained
from the central point source models in a static, spherical medium,
while the dashed lines represent the spectra obtained from the uniformly
distributed source models.}
\end{figure}

If \ion{Mg}{2} originates from the Mg$^{+}$ gas surrounding fully-ionized
\ion{H}{2} regions, the \ion{Mg}{2} photons will experience a relatively
large number of resonance scatterings both in the \ion{H}{2} regions
and WNM. The continuum UV radiation near the wavelengths of \ion{Mg}{2},
which will be mostly emitted from bright young stars, will also undergo
a similarly large number of resonance scatterings and thus produce
\ion{Mg}{2} absorption, emission, or both, depending on the geometry
and kinematics of the gas. In contrast, in a configuration where \ion{Mg}{2}
is produced in the diffuse WNM, the resonance scattering effect would
be relatively weak compared to the case of compact sources with the
same amount of Mg$^{+}$ gas. Figure \ref{fig21} compares the spectra
predicted from the models where the \ion{Mg}{2} source and Mg$^{+}$
gas are well mixed with those obtained from the models with a central
source. Both model types assume a spherical medium. The figure clearly
demonstrates that the spatially extended source (dashed lines) yields
weaker resonance-scattering signatures than the point
source (solid lines). This difference is attributed to the geometrical
effects that the optical depth measured from an outer radius is smaller
than that from the center.

The fact that galaxies with strong \ion{Mg}{2} emission tend to exhibit
bluer UV spectral slopes compared to those showing absorption \citep{Finley2017,Feltre2018}
suggests that the WNM may play a significant role in generating the
\ion{Mg}{2} emission line in these galaxies. This inference is drawn
from the correlation between bluer UV spectral slopes and increased
UV radiation around $\lambda\sim1620$\AA. In this case, \ion{Mg}{2}
lines in these galaxies are likely to experience relatively weaker
resonance effects. Conversely, in galaxies with redder UV slopes,
the contribution of the WNM may be relatively minor. In such cases,
\ion{Mg}{2} emission from \ion{H}{2} regions and the stellar continuum
near \ion{Mg}{2} wavelengths becomes more prominent. The \ion{Mg}{2}
emission line and continuum photons will undergo somewhat stronger
resonance effects. The resonantly scattered continuum can then give
rise to absorption lines, potentially dominating the emission from
\ion{H}{2} regions.

\subsection{Alkali-Metal-Like Resonance Lines}

Recently, resonance doublet lines from alkali metal-like ions, whose
ground states consist of a single `s' electron outside a closed shell,
have gained interest in the literature and have been observed in high-redshift
galaxies. This is because star-forming galaxies are believed to have
been the primary source of the reionization of the IGM at high redshifts.
It is difficult to measure the LyC directly, so researchers have attempted
to find indirect tracers of LyC instead. For this purpose, it has
been suggested that the \ion{C}{4} $\lambda\lambda1548,\:1551$ doublet
could potentially be a valuable indicator of LyC photon escape at
low redshift ($z\sim0.3-0.4$; \citealp{Schaerer2022}) and high-redshift
($z>6$; \citealp{Saxena2022}). \citet{Berg2019} also reported the
observations of intense nebular \ion{He}{2} and double-peaked \ion{C}{4}
emission from two galaxies at $z\sim5-7$. These results suggest that
in these galaxies, a significant fraction of high-energy LyC photons
can escape through paths of highly ionized gas with low column densities.

In the context of probing the LyC escape, \citet{Henry2018} found
a tight correlation between the Ly$\alpha$ escape fraction in local
compact star-forming galaxies and the \ion{Mg}{2} escape fraction.
They used one-dimensional photoionization models to find the correlation
between the intrinsic \ion{Mg}{2} emission line flux and oxygen lines
[\ion{O}{3}] 5007\AA\ and [\ion{O}{2}] 3727\AA, often used to
trace LyC leakage. \citet{Guseva2020} detected strong resonant \ion{Mg}{2}
emission lines and non-resonant, fluorescence \ion{Fe}{2}$^{*}$
$\lambda\lambda$2612, 2626 emission lines in the spectra of LyC leakers
at $z\sim0.3-0.4$. These results suggest that \ion{Mg}{2} emission
lines may be a helpful indicator of escaping Ly$\alpha$ and LyC emission.

Therefore, it is suggestive that both \ion{C}{4} and \ion{Mg}{2}
lines can help in understanding the porosity condition of the ISM
and CGM, through which LyC photons escape, in high-redshift galaxies.
However, it should be noted that \ion{C}{4} is unlikely to originate
from ordinary \ion{H}{2} regions or the WNM, as the ionization potential
of C$^{+3}$ is 47.89 eV, which is much higher than that of a He$^{0}$
atom. The hot gas with a temperature $\gtrsim10^{5}$ K from which
\ion{C}{4} originates would primarily be produced by supernova shocks
and/or by the hard radiation from X-ray binaries (and AGNs if present).
Other high-ionization doublet lines, such as \ion{O}{6} $\lambda\lambda$1032,
1038 and \ion{N}{6} $\lambda\lambda$1239, 1243, would also serve
as valuable tracers of low-density, high-temperature gas phases. These
high-ionization lines have been studied both theoretically and observational
in the Milky Way \citep[e.g.,][]{Shelton2018,Jo2019}. In particular,
\citet{Jo2019} have made a sky survey map of the \ion{C}{4} emission
line and found that the hot gas in our Galaxy has a scale height of
$\sim$ kpc. However, their analysis did not consider the resonance
scattering of \ion{C}{4}. Although our Galaxy might differ from compact
star-forming galaxies in high redshift, a detailed understanding of
the map would help us understand the nature of hot gas in high redshift
galaxies.

In contrast to the high-ionization lines, \ion{Mg}{2} traces a relatively
warm and neutral gas phase. The typical Doppler parameter $b$ of
\ion{Mg}{2} absorption systems is found to be $\approx5$ km s$^{-1}$,
constraining that the typical temperature of Mg$^{+}$ gas to be $\approx30,000$
K or less \citep[e.g.,][]{Rigby2002,Ding2005,Churchill2020}. The substantial difference in the line widths between the individual absorption systems ($\sim 5$ km s$^{-1}$) and compact star-forming galaxies ($\sim 90$ km s$^{-1}$) seems to indicate that the line width in the compact galaxies is predominantly influenced by relatively large-scale gas motion, whereas that of the absorption systems arises from relatively local, small-scale motion within individual clouds.
A clear understanding of the formation mechanisms and physical properties
of these diffuse warm and hot gases on a galactic scale may be necessary.
In particular, their volume-filling fractions are crucial factors
determining the escape of LyC photons from galaxies.

\section{SUMMARY}

\label{sec:5}

This paper investigated the RT of \ion{Mg}{2} doublet lines in two
simple geometries (sphere and cylinder), providing valuable insights
for interpreting observational data. Future research is expected to
develop models that adopt more complex and clumpy media. The main
results of this paper are summarized as follows:
\begin{itemize}
\item In spherical models without dust, the doublet flux ratio and escape
fraction of \ion{Mg}{2} are always 2 and 1, respectively.
\item When studying resonance doublet emission lines, it has been generally
assumed that the flux ratio of doublet from optically thick media
is always lower than their optically thin value (e.g., $F_{2796}/F_{2803}\le2$).
Therefore, doublet ratios lower than two have been considered evidence
of resonance scattering and the existence of an optically thick medium
near or surrounding the emitting gas. However, in cylindrical models,
the doublet flux ratio can also be much higher than the intrinsic
value 2 when flat media are viewed face-on. Additionally, the escape
fraction can be larger than 1 when observed face-on. In contrast,
when a geometrically and optically thin disk is viewed edge-on, the
doublet ratio can be as low as $\sim$1.2.
\item When dust is included, the doublet flux ratio and escape faction are
reduced; however, the dust effects become noticeable only when the column
density of Mg$^{+}$ is $N_{\text{Mg}^{+}}\gtrsim10^{15}$ cm$^{-2}$
($\tau_{0}\gtrsim28$), corresponding to the dust extinction optical
depth of $\tau_{{\rm dust}}\gtrsim0.06$, in static media, or when the dust-to-Mg$^{+2}$ gas ratio is substantially higher than that in the Milky Way.
\item The EWs of the absorption and emission lines resulting from resonance
scattering of the stellar continuum can be qualitatively interpreted
using the curve of growth theory for pure absorption lines. The absorption
and emission features somewhat match and compensate in low column-density
media but are separated in high column-density and expanding media.
In the presence of dust, it is found that the EW of continuum-pumped
\ion{Mg}{2} emission is significantly reduced compared to that of
absorption. However, the reduction in the emission line is less than
what would be expected if one ignores the dust scattering effect of
the continuum.
\item It is important to note that, in cylindrical models, pure absorption
and pure emission spectra due to the stellar continuum can emerge
depending on the viewing angle. In an edge-on view, the spectra show
pure absorption, while a face-on view gives rise to pure emission
spectra.
\end{itemize}
The spatial variations in the surface brightness and doublet ratio of the \ion{Mg}{2} halo due to a central \ion{Mg}{2} emission line source or a continuum source were also investigated, and the results are summarized as follows:
\begin{itemize}
\item The radial surface brightness profile of the \ion{Mg}{2} halo is, in general, steeper in expanding media than in static cases, regardless of the source type.
\item The doublet flux ratio of the \ion{Mg}{2} halos shows rather complex profiles. However, the ratio generally increases gradually with radius when the column density is low ($N_{\text{Mg}^+} \lesssim 10^{15}$ cm$^{-2}$), but tends to decrease slightly at large radii when the column density is high ($N_{\text{Mg}^+} \gtrsim 10^{15}$ cm$^{-2}$). For the intrinsic \ion{Mg}{2} source, the doublet ratio can be higher than 2, reaching up to 3 in a rapidly expanding, low-density medium. However, the doublet ratio typically falls within the range of $\approx 0.75-1.75$ and does not exceed 2 in the case of a continuum source.
\item In actual observations, aperture effects (such as size and offset from the galaxy center, etc.) may significantly affect the measured radial profiles of the doublet ratio, reducing its spatial variation.
\end{itemize}
The following summarizes the observational implications of the results
and the related topics discussed in this paper.
\begin{itemize}
\item The doublet flux ratios of \ion{Mg}{2}, as low as observed in star-forming
galaxies showing \ion{Mg}{2} emission lines but no signatures of
velocity shifts and double peaks, cannot be accounted for by spherically
symmetric models, whether or not dust is included when considering
only the RT of the intrinsic \ion{Mg}{2} emission line.
\item Instead of spherical models, they may be reasonably well explained
when the galaxies are geometrically thin disks viewed highly inclined
($\beta_{{\rm inc}}\gtrsim80^{\circ}$) or contain large and relatively
flat Mg$^{+2}$ gas clouds situated edge-on. The continuum-pumped
emission line will also be necessary to explain various spectral shapes
and doublet flux ratios.
\item It is discussed that \ion{Mg}{2} emission originating from the diffuse
WNM may be important when the UV spectral slope of a galaxy is relatively
blue.
\item It is also pointed out that the optical depth derived using the foreground
screen model should be regarded as effective rather than the actual
value.
\end{itemize}
\begin{acknowledgements}
This work was partially supported by a National Research Foundation
of Korea (NRF) grant funded by the Korea government (MSIT) (No. 2020R1A2C1005788)
and by the Korea Astronomy and Space Science Institute grant funded
by the Korea government (MSIT; No.\ 2024183100 and 2024186900).
\end{acknowledgements}

\end{document}